\documentclass[10pt,twocolumn]{article}
\pdfoutput=1
\usepackage{authblk}
\usepackage{abstract}
\usepackage[footnotesize]{caption}
\usepackage{subcaption}
\usepackage{cite}
\usepackage{amsmath}
\usepackage{amssymb}
\usepackage{graphicx} \graphicspath{ {images/} }
\usepackage[latin1]{inputenc}
\usepackage{mathrsfs}
\usepackage[dvipsnames]{xcolor}
\usepackage{hyperref}
\usepackage[hmarginratio=1:1,top=32mm,columnsep=20pt]{geometry}
\usepackage{paralist}
\usepackage{multirow}

\newcommand{\bea}{\begin{eqnarray}}
\newcommand{\eea}{\end{eqnarray}}
\newcommand{\be}{\begin{equation}}
\newcommand{\ee}{\end{equation}}
\newcommand{\ba}{\begin{array}}
\newcommand{\ea}{\end{array}}

\def\gsim{\mathrel{\rlap{\lower4pt\hbox{\hskip1pt$\sim$}}
    \raise1pt\hbox{$>$}}}

\renewcommand\Affilfont{\normalsize\itshape}

\title{\fontsize{16pt}{10pt}\selectfont
		\textbf{Sterile neutrino searches at future $e^-e^+$, $pp$, and $e^-p$ colliders}}
\author[1,2]{Stefan~Antusch\thanks{E-mail: \texttt{stefan.antusch@unibas.ch}}}
\author[1]{Eros~Cazzato\thanks{E-mail: \texttt{e.cazzato@unibas.ch}}}
\author[1]{Oliver~Fischer\thanks{E-mail: \texttt{oliver.fischer@unibas.ch}}}

\affil[1]{\Affilfont Department of Physics, University of Basel, \authorcr 
 		  \Affilfont Klingelbergstr.\ 82, CH-4056 Basel, Switzerland \authorcr\mbox{}}
\affil[2]{\Affilfont Max-Planck-Institut f\"ur Physik (Werner-Heisenberg-Institut),\authorcr
		  \Affilfont F\"ohringer Ring 6, D-80805 M\"unchen, Germany}

\date{}

\begin{document}

\twocolumn
[

\maketitle

\setlength{\absleftindent}{50pt}
\setlength{\absrightindent}{50pt}
\vspace{-15pt}
\begin{onecolabstract}

\noindent Sterile neutrinos are among the most attractive extensions of the SM to generate the light neutrino masses observed in neutrino oscillation experiments. 
When the sterile neutrinos are subject to a protective symmetry, they can have masses around the electroweak scale and potentially large neutrino Yukawa couplings, which makes them testable at planned future particle colliders.
We systematically discuss the production and decay channels at electron-positron, proton-proton and electron-proton colliders and provide a complete list of the leading order signatures for sterile neutrino searches. Among other things, we discuss several novel search channels, and present a first look at the possible sensitivities for the active-sterile mixings and the heavy neutrino masses. We compare the performance of the different collider types and discuss their complementarity.

\end{onecolabstract}
\hrulefill
\vspace{1cm}
]

\saythanks

\noindent

\section{Introduction}
The main physics goal of future particle colliders is to test the current Standard Model (SM) of elementary particles and to search for effects of new physics, i.e.\ new particles and new interactions not present in the SM. The existence of such new physics is guaranteed by various observations from particle physics as well as from cosmology, which can not be explained within the SM (and the present cosmological model based on general relativity as the theory of gravity). 

The discovery of the/a Higgs boson by the ATLAS \cite{Aad:2012tfa} and CMS \cite{Chatrchyan:2012xdj} experiments at the LHC plays an important r\^ole for the search strategies for new physics at future colliders. To start with, although the discovered Higgs boson is compatible with the one of the SM, its properties are not really well tested to date. This leaves a lot of room for deviations of the Higgs properties from the SM predictions. Indeed, various extensions of the SM predict a non-minimal Higgs sector. 
Furthermore, even for a minimal Higgs sector important questions regarding the Higgs boson, related to its ``task'' of giving mass to elementary particles, are currently unanswered. For example, it is not tested that the Higgs is indeed generating its own mass, which requires the measurement of the Higgs self-coupling. 

Another puzzle related to elementary particle masses and the Higgs sector is the origin of neutrino masses. Generating masses for the light neutrinos would require either an extended Higgs sector or the addition of extra neutral fermions (which are gauge singlets and therefore often referred to as ``sterile'' neutrinos) plus a coupling of the neutrinos to the discovered Higgs boson. A precision study of Higgs properties at future colliders may thus shed light on the question how to extend the SM in order to include the observed neutrino masses.  
     
In addition to their interactions with the Higgs boson, the sterile neutrinos, or, more precisely, the heavy neutrino mass eigenstates, which are an admixture of the active neutrinos of the SM and the extra sterile states, also interact with the weak gauge bosons. This leads to various production and decay channels, and corresponding signatures at future electron-positron ($e^-e^+$), proton-proton ($pp$) and electron-proton ($e^- p$) colliders. 

The aim of this article is to provide a systematic discussion and assessment of these signatures for sterile neutrino searches. We discuss a list of the leading order signatures, comprising the production and decay channels and their dependencies on the active-sterile neutrino mixing parameters, lepton-number-violating (LNV) and lepton-flavour-violating (LFV) effects, vertex displacement and non-unitarity effects. We discuss several novel search channels, for which we present ``first looks'' at the sensitivities for the active-sterile mixing parameters and sterile neutrino masses, as well as updated sensitivity estimates. We summarize the estimated sensitivites for the FCC-ee, CEPC, HL-LHC, FCC-hh/SppC, LHeC and FCC-eh and compare them for the different collider types.

\begin{figure}
\begin{center}
\includegraphics[width=0.45\textwidth]{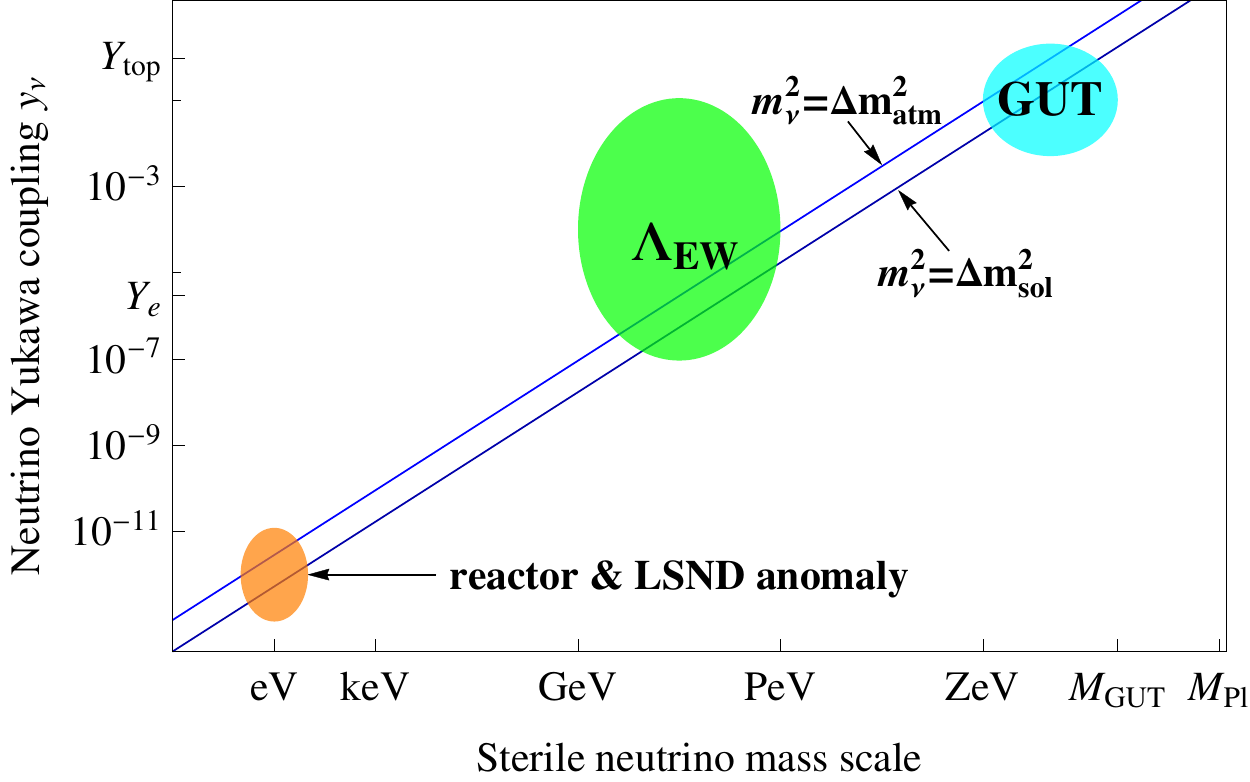}
\caption{Sketch of the landscape of sterile neutrino extensions of the SM. EW scale neutrino models with a protective ``lepton number''-like symmetry, such as the used SPSS benchmark model \cite{Antusch:2015mia}, can have sterile neutrino masses in the relevant range for particle collider experiments, shown by the green area, with Yukawa couplings above the na\"ive expectation, which is denoted by the blue lines.}
\label{fig:landscape}
\end{center}
\end{figure}

For the sensitivity estimates we consider low scale seesaw scenarios with a protective ``lepton number''-like symmetry, using the Symmetry Protected Seesaw Scenario (SPSS) as benchmark model (cf.\ section \ref{sec:SPSS}), where the masses of the sterile states can be around the electroweak scale (cf.\ fig.\ \ref{fig:landscape}). 

\section{Theoretical framework}
Mass terms for SM neutrino masses can be introduced when right-handed (i.e.\ sterile) neutrinos are added to the field content of the SM. These sterile neutrinos are singlets under the gauge symmetries of the SM, which means they can have a direct (so-called Majorana) mass term, that involves exclusively the sterile neutrinos, as well as Yukawa couplings to the three active (SM) neutrinos contained in the SU(2)$_\mathrm{L}$-lepton doublets and the Higgs doublet. 

In the simplistic case of only one active and one sterile neutrino, with a large mass $M$ and a Yukawa coupling $y$ such that $M \gg y_\nu\, v_\mathrm{EW}$, where $v_\mathrm{EW}$ denotes the vacuum expectation value (vev) of the neutral component of the Higgs SU(2)$_L$-doublet, the mass of the light neutrino $m$ is given by $m \approx y_\nu^2\, v^2_\mathrm{EW}/M$, while the heavy state has a mass $\sim M$. The prospects for observing such a sterile neutrino at colliders are not very promising, since in order to explain the small mass of the light neutrinos (below, say, $0.2$ eV), the mass of the heavy state would either have to be of the order of the Grand Unification (GUT) scale, for a Yukawa coupling of ${\cal O}(1)$, or the Yukawa coupling would have to be tiny and the active-sterile mixing would be highly suppressed.
 
However, in the realistic case of three active neutrinos and two\footnote{With two mass differences observed in oscillations of the light neutrinos, at least two sterile neutrinos are required to give mass to at least two of the active neutrinos.}  or more sterile neutrinos, the simple relation from above no longer holds and the possible values of the masses of the sterile neutrinos and the Yukawa couplings have to be reconsidered. In particular, if the theory comprises for instance an approximate ``lepton number''-like symmetry or a suitable discrete symmetry, one finds that sterile neutrinos with masses around the electroweak (EW) scale and unsuppressed (up to ${\cal O}$(1)) Yukawa couplings are theoretically allowed, and due to the protective ``lepton number''-like symmetry the scenario is stable under radiative corrections. 

This scenario has the attractive features that the new physics scale lies not (much) above the EW scale -- which avoids an explicit hierarchy problem -- and that no unmotivated tiny couplings have to be introduced. Various models of this type are known in the literature (see e.g.\ \cite{Wyler:1982dd,Mohapatra:1986bd,Shaposhnikov:2006nn,Kersten:2007vk,Gavela:2009cd,Malinsky:2005bi}). One example is the so-called ``inverse seesaw'' \cite{Wyler:1982dd,Mohapatra:1986bd}, where the relation between the masses of the light and sterile neutrinos are schematically given by $m \,\approx \,\epsilon \,y_\nu^2 v^2_\mathrm{EW}/M^2$, where $\epsilon$ is a small quantity that parametrizes the breaking of the protective symmetry. 
As $\epsilon$ controls the magnitude of the light neutrino masses, the coupling $y_\nu$ can in principle be large for any given $M$.

\subsection{Sterile neutrinos with EW scale masses}\label{sec:SPSS}
The relevant features of seesaw models with such a protective ``lepton number''-like symmetry were for instance discussed in refs.\ \cite{Wyler:1982dd,Mohapatra:1986bd,Shaposhnikov:2006nn,Kersten:2007vk,Gavela:2009cd,Malinsky:2005bi}), and may be represented by the benchmark model that was introduced in \cite{Antusch:2015mia}, referred to as the Symmetry Protected Seesaw Scenario (SPSS) in the following.
The Lagrangian density of the SPSS, considering a pair of sterile neutrinos $N_R^1$ and $N_R^2$, is given in the symmetric limit ($\epsilon = 0$) by
\be
\mathscr{L} = \mathscr{L}_\mathrm{SM} -  \overline{N_R^1} M N^{2\,c}_R - y_{\nu_{\alpha}}\overline{N_{R}^1} \widetilde \phi^\dagger \, L^\alpha
+\mathrm{H.c.} 
+ \dots  \;,
\label{eq:lagrange}
\ee
where $\mathscr{L}_\mathrm{SM}$ contains the usual SM field content and with $L^\alpha$, $(\alpha=e,\mu,\tau)$, and $\phi$ being the lepton and Higgs doublets, respectively. The dots indicate possible terms for additional sterile neutrinos, which we explicitly allow for provided that their mixings with the other neutrinos are negligible, or that their masses are very large, such that their effects are irrelevant for collider searches. The $y_{\nu_{\alpha}}$ are the complex-valued neutrino Yukawa couplings, and the mass $M$ can be chosen real without loss of generality. 

As explained above, masses for the light neutrinos are generated when the protective symmetry gets broken. 
In this rather general  framework, the neutrino Yukawa couplings $y_{\nu_{\alpha}}$ and the sterile neutrino mass scale $M$ are essentially free parameters, and $M$ can well be around the EW scale.\footnote{In specific models there are correlations among the $y_{\nu_{\alpha}}$. The strategy of the SPSS is to study how to measure the $y_{\nu_{\alpha}}$ independently, in order to test (not a priori assume) such correlations.   
}

From eq.~(\ref{eq:lagrange}), we obtain the mass matrix ${\cal M}$ of the relevant neutral fermions, i.e.\ the active neutrinos and the two sterile neutrinos $N_R^1$ and $N_R^2$, after EW symmetry breaking. It can be diagonalised with the unitary 5 $\times$ 5 leptonic mixing matrix $U$:
\be
 U^T\, {\cal M}\, U \cong \text{Diag}\left(0,0,0,M,M\right) 
\,.
\label{eq:diagonalisation}
\ee
The resulting mass eigenstates are the three light neutrinos $\nu_i$ $(i=1,2,3)$, which are massless in the symmetric limit, and two heavy neutrinos $N_j$ $(j=1,2)$ with approximately degenerate mass eigenvalues $M$ (in the symmetric limit).
The mixing of the active and sterile neutrinos can be quantified by the mixing angles and their magnitude:
\be
\theta_\alpha = \frac{y_{\nu_\alpha}^{*}}{\sqrt{2}}\frac{v_\mathrm{EW}}{M}\,, \qquad |\theta|^2 := \sum_{\alpha} |\theta_\alpha|^2\,,
\label{def:thetaa}
\ee
with  $v_\mathrm{EW} = 246.22$ GeV. 
Using the mixing angles $\theta_\alpha$ we can express the leptonic mixing matrix $U$ in eq.\ \eqref{eq:diagonalisation}, in the limit of exact symmetry, as (cf.\ \cite{Antusch:2015mia}):
\be
U = \left(\begin{array}{ccccc} 
{\cal N}_{e1}	& {\cal N}_{e2}	& {\cal N}_{e3}	& - \frac{\mathrm{i}}{\sqrt{2}}\, \theta_e & \frac{1}{\sqrt{2}} \theta_e 	\\ 
{\cal N}_{\mu 1}	& {\cal N}_{\mu 2}  	& {\cal N}_{\mu 3}  	& - \frac{\mathrm{i}}{\sqrt{2}}\theta_\mu & \frac{1}{\sqrt{2}} \theta_\mu  \\
{\cal N}_{\tau 1}	& {\cal N}_{\tau 2} 	& {\cal N}_{\tau 3} 	& - \frac{\mathrm{i}}{\sqrt{2}} \theta_\tau & \frac{1}{\sqrt{2}} \theta_\tau \\  
0	   	& 0		& 0	&  \frac{ \mathrm{i}}{\sqrt{2}} & \frac{1}{\sqrt{2}}\\
-\theta^{*}_e	   	& -\theta^{*}_\mu	& -\theta^{*}_\tau &\frac{-\mathrm{i}}{\sqrt{2}}(1-\tfrac{1}{2}\theta^2) & \frac{1}{\sqrt{2}}(1-\tfrac{1}{2}\theta^2)
\end{array}\right)\,.
\label{eq:mixingmatrix}
\ee
The leptonic mixing matrix $U$ is unitary up to second order in $\theta_\alpha$. The elements of the non-unitary $3\times 3$ submatrix ${\cal N}$, which is the effective mixing matrix of the three active neutrinos, i.e.\ the Pontecorvo--Maki--Nakagawa--Sakata (PMNS) matrix relevant for neutrino oscillation experiments, are given as 
\be
{\cal N}_{\alpha i} = (\delta_{\alpha \beta} - \tfrac{1}{2} \theta_{\alpha}\theta_{\beta}^*)\,(U_\ell)_{\beta i}\,,
\label{eq:matrixN}
\ee
with $U_\ell$ being a unitary $3 \times 3$ matrix. 

When the Higgs boson develops its vacuum expectation value the light and heavy neutrino mass eigenstates emerge as admixtures of the active and sterile neutrinos. The weak currents in the mass basis are given by
\bea
j_\mu^\pm & = & \sum\limits_{i=1}^5 \sum\limits_{\alpha=e,\mu,\tau}\frac{g}{\sqrt{2}} \bar \ell_\alpha\, \gamma_\mu\, P_L\, U_{\alpha i}\, \tilde n_i\, + \text{ H.c.}\,, \\
j_\mu^0 & = & \sum\limits_{i,j=1}^5 \sum\limits_{\alpha=e,\mu,\tau}\frac{g}{2\,c_W} \overline{\tilde n_j}\, U^\dagger_{j\alpha}\, \gamma_\mu\, P_L\, U_{\alpha i}\, \tilde n_i\,, 
\label{eq:weakcurrentmass}
\eea
with $U$ the leptonic mixing matrix in eq.\ \eqref{eq:mixingmatrix}, $g$ being the weak coupling constant, $c_W$ the cosine of the Weinberg angle and $P_L = {1 \over 2}(1-\gamma^5)$ the left-chiral projection operator, and where we introduced the neutrino  mass eigenstates
\be
\tilde n_j = \left(\nu_1,\nu_2,\nu_3,N_4,N_5\right)^T_j = U_{j \alpha}^{\dagger} n_\alpha\,,
\ee
with the definition
\be
n = \left(\nu_{e_L},\nu_{\mu_L},\nu_{\tau_L},(N_R^1)^c,(N_R^2)^c\right)^T\,.
\ee
The weak currents involving the heavy neutrinos can be expressed as
\bea
j_\mu^\pm & \supset &  \frac{g}{2} \, \theta_\alpha \, \bar \ell_\alpha \, \gamma_\mu P_L \left(-\mathrm{i} N_4 + N_5 \right) + \text{H.c.} \,, \label{eq:weakcurrent1}\\
j_\mu^0 & = & \frac{g}{2\,c_W} \sum\limits_{i,j=1}^5 \vartheta_{ij} \overline{ \tilde n_i} \gamma_\mu P_L \tilde n_j\,,
\label{eq:weakcurrent2}
\eea
with the definition:
\be
\vartheta_{ij} = \sum_{\alpha=e,\mu,\tau} U^\dagger_{i\alpha}U_{\alpha j}^{}\,.
\ee
The Yukawa part of the Lagrangian density in the mass basis to leading order in the active-sterile mixing angle, is
\begin{align}
\frac{M}{v_\mathrm{EW}} \sum\limits_{i=1}^3 \left(\vartheta_{i4}^* \overline{N_4^c}+ \vartheta_{i5}^*\overline{N^c_5}\right) h\, \nu_i +\text{ H.c.}  \,,
\label{eq:Lykawa}
\end{align}
with $h = \sqrt{2} \,\mbox{Re}{(\phi^0)}$ being the real scalar Higgs boson.

In the limit of exact symmetry, the SPSS benchmark model  introduces seven additional parameters to the theory, the moduli of the neutrino Yukawa couplings ($|y_{\nu_e}|$, $|y_{\nu_\mu}|$, $|y_{\nu_\tau}|$), their respective phase, and the mass $M$. The phases are difficult to measure at colliders. They may be accessible in neutrino oscillation experiments (see e.g.\ \cite{FernandezMartinez:2007ms,Antusch:2009pm}). In the following we will restrict ourselves to the four parameters $|y_{\nu_e}|$, $|y_{\nu_\mu}|$, $|y_{\nu_\tau}|$ and $M$.

\subsection{Heavy neutrino production and decay in $e^-e^+, pp$ and $e^-p$ collisions}
\label{sec:production-and-decay}
In this section we discuss the dominant production channels of the heavy neutrino mass eigenstates in $e^-e^+,$ $pp$ and $e^-p$ collisions, and their subsequent decays at the leading order. In this line, we address the dependency on active-sterile mixing angles for the different processes, and we comment on the occurrence of observable lepton number violating (LNV) and lepton flavour violating (LFV) effects.

\subsubsection{Production processes}
\label{sec:production processes}
\begin{figure*}[t]
\includegraphics[width=\textwidth]{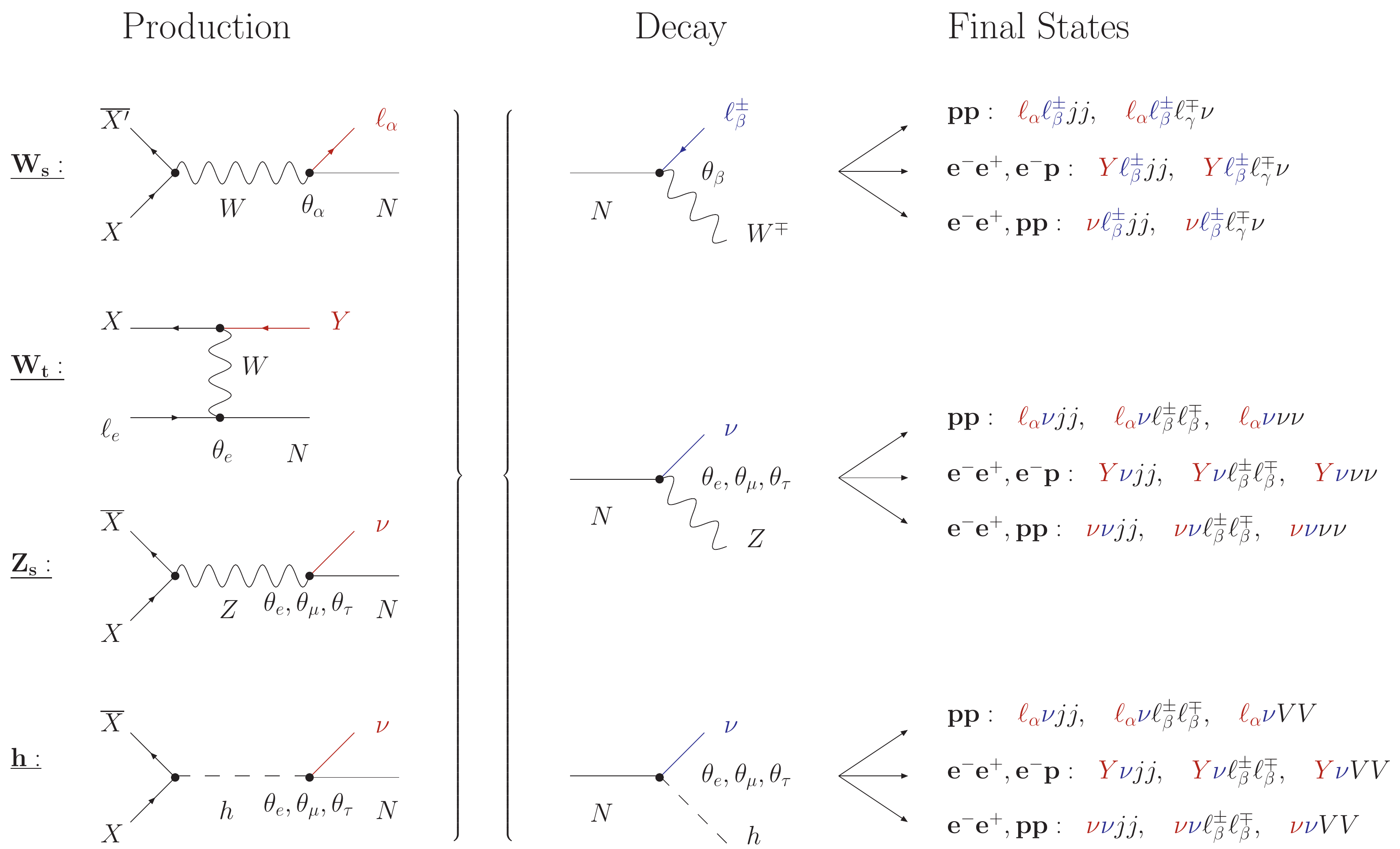}
\caption{Pictographic representation of the different heavy neutrino production and decay channels at leading order, including the dependency of the active-sterile mixing parameters. These production and decay channels yield possible final states for sterile neutrino searches at different collider types.}
\label{fig:production_and_decay}
\end{figure*}

The heavy neutrino states can be produced in high energy collisions by the weak interaction (see eq.~\eqref{eq:weakcurrent1} and eq.~\eqref{eq:weakcurrent2}) or the Higgs boson (see eq.~\eqref{eq:Lykawa}). The production processes for heavy neutrino mass eigenstates at leading order in the active-sterile mixing angles and in the weak coupling constant, are given in the first column of fig.~\ref{fig:production_and_decay}. We now specify the relevant production channels for the $e^-e^+,$ $pp$ and $e^-p$ colliders (for a summary cf.~tab.~\ref{tab:summary}):

\begin{itemize}
\item \textbf{$e^-e^+$ colliders:} There are two dominant production channels. One is given by the exchange of a $W$ boson in the $t$-channel, labelled with $\mathbf{W_t}$ in fig.~\ref{fig:production_and_decay}, where $X=\overline{\ell}_e$ in the initial state is the anti particle to $\ell_e=e^-,e^+$ and $Y=\nu$ (where we suppressed the indices of the light neutrino mass eigenstates for simplicity). Another production channel is depicted by the diagram labelled $\mathbf{Z_s}$, 
where the initial states $\{X,\overline X\}$ are the electron positron pair $\{\ell_e,\overline{\ell}_e\}$. 
A sub-dominant channel is given by Higgs boson decays into heavy and light neutrinos, given by the diagram labelled $\mathbf{h}$.
The Higgs boson can be produced for instance via Higgs strahlung or $WW$ boson fusion. We note that its production from the $e^-e^+$ pair is usually negligible, due to the smallness of the electron Yukawa coupling. The sub-dominant channel via the Higgs can be relevant when the heavy neutrino mass $M$ is below the Higgs boson mass $m_h$.

\item \textbf{$pp$ colliders:} The dominant production channels for heavy neutrinos in proton-proton collisions are Drell-Yan processes. In fig.~\ref{fig:production_and_decay} they are denoted by the diagrams labelled $\mathbf{W_s}$, with $\{X,\overline{X^\prime}\}=\{q_u,\overline{q_d}\}$ or $\{q_d,\overline{q_u}\}$, and $\mathbf{Z_s}$, with $\{X,\overline{X}\}=\{q,\overline{q}\}$, where $q_u,\,q_d,\,q$ are up-type quarks, down-type quarks, and constituents of the proton, respectively.
A sub-dominant process at higher order is given by $W\gamma$ fusion with initial states $\{q,\gamma\}$, which is further suppressed by the photon's parton distribution function (PDF).
Also at $pp$ colliders, the production of heavy neutrinos from diagram $\mathbf{h}$ are sub-dominant. The Higgs boson can be produced, for instance, via vector boson fusion (including gluons).

\item \textbf{$e^-p$ colliders:} 
The dominant production channel for heavy neutrinos is given by the diagram $\mathbf{W_t}$ in fig.~\ref{fig:production_and_decay}. In electron-proton collisions, $X$ is a proton constituent (e.g.\ a quark) and $Y$ is the isospin partner of $X$. Another leading order production channel is given by $W\gamma$ fusion, labelled $W_t^{(\gamma)}$, with $X=\gamma$ and $Y=W^-$ which is, contrary to the $pp$ colliders, only suppressed by the photon's PDF.
Furthermore, for $M<m_h$ the production via the Higgs boson is possible, when the latter is produced via vector boson fusion, which is, however a process of higher order.
\end{itemize}

\subsubsection{Signal channels}
For the here considered sterile neutrino masses, all the heavy neutrino mass eigenstates will decay according to the second column of fig.~\ref{fig:production_and_decay}. Also the $Z,W$ and Higgs bosons decay further into SM particles.
The possible final states from the production and decay of heavy neutrinos constitute the different signatures for sterile neutrino searches. We display them in the third column of fig.~\ref{fig:production_and_decay} and denote for which collider they have a production channel at leading order.

\subsubsection{Remarks on lepton number violation (LNV)} \label{sec:LNV}

We note that in our benchmark model there is no LNV due to the exact protective symmetry. When neutrino masses are generated from small perturbations of the structure in eq.\ (\ref{eq:lagrange}), LNV effects will be introduced, however suppressed by the (approximate) protective symmetry. For the following comment, we assume such small perturbations to be present. 

When light neutrinos are in the final states, it is in general difficult to experimentally measure LNV, since the light neutrinos escape the detector without revealing their lepton number.\footnote{We also remark that, of course, the small Majorana masses of the light neutrinos, induced by small perturbations of the structure in eq.\ (\ref{eq:lagrange}) as mentioned above, also violate lepton number. However for collider phenomenology this is subdominant compared to the lepton number violation from the perturbed heavy neutrino sector, and can safely be neglected. In this sense one can attribute a lepton number to the light neutrinos produced in a given process.} 
An overview of the lepton number violating processes for the different colliders is shown in tab.~\ref{tab:summary}. Therein, we only indicate channels with LNV if no light neutrinos are present in the final state, i.e.\ if there exists an unambiguous signature for LNV.

In $e^-e^+$ collisions, at leading order, LNV processes always give rise to a light neutrino in the final state such that, as mentioned above, there is no unambiguous LNV signature. 
We remark that, using the different kinematics of signature and background final states with different light neutrino lepton number, it might nevertheless be possible to find a signal of LNV at $e^-e^+$ colliders.

 In $pp$ collisions the final state $\ell_\alpha^\pm \ell_\beta^\pm jj$, where $j$ denotes a hadronic jet, unambiguously violates lepton number by two units and is often called ``same sign di-lepton''. The final state $\ell_\alpha^\pm \ell_\beta^\pm \ell_\gamma^\mp \nu$ is produced by both, lepton number violating and conserving processes, and is often called ``trilepton''. 

In $e^-p$ collisions, the final state $\ell_\beta^+jjj$ unambiguously violates lepton number by two units, while $j \ell_\beta^\pm\ell_\gamma^\mp \nu$ can be produced from lepton number conserving or violating processes. 
Furthermore, additional LNV signatures exist with heavy neutrino production via $W\gamma$ fusion, as will be discussed in section \ref{sec:ep_colliders}.

\subsubsection{Remarks on lepton flavour violation (LFV)}\label{sec:LFV}

In the SM, lepton flavour (LF) is conserved since neutrinos are massless and the neutrino flavour eigenstates are always produced in conjunction with the corresponding charged lepton flavour (and mass) eigenstates. In the presence of sterile neutrinos, this is no longer the case and LFV final states are to be expected.

As for LNV, also here we note that when light neutrinos are in the final states, it can be difficult to experimentally measure LFV, since the light neutrinos escape the detector without revealing their flavour composition. However, for final states without neutrinos, or with less neutrinos involved than charged leptons, there can nevertheless be unambiguous signatures for LFV at the parton level. We remark that SM processes that lead to final states with additional neutrinos provide an important background at the reconstructed level.  

 At $pp$ and $e^- p$ colliders, it is possible to have unambiguous signals for LFV at the parton level, without neutrinos in the final state, for instance the signature $\ell_\alpha \ell_\beta jj$ with $\alpha \not= \beta$ at $pp$ colliders and $\ell_\alpha jjj$ with $\alpha \not= e$ at $e^- p$ colliders. Unambiguous parton level signatures with one neutrino in the final state are $\ell_e \ell_\mu \ell_\tau \nu$ at $pp$ colliders and $j \ell^-_\alpha \ell^+_\beta \nu$ with $\alpha \not= e$ and $\alpha\neq\beta$ at $e^- p$ colliders. Further LFV signatures exist with $W\gamma$ fusion and at higher order. 

In $e^-e^+$ collisions, there are no unambiguous signatures at tree-level, but efficient searches are in principle possible via LFV $Z$ and Higgs boson decays, which, in the presence of sterile neutrinos, are induced at loop level.

\begin{table}
\begin{center}
\begin{tabular}{|c|c|c|c|}\cline{2-4}
\multicolumn{1}{c|}{} & $e^-e^+$ & $pp$ & $e^-p$ \\ \hline
$\mathbf{W_s}$ & $\times$ & \checkmark +\,LNV/LFV & $\times$\\ 
$\mathbf{W_t}$ & \checkmark  & $\times$ & \checkmark \,+LNV/LFV\\ 
$\mathbf{Z_s}$ & \checkmark  & \checkmark & $\times$\\ 
$\mathbf{h}$ & (\checkmark) & (\checkmark) & (\checkmark)\\ \hline
\end{tabular}
\caption{Summary of the production channels for the different colliders at leading order, indicating with the labels LNV and/or LFV whether there exists an unambiguous signature for probing LNV and/or LFV. For $e^-e^+$ colliders, there are always light neutrinos in the final state such that it is experimentally difficult to detect the LNV. We remark that the $Z$ pole run and the Higgs physics run at $e^-e^+$ colliders can be sensitive to loop-induced LFV. The checkmarks in brackets for the production processes via $h$ indicate that this is a sub-dominant channel compared to the $W$ and $Z$ channels.
}
\label{tab:summary}
\end{center}
\end{table}

\begin{table}
\begin{center}
\renewcommand{\arraystretch}{2.3}
\begin{tabular}{|c|c|c|c|c|} \cline{3-4}
\multicolumn{1}{c}{}& \multicolumn{1}{c|}{} & \multicolumn{2}{c|}{Decay channel}   \\ \cline{3-4}
\multicolumn{1}{c}{}& \multicolumn{1}{c|}{} & \multicolumn{1}{c}{$W$} & $Z (h)$   \\ \hline
\multirow{3}{*}{\rotatebox[origin=c]{90}{\parbox[c]{2cm}{\centering Production channel}}}
&$\mathbf{W_s}$ & $\displaystyle\frac{|\theta_\alpha \theta_\beta|^2}{ |\theta|^2}$ & $|\theta_\alpha|^2$ \\\cline{3-4}
&$\mathbf{W_t}$ & $\displaystyle\frac{|\theta_e \theta_\beta|^2}{ |\theta|^2}$ & $|\theta_e|^2$ \\\cline{3-4}
&$\mathbf{Z_s (h)}$ & $|\theta_\beta|^2$ $\displaystyle\vphantom{\frac{1}{\sum |\theta_\gamma|^2}}$ & $|\theta|^2$\\ \hline
\end{tabular}
\renewcommand{\arraystretch}{1}
\caption{Summary of the dependencies on the active-sterile mixing parameters from the different production and decay channels on the cross section level in the narrow width approximation (and in leading order). The active-sterile mixing parameter $|\theta|^2$ is defined in eq.\ \eqref{def:thetaa}.}
\label{tab:dependency}
\end{center}
\end{table}

\subsubsection{General remarks}
Here we discuss some features from sterile neutrinos, and we 
specify the dependency of the different decay channels on the active-sterile mixing parameters $\theta_{\alpha},\,\alpha= e,\mu,\tau$.

\begin{itemize}
\item Indirect effects from PMNS non-unitarity: Even if the heavy neutrinos are not produced directly in collisions, they can nevertheless have indirect effects from the induced non-unitarity of the PMNS matrix, i.e.\ of the matrix ${\cal N}$ in eq.~(\ref{eq:mixingmatrix}). We will discuss these effects in more detail in the following sections.

\item Reconstruction of the sterile neutrino mass $M$: The semi-leptonic decays of the heavy neutrino ($N \to \ell jj$) feature a ``bump'', corresponding to the mass of the heavy neutrino, in the invariant mass spectrum of the $\ell jj$ system. Although at $pp$ and $e^- p$ colliders only transverse observables are available, one can in principle use this for reconstructing the sterile neutrino mass parameter $M$.

\item Vertex displacement: Heavy neutrinos with masses below the $W$ boson mass and with very small active-sterile mixings can have a long lifetime that leads to a displacement of the decay products from the interaction point. This is an exotic signature and can be searched for at all three particle collider types.

\item Dependencies on the mixing angles: 
The dependencies of the different production and decay processes to the $|\theta_\alpha|$ can be read off of the first and second columns of fig.\ \ref{fig:production_and_decay}, respectively.
We summarize the possible combinations in tab.~\ref{tab:dependency}, which correspond to the signatures on the cross section level in narrow width approximation (and in leading order).

The production of heavy neutrinos at $e^-e^+$ colliders via the process $\mathbf{Z_s}$ and at $pp$ colliders via the processes $\mathbf{W_s},\,\mathbf{Z_s}$ is sensitive to the sum $|\theta|^2$. Conversely, the heavy neutrino production process at $e^-e^+$ and $e^-p$ colliders via $\mathbf{W_t}$ is sensitive to $|\theta_e|^2$.

The relative strengths of the mixing angles $|\theta_\alpha|$ and their combinations $|\theta_\alpha\theta_\beta|$ can be inferred in principle via the channels involving charged leptons.

\end{itemize}

\subsection{Present constraints}\label{sec:present_constraints}
The present constraints from past and ongoing experiments for sterile neutrinos have been presented and discussed in \cite{Antusch:2015mia,Antusch:2014woa,Antusch:2015gjw}, using the SPSS as benchmark scenario.

A general overview of sterile neutrino extensions of the SM and their observable effects can be found, e.g., in the review \cite{Abazajian:2012ys}. 
Further observable features of various models with right-handed neutrinos were studied, e.g., in refs.\ \cite{Drewes:2013gca,Drewes:2015iva,Abada:2015oba,Drewes:2016gmt,Heurtier:2016iac,Ferrari:2000sp,Humbert:2015epa,Duarte:2015iba,Kang:2015uoc,Dev:2016dja,Baglio:2016ijw,Thuc:2016qva,Borah:2016iqd,Fernandez-Martinez:2016lgt,Drewes:2016jae,Hambye:2016sby}.

In this section, we give a short summary of the present constraints on the three neutrino Yukawa couplings $y_{\nu_\alpha}$  (or, equivalently, the active-sterile mixing parameters  $\theta_\alpha$) for masses $M$ between 10 GeV and 250 GeV that are taken from ref.~\cite{Antusch:2015mia}.

\subsubsection{Indirect searches in electroweak precision data}
In the presence of sterile neutrinos, the mixing matrix of the three active neutrinos, i.e.\ ${\cal N}$ in eq.\ \eqref{eq:mixingmatrix} and \eqref{eq:matrixN}, is effectively non-unitary (see e.g.\ \cite{Langacker:1988ur,Antusch:2006vwa}). The subsequent modification of the weak currents then leads to modified predictions for electroweak observables compared to the SM, which constitutes an ``indirect'' effect and allows to test these models via precision measurements.

For instance, the Fermi constant $G_F$ that prominently enters many theory predictions for the EWPOs, is (one of) the main source(s) for the non-unitarity effects.
It is inferred from muon decays and due to the modification of the weak currents, the theory prediction for $G_F$ is sensitive to the active-sterile mixing parameters. Assuming $M > m_\mu$, one has on the cross-section level
\be
\sigma^{}_{\mu^- \to e^- \nu \bar \nu} =  \left( \cal N^{} N^\dag\right)_{ee}\left( \cal N^{} N^\dag\right)_{\mu\mu} \cdot \sigma^{\rm SM}_{\mu^- \to e^- \nu \bar \nu}\,,
\label{eq:xsectionrel}
\ee
such that one obtains the modified Fermi constant $G_\mu$:
\be
G_F^2\; \to\; G_\mu^2 =\; G_F^2\cdot({\cal NN}^\dagger)_{ee}({\cal NN}^\dagger)_{\mu \mu}\;.
\label{eq:GF}
\ee
As a consequence this affects the theory prediction for other electroweak observables, such as the weak mixing angle $\theta_W$, the $W$ boson mass, etc.

\begin{figure}[t]
\begin{center}
\includegraphics[scale=0.5]{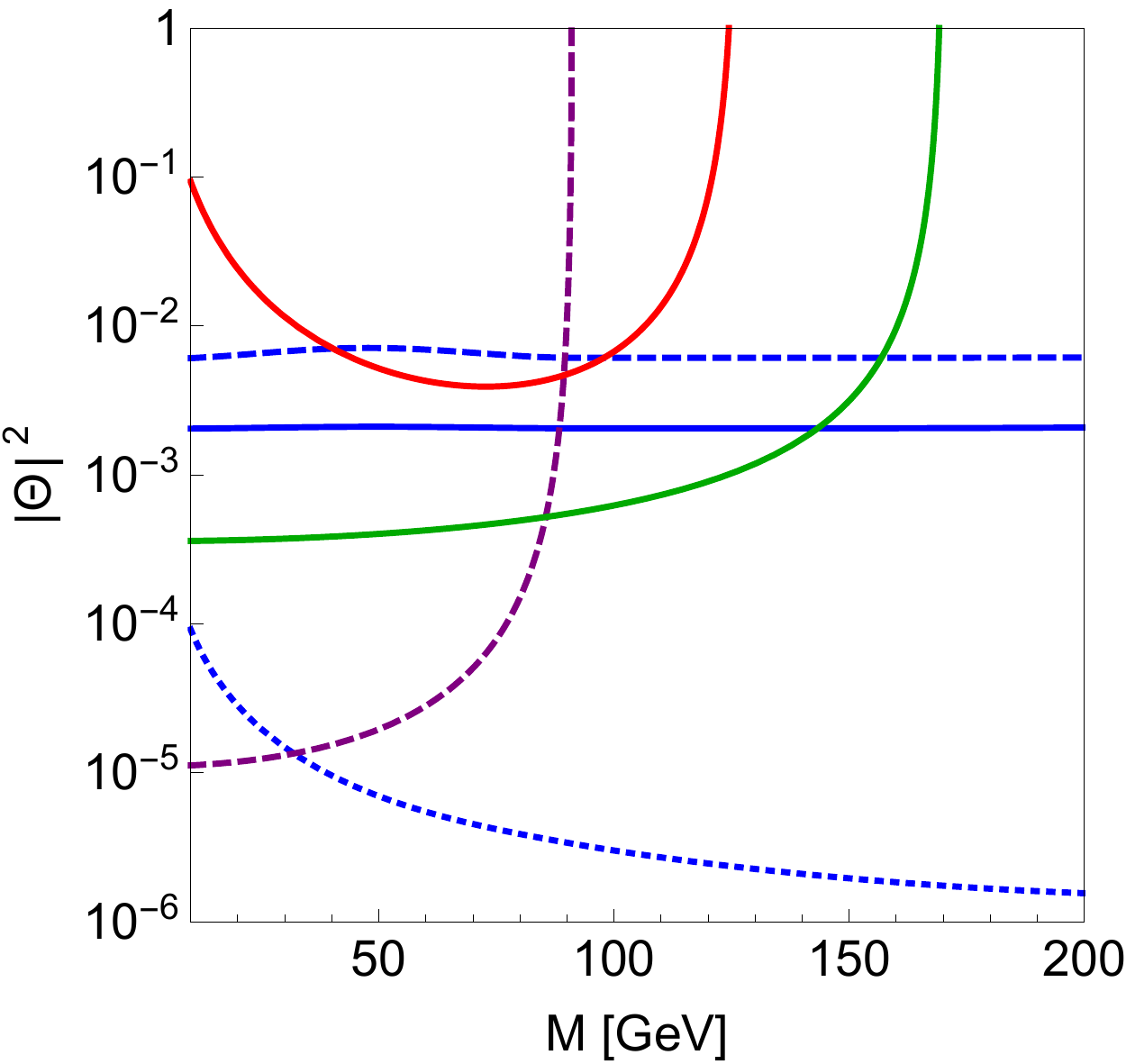}

\vspace{0.5cm}
\includegraphics[scale=0.825]{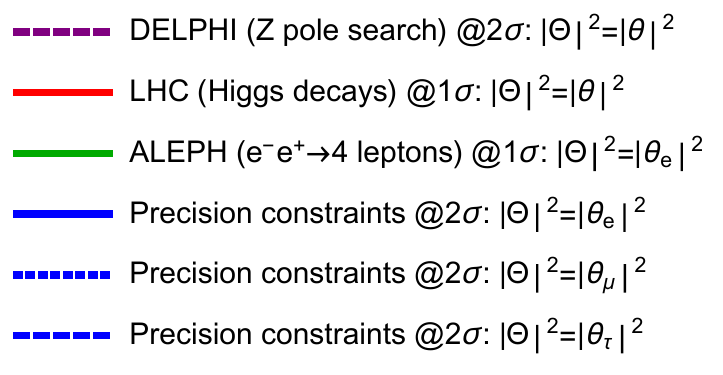}

\end{center}
\caption{Summary of present constraints on sterile neutrino properties taken from ref.~\cite{Antusch:2015mia}.}
\label{fig:present}
\end{figure}

Further observables that are affected by the modification of the weak currents are lepton-universality, rare charged lepton flavour violating decays, CKM unitarity tests, and low energy measurements of the weak mixing angle.
A detailed list of the modified predictions can be found in refs.\ \cite{Antusch:2014woa,Antusch:2015mia} and references therein, and also for a similar setup in ref.\ \cite{Fernandez-Martinez:2016lgt}.
We show the resulting experimental constraints on the three active-sterile mixing parameters, $|\theta_\alpha|,\, (\alpha=e,\,\mu,\,\tau$) from ref.~\cite{Antusch:2015mia} as blue lines in fig.\ \ref{fig:present}.

\subsubsection{Indirect searches in Higgs boson decays}
The measurement of the branching ratios of the Higgs boson at the LHC allows to constrain the contribution from its decays to neutrinos. Higgs boson decays into neutrinos are possible due to the neutrino Yukawa term in eq.\ \eqref{eq:lagrange} for $M \leq m_h$, which modify the total Higgs boson decay width and branching ratios.
From the combined measurements for the branching ratios of $h\to VV$ ($V=\gamma,Z,W$) from ATLAS and CMS, $h\to \gamma\gamma$ is found to be most sensitive to the active-sterile mixing parameters.
Its bound is used to limit the partial decay width $\Gamma(h\to \nu N)$, which constrains the sum of the squared sterile mixing angles $|\theta|^2$, see in fig.~\ref{fig:present} as was derived in ref.~\cite{Antusch:2015mia}. 
Further constraints from the LHC measurements of the Higgs boson couplings were studied in refs.\ \cite{Bizot:2015zaa,Graesser:2007yj}.

\subsubsection{Direct searches at LEP and the LHC}
Direct searches for sterile neutrinos comprise processes in which a heavy neutrino is produced on-shell and then decays into SM particles. The presently most sensitive direct constraints are given by sterile neutrino production in $Z$ boson decays (the production channel $\mathbf{Z_s}$ for $M < m_Z$), four lepton final states from $WW$ decays, and lepton-number-violating signatures.

From the analyses for heavy neutral lepton searches at the $Z$-pole conducted by the LEP-I collaborations DELPHI \cite{Abreu:1996pa}, OPAL \cite{Akrawy:1990zq}, ALEPH \cite{Decamp:1991uy} and L3 \cite{Adriani:1993gk}, one gathers that the most stringent bound is on the branching ratio $Z \to \nu \, N$ from DELPHI, given at 95\% C.L.\ as
\be
Br(Z \to \nu \, N) < 1.3 \times 10^{-6}\,.
\label{eq:eetonuN}
\ee
This bound constrains the sum of the squared active-sterile mixing angles, $|\theta|^2$, and is shown in fig.\ref{fig:present} (cf.\ ref.~\cite{Antusch:2015mia}).

The cross section for $WW$ production was studied by analyzing four lepton final states at and beyond the $WW$ threshold at LEP-II. The ALEPH collaboration has set bounds on deviations of the SM cross section for four lepton final states at the $1\sigma$ C.L. from which a constraint for the active-sterile mixing parameter $|\theta_e|$ can be inferred, cf.~ref.~\cite{Antusch:2015mia}, as is shown in fig.~\ref{fig:present} by the green line.

We remark that the lepton number violating ``same sign di-lepton'' signature also provides a promising search channel at the LHC.\footnote{\label{fntLNV} We note that often in LHC analyses only one sterile neutrino is considered. In this case the na\"ive seesaw formula applies which means that for a sterile neutrino with mass $M$ the mixing $\theta^2$ should be smaller than ${\cal O}(m_\nu/M)\approx 10^{-12} ({100\:\mbox{GeV}}/{M})$. In scenarios with protective ``lepton number''-like symmetry larger active-sterile mixing angles are possible and consistent with small neutrino masses, however then the LNV channels are suppressed by the approximate ``lepton number''-like symmetry.} Presently, the resulting constraints on the active-sterile mixing parameters are, however, not competitive with the indirect constraints from LEP (cf.\ ref.~\cite{Das:2014jxa}).

\section{Searches at future $e^-e^+$colliders}
\begin{figure*}[t]
\includegraphics[scale=0.8]{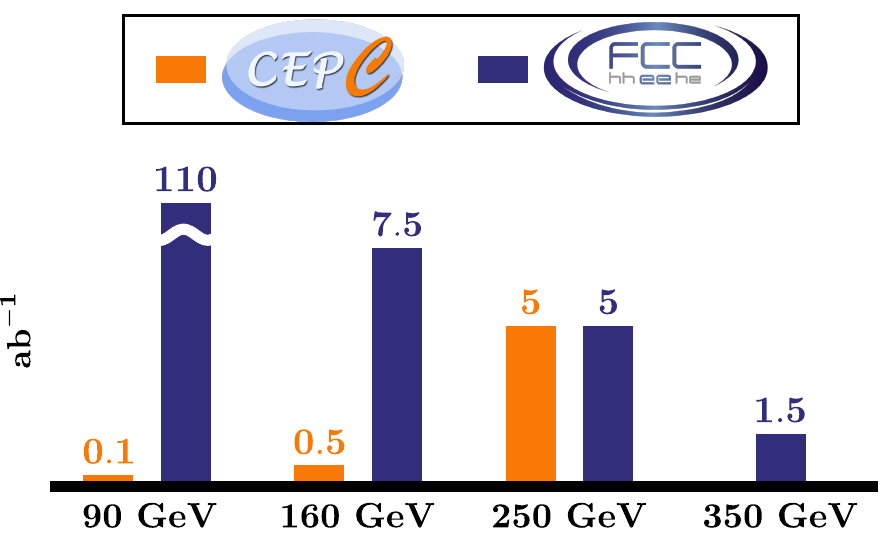}
\includegraphics[scale=0.8]{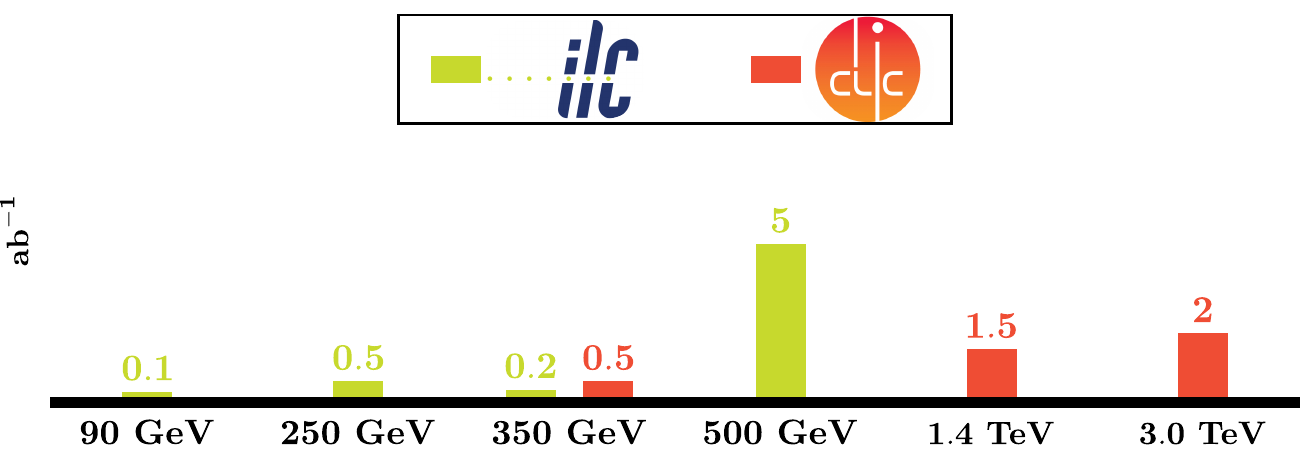}
\caption{\footnotesize Physics programs of the different future lepton colliders given by the center-of-mass energy and envisaged integrated luminosity.
\\
\textbf{Circular future lepton colliders (left):} 
For the CEPC we use the exemplary integrated luminosities from the preCDR \cite{CEPC-SPPCStudyGroup:2015csa}. For the FCC-ee \cite{Gomez-Ceballos:2013zzn} we use the product of the target instantaneous luminosities from \cite{FCCweb} (for two interaction points) and the envisaged run-times, and the Higgs run with a center-of-mass energy of 240 GeV. \textbf{Linear future lepton colliders (right):} For the ILC \cite{Baer:2013cma} we consider the G-20 operation scenario from ref.\ \cite{Brau:2015ppa}, and we further include the Giga-Z operation.
For the CLIC \cite{Linssen:2012hp} we consider the discussed physics runs in \cite{Abramowicz:2013tzc}.}
\label{fig:modioperandi}
\end{figure*}
Future lepton colliders aim at providing high-precision measurements of collision data. They are designed to study rare $Z$, $W$, Higgs boson and top decays, which may give insight into physics beyond the SM.

Presently, two types of lepton collider layouts are pursued: circular accelerators with particularly high luminosities at energies below $\sim 350$ GeV and the linear accelerators that can achieve center-of-mass energies up to $\sim 1$ TeV and beyond, and which also allow for highly polarized beams. 
The presently discussed linear accelerators are the International Linear Collider (ILC), which has completed its technical design, and the Compact LInear Collider (CLIC), which is currently undergoing its technical design phase.

The two presently discussed circular accelerators, the Circular Electron Positron Collider (CEPC) and the Future Circular Collider (FCC-ee), are in their conceptual phase.

Each of these planned colliders has its own unique physics program, defined by a target luminosity for specific center-of-mass energies. A physics program can encompass a $Z$-pole run, a $WW$ threshold run, SM Higgs physics, a top threshold run, and high energy runs, which may allow testing rare SM processes. The physics program for the circular and linear colliders is shown in the left and right panel of fig.~\ref{fig:modioperandi}, respectively.

Extensive searches for sterile neutrinos have been performed at LEP, as discussed in section \ref{sec:present_constraints} on present constraints. In this section we discuss the different search channels for sterile neutrinos at lepton colliders, including novel ones, and present updated sensitivities. For the sensitivity estimates for future colliders, we focus on the CEPC and FCC-ee. We will present the sensitivity estimates for the ILC and CLIC elsewhere.

\subsection{Production mechanism}
\begin{figure}[t]
\begin{center}
\includegraphics[width=0.3\textwidth]{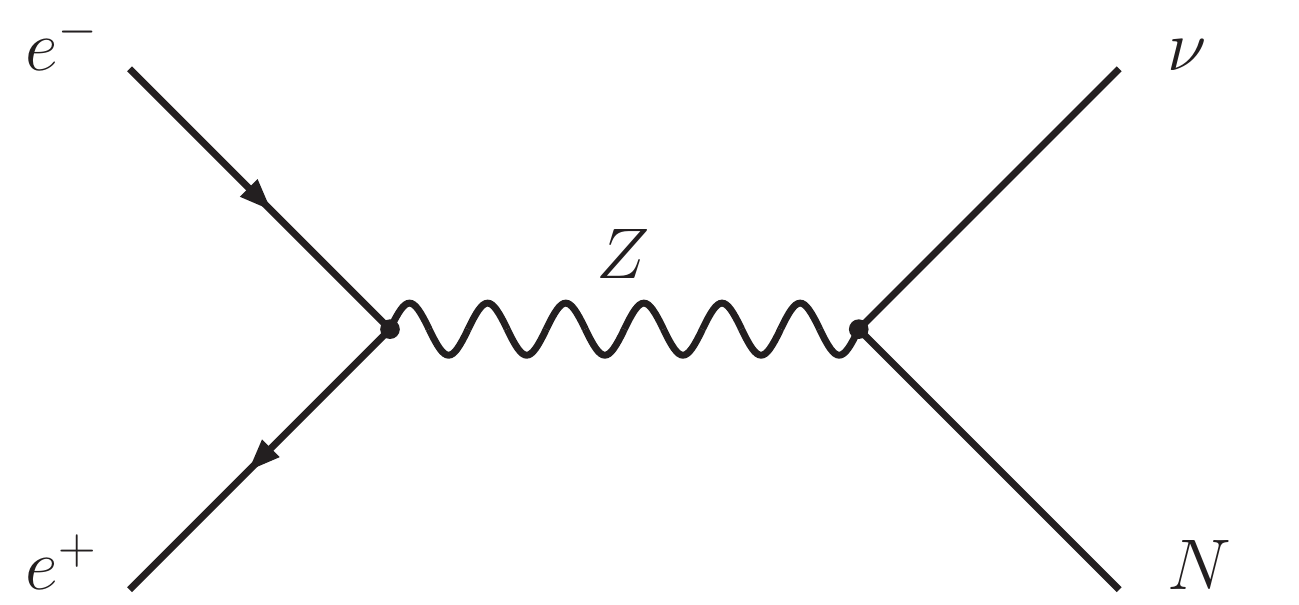}

production channel: $\mathbf{Z_s}$
\vspace{10pt}

\includegraphics[width=0.3\textwidth]{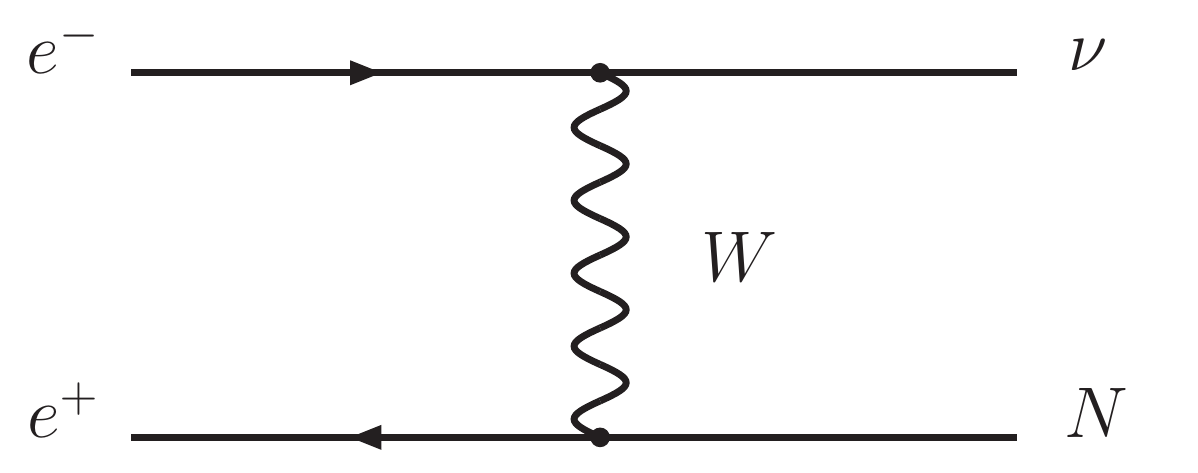}

production channel: $\mathbf{W_t}$
\end{center}
\caption{Dominating Feynman diagrams for the production of heavy neutrinos. Heavy neutrino production via the s-channel $Z$ boson is dominant at the $Z$-pole. For center-of-mass energies above the $Z$-pole the dominant production stems from the t-channel exchange of a $W$ boson.}
\label{fig:Nproduction}
\end{figure}

As discussed in sec.~\ref{sec:production processes} heavy neutrino production proceeds via the channels $\mathbf{Z_s}$ and $\mathbf{W_t}$, with the corresponding Feynman diagrams depicted in fig.\ \ref{fig:Nproduction}, which are both proportional to the active-sterile mixings $|\theta|^2$ and $|\theta_e|^2$, respectively, when summed over all light neutrinos. 
Both production mechanisms give rise to a heavy neutrino that is accompanied by a light neutrino.

\begin{figure*}[t]
\begin{center}
\includegraphics[height=0.3\textwidth]{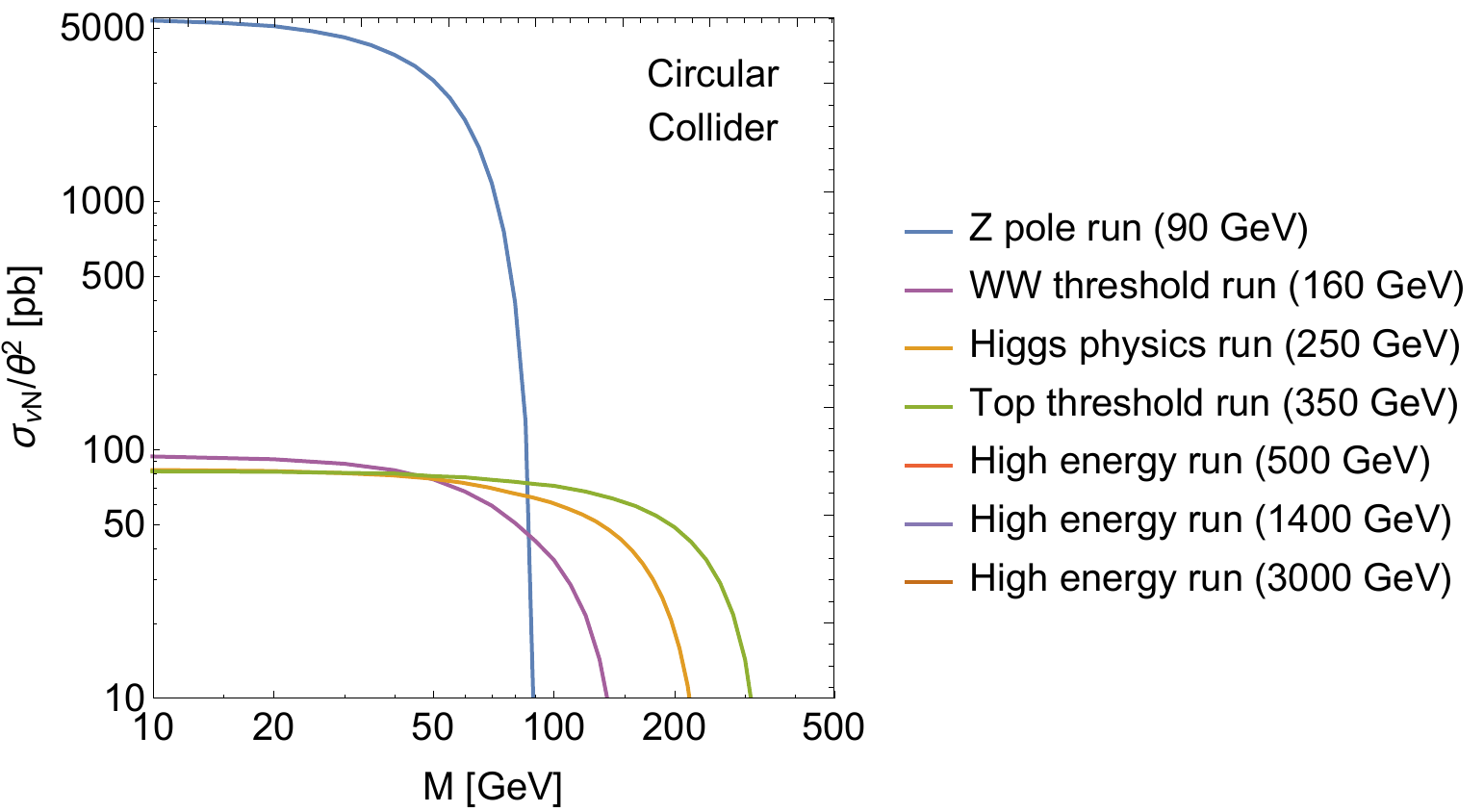}
\includegraphics[height=0.3\textwidth]{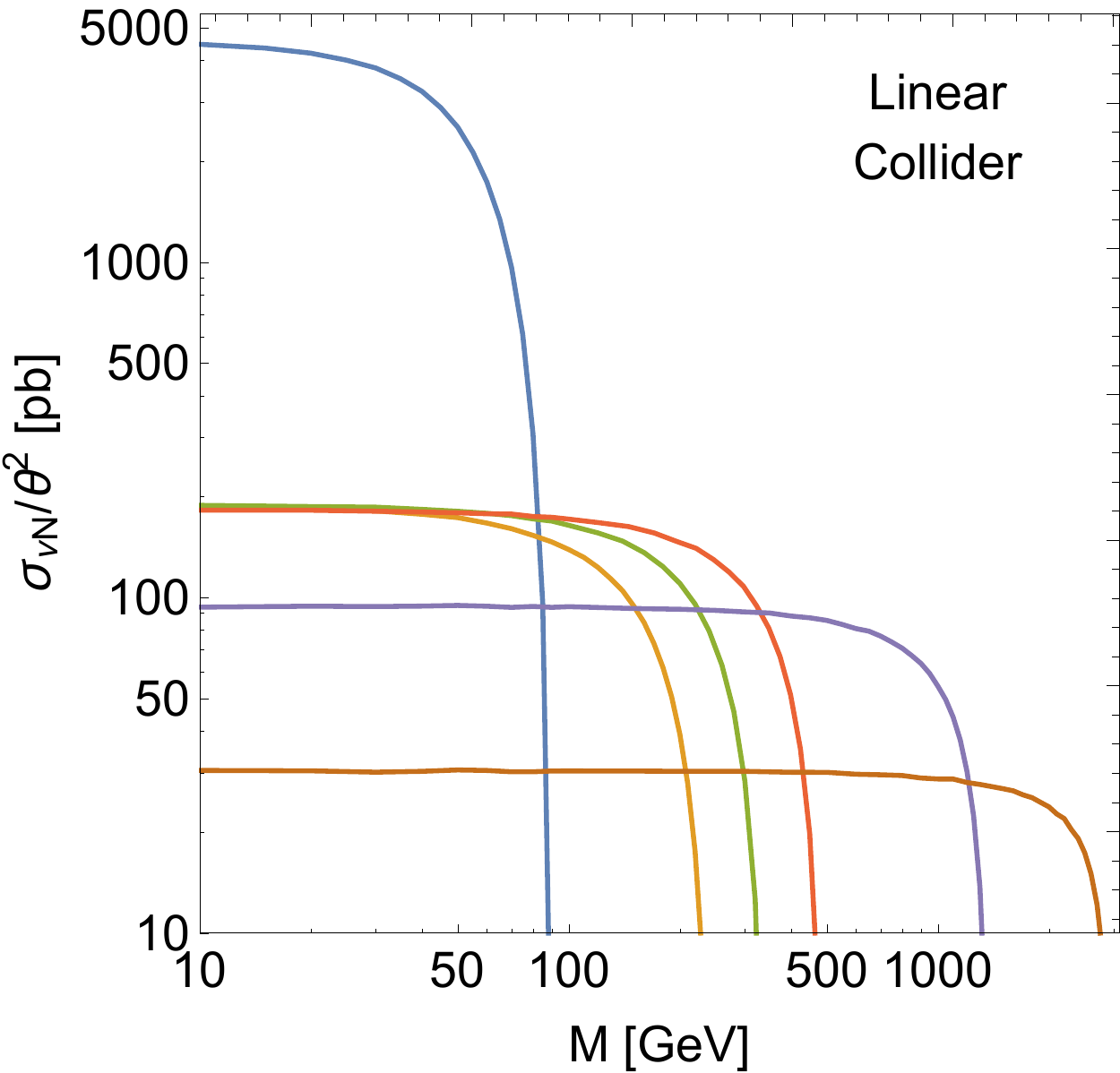}
\end{center}
\caption{Production cross section for heavy neutrinos at different center-of-mass energies, divided by the square of the active-sterile mixing angle. For all the lepton colliders initial state radiation is included, and for the linear colliders we also included beamstrahlung effects and use a $(L,R)$ beam polarisation of (80\%,30\%). 
For the cross section calculation we have applied the following cut: $|\cos(\theta) | \le 0.99$, with $\theta$ being the angle between the heavy neutrino and the lepton beams.
}
\label{fig:cross-section-ee}
\end{figure*}

We define the heavy neutrino production cross section, to leading order in the small active-sterile mixing, by
\be
\sigma_{\nu N} = \sum_{i,j} \sigma(e^- e^+ \to N_j\,\nu_i)\,,
\label{eq:sigmanuN}
\ee
where the sum is taken over all the light neutrino ($i=1,2,3$) and heavy neutrino ($j=1,2$) mass eigenstates.
We display the dependency of the cross sections on the sterile neutrino mass $M$ for the different physics runs and for the different accelerator layouts in fig.\ \ref{fig:cross-section-ee}. The cross sections were evaluated by implementing the SPSS via Feynrules \cite{Alloul:2013bka} into the Monte Carlo event generator WHIZARD \cite{Kilian:2007gr,Moretti:2001zz}, where initial state radiation and only for the linear colliders lepton beam polarisation has been included. 

We remark that for $\sqrt s \simeq m_z$ heavy neutrino production proceeds dominantly via $\mathbf{Z_s}$, while for $\sqrt s =$ 160~GeV and above it is dominated by $\mathbf{W_t}$. This allows for a separate assessment of the two different production channels via the center-of-mass energy or, respectively,  the physics program.

It is interesting to note that we can expect up to $\mathcal{O}(10^{4})$ heavy neutrinos per ab$^{-1}$ for values of $|\theta|^2$ consistent with the present constraints.

\subsection{Signatures and searches}
\begin{table*}[t]
\begin{center}
\begin{tabular}{|l|c|c|c|c|}
\hline
Name & Final State & Channel [production,decay] & $|\theta_\alpha|$ dependency  \\ \hline\hline
lepton-dijet & $\ell_\alpha\nu jj$ & [$\mathbf{W_t},W$], [$\mathbf{Z_s},W$] 
 & ${\dfrac{|\theta_e\theta_\alpha|^2}{\theta^2}}^{(**)}, {|\theta_\alpha|^2}^{(**)}$
 \\ \hline
dilepton
& $\ell_\alpha\ell_\beta\nu\nu$ & [$\mathbf{W_t},\{W,Z (h)\}$], [$\mathbf{Z_s},\{W,Z (h)\}$]
 & ${\left\{\dfrac{|\theta_e\theta_\alpha|^2}{\theta^2}^{(*)}
,{|\theta_e|^2}^{(*)}\right\}}^{(**)}$, ${\left\{{|\theta_\alpha|^2}^{(*)}, |\theta|^2 \right\}}^{(**)}$
 \\ \hline
dijet
& $\nu\nu jj$ &  [$\mathbf{W_t}, Z (h)$], [$\mathbf{Z_s},Z (h)$]
 & ${|\theta_e|^2}^{(**)}$, ${|\theta|^2}^{(**)}$ $\vphantom{\dfrac{{}^1}{{}^1}}$ 
 \\ \hline
invisible & $\nu\nu\nu\nu$ & [$\mathbf{W_t},Z $], [$\mathbf{Z_s},Z $]
 & ${|\theta_e|^2}^{(**)}$, ${|\theta|^2}^{(**)}$ $\vphantom{\dfrac{{}^1}{{}^1}}$ 
 \\ \hline
\end{tabular}
\end{center}
\caption{Signatures of sterile neutrinos at leading order for $e^-e^+$ colliders with their corresponding final states, production and decay channels (cf.\ section \ref{sec:production-and-decay}), dependency on the active-sterile mixing parameters.\\
$^{(*)}:$ The dependency on the active-sterile mixing can be inferred when the origin of the charged leptons can be reconstructed.\\
$^{(**)}:$ The dependency on the active-sterile mixing is determined by the center-of-mass energy, i.e.\ by the physics run of the given $e^-e^+$ collider.
}
\label{tab:signatures_ee}
\end{table*}

In this section, we discuss observable effects from sterile neutrinos at $e^-e^+$ colliders, which manifest themselves in specific final states with the related production and decay channels, cf.~fig.~\ref{fig:production_and_decay}, and the dependency on the active-sterile mixing angles.
We refer to these effects as signatures and list them in tab.\ \ref{tab:signatures_ee}.

In the following, we discuss these and other signatures for sterile neutrinos at future lepton colliders, thereby updating several estimates for the sensitivities of the CEPC and FCC-ee.

\subsubsection{Lepton-dijet}
The heavy neutrino decays via the charged current together with the hadronic decays of the $W$ boson yield the final state $\ell_\alpha \nu jj$, with the invariant mass of the two jets being consistent with $m_W$.
The invariant mass of the visible final states allows to infer the heavy neutrino mass $M$.
For center-of-mass energies above the $Z$-pole this signature is mainly dependent on $|\theta_e\theta_\alpha|^2/|\theta|^2$. 
We show our estimates for the $1\sigma$ sensitivity of this signature at 250 and 350 GeV for the CEPC and FCC-ee, respectively, by the orange line in fig.~\ref{fig:sensitivity_ee}, where we use $|\theta_\alpha|=|\theta_e|$ and $|\theta_\mu|=|\theta_\tau|=0$. For details on the calculation of the sensitivity we refer the reader to section \ref{app:distributions-ee} in the appendix.

For $\sqrt s \simeq$ 90~GeV, this signature is dependent on $|\theta_\alpha|^2$ and has been investigated at LEP in ref.~\cite{Abreu:1996pa}. Its sensitivity is included in the dashed purple line, which was obtained from a trivial rescaling of the DELPHI results by the luminosity, in fig.~\ref{fig:sensitivity_ee}.
This signature has been studied for the ILC in refs.\ \cite{Das:2012ze,Banerjee:2015gca}.

We note that this final state can be produced by both lepton-number-conserving and lepton-number-violating processes and it might be possible to infer lepton number violation from the kinematic distributions.

\subsubsection{Dilepton}
The dilepton final state $\ell_\alpha \ell_\beta \nu \nu$ can be achieved from heavy neutrinos that decay leptonically via the $W$ boson or that decay into two charged leptons via the $Z$ or Higgs boson.

For the $W$ boson decay channel at energies above the $Z$-pole, the resulting process $e^- e^+ \xrightarrow{\mathbf{W_t}}  N\nu \xrightarrow{W}  \ell_\alpha^\pm W^\mp\nu \to  \ell_\alpha^\pm\ell_\beta^\mp\nu\nu$  can be mistaken for the SM process $e^- e^+ \to W^+ W^-$ and thus lead to a modified $WW$-production cross section. 
In ref.\ \cite{Antusch:2015mia} the sensitivity of this channel was estimated for $|\theta_e|\neq 0$ and $|\theta_\mu|=|\theta_\tau|=0$ using statistical uncertainties of the $WW$ production cross section (considering only leptonic final states), shown by the green line in fig.\ \ref{fig:sensitivity_ee}.

When the heavy neutrino decays proceed via the $Z$ boson, the invariant mass of the charged lepton pair is compatible with $m_Z$. For the physics runs above the $Z$-pole, the channel $e^- e^+ \xrightarrow{\mathbf{W_t}} N\nu \xrightarrow{Z} Z\nu \nu \to  \ell_\alpha^\pm\ell_\beta^\mp\nu\nu$, where $\alpha =\beta$ at tree level, constitutes a signal with the SM background given by $e^- e^+ \to ZZ$.
The signature yields a ``mono-$Z$ boson'' candidate that may cause deviations of the SM predicted mono-$Z$-production cross section.
To the best of our knowledge, this signature has not yet been investigated with respect to sterile neutrino searches.

Similarly, for masses $M$ above $m_h$ the heavy neutrino can decay via a Higgs boson, which in turn decays into a pair of $\tau$ leptons. This yields the signature of a ``mono-Higgs'' candidate, similar to the one discussed in ref.\ \cite{Antusch:2015gjw} for a dijet final state (see subsection \ref{sec:dijet_ee}).

At center-of-mass energies around the $Z$ pole, the dilepton signature is generated from the decays of the heavy neutrino via off-shell $W,Z$ and Higgs bosons. Its sensitivity is included in the dashed purple line in fig.~\ref{fig:sensitivity_ee}.

\subsubsection{Dijet}\label{sec:dijet_ee}
The heavy neutrino decays via $Z$ and Higgs boson, which in turn decay hadronically, yields the final state with two jets that have an invariant mass compatible with $m_Z$ and $m_h$, respectively, and missing energy due to the light neutrinos escaping detection.

At center-of-mass energies above the $Z$ pole the signal strength is proportional to $|\theta_e|^2$ and this signature includes hadronic ``mono-Z'' and ``mono-Higgs'' candidates, which contribute to the overall $Z$ and Higgs boson production when the sterile neutrino mass $M$ is larger than $m_Z$ and $m_h$, respectively. 
The sensitivities of the mono-Higgs signature at circular lepton colliders have been investigated in ref.\ \cite{Antusch:2015gjw}. We refer to this search channel as ``mono-Higgs'' and denote it by the solid and dashed yellow lines in fig.\ \ref{fig:sensitivity_ee}.

At center-of-mass energies around the $Z$ pole the signal strength is proportional to $|\theta|^2$ and the dijet signature is generated from the decays of the heavy neutrino via off-shell $Z$ and Higgs bosons. Its sensitivity is included in the dashed purple line in fig.~\ref{fig:sensitivity_ee}.

\subsubsection{Invisible}
A completely invisible decay of the heavy neutrinos is possible via $Z$ boson decays into light neutrinos. 
This invisible signature can in principle be tested via the spectrum of initial state radiated photons.
We remark that the ``radiative return'' to the $Z$ pole allows to measure the invisible decay width of the $Z$ boson, which is dominated by the decays into light neutrinos and allows for a measurement of the non-unitarity effects from sterile neutrinos with masses larger than $m_Z$ (cf.\ ref.~\cite{Carena:2003aj}).

We note that the decays via the Higgs boson into two light neutrinos are possible but suppressed by one more order in the $\theta_\alpha$. Presently, they are constrained by Higgs boson branching ratios.

\subsubsection{Indirect searches via electroweak precision observables}
Sterile neutrinos can be searched for ``indirectly'' via the electroweak precision observables (EWPOs). Possible future sensitivities can be inferred from the precision of the future experimental measurements and the theory prediction. They are dependent on the parameters $|\theta_e|^2+|\theta_\mu|^2$ and $|\theta_\tau|^2$ separately.
In ref.~\cite{Antusch:2014woa,Antusch:2015mia,Antusch:2015rma,Antusch:2016brq} it was shown that the FCC-ee and the CEPC are sensitive to sterile neutrino masses up to about 60 and 40 TeV, respectively, via the combination $|\theta_e|^2+|\theta_\mu|^2$, using eq.~(\ref{def:thetaa}) for neutrino Yukawa couplings ${\cal O}(1)$. We refer to these estimates for the sensitivities on the active-sterile mixings as indirect searches via the EWPOs and denote them by the solid and dashed blue lines in fig.\ \ref{fig:sensitivity_ee}, which are taken from ref.~\cite{Antusch:2015mia}.

\subsubsection{Indirect searches via Higgs boson branching ratios}\label{sec:HiggsBranchingRatios}
In the presence of sterile neutrinos with non-vanishing neutrino Yukawa couplings, the total Higgs boson decay width may be enlarged due to the extra decay channels into light and heavy neutrinos, which are dependent on $|\theta|^2$. Such an enlarged total decay width may decrease the Higgs boson branching ratios into SM particles, which can be measured at future lepton colliders at the Higgs threshold and beyond as is discussed in ref.~\cite{Antusch:2015mia}. The Higgs branching ratio into two $W$ bosons is expected to be the one with the highest precision \cite{Ruan:2014xxa}, and we show the corresponding sensitivity by the red line in fig.\ \ref{fig:sensitivity_ee} from ref.~\cite{Antusch:2015mia}.

Furthermore, sterile neutrinos contribute an invisible decay channel for the Higgs boson, via their decays into three light neutrinos, or when they are sufficiently long-lived to escape the detector.
The invisible decay channel of the Higgs boson leads gives rise to the Higgs-to-invisible branching ratio with tiny SM background that can in principle be measured e.g.\ in the recoil-spectrum of the associated $Z$ boson.

\subsubsection{Displaced vertex searches}
\begin{figure*}[t!]
\begin{minipage}{0.32\textwidth}
\includegraphics[width=\textwidth]{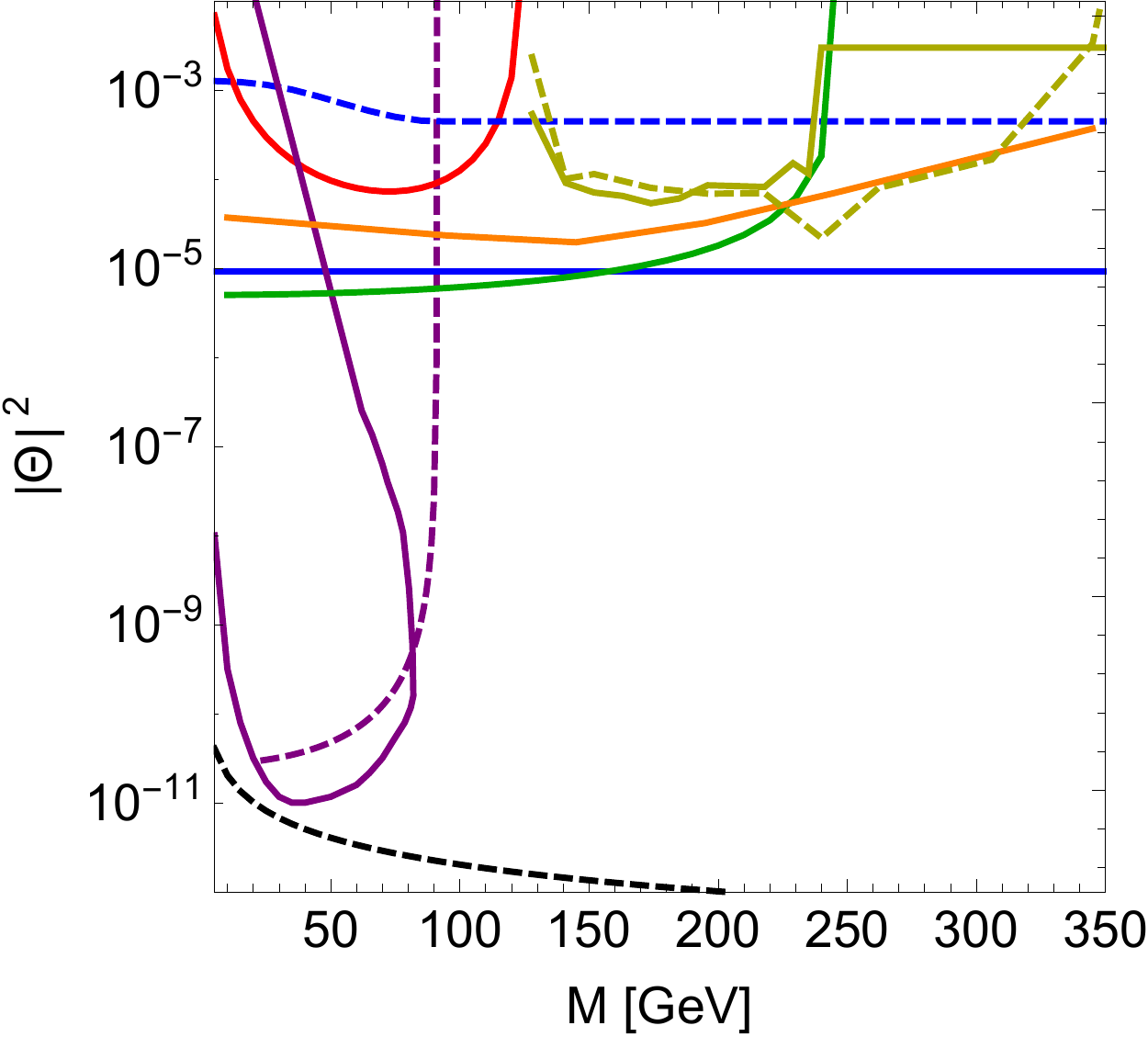}

\vspace{-50pt}\hspace{120pt}{\bf FCC-ee}
\vspace{50pt}
\end{minipage}
\begin{minipage}{0.34\textwidth}
\begin{center}
\vspace{-30pt}
\includegraphics[width=1.0\textwidth]{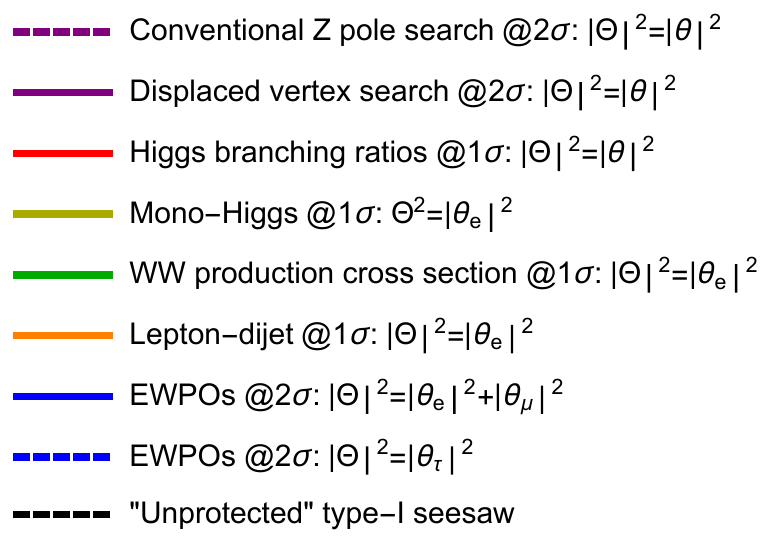}
\end{center}
\end{minipage}
\begin{minipage}{0.32\textwidth}
\includegraphics[width=\textwidth]{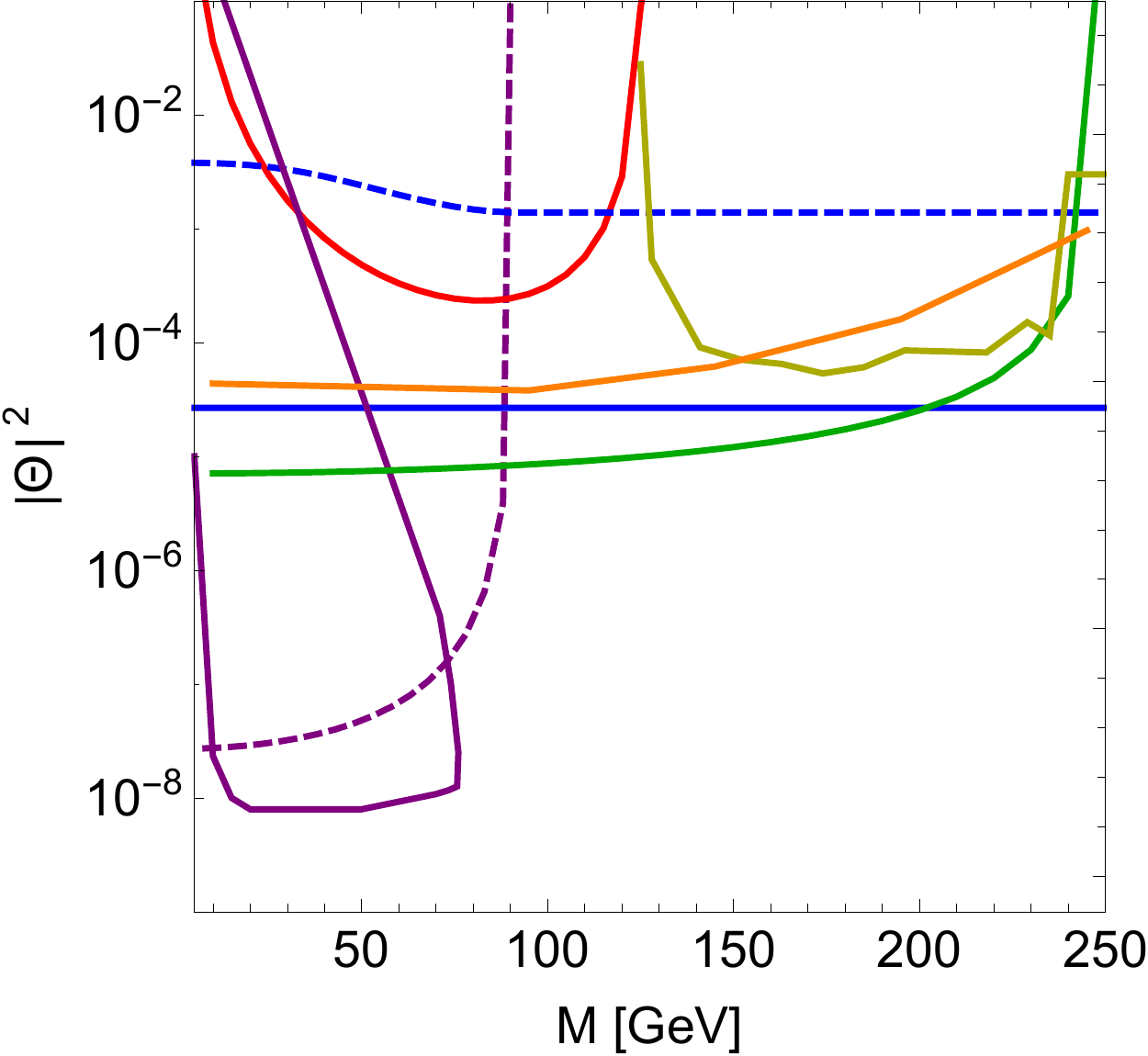}

\vspace{-50pt}\hspace{120pt}{\bf CEPC}
\vspace{50pt}
\end{minipage}
\caption{Sensitivities of the different signatures to the active-sterile mixing and masses of sterile neutrinos at the FCC-ee and the CEPC. For details on the signatures see text and tab.\ \ref{tab:signatures_ee}, for the considered modi operandi see fig.~\ref{fig:modioperandi}. See section \ref{sec:summary-ee} for a summary on the references that were used. }
\label{fig:sensitivity_ee}
\end{figure*}

Heavy neutrinos with masses below the $W$ boson mass and with very small mixings may be sufficiently ``long-lived'' to give rise to displaced vertices, i.e.\ the heavy neutrino lifetime allows for decays that feature a visible displacement from the interaction point. Via virtual $W$, $Z$ and $h$ they decay into the kinematically available SM particles.
This displaced secondary vertex constitutes an exotic and powerful signature to search for and is discussed in refs.~\cite{Antusch:2016vyf,Blondel:2014bra}. We refer to this search channel as the ``displaced vertex search'' and show its sensitivity for the $Z$-pole run, which is sensitive to $|\theta|^2$, by the solid purple line in fig.\ \ref{fig:sensitivity_ee} (taken from ref.~\cite{Antusch:2016vyf}).
We note that also for the high energy runs displaced vertex searches are possible. There, however, the production is mostly sensitive to $|\theta_e|^2$ whereas the decays depend on all the $|\theta_\alpha |$.

\subsubsection{LFV signatures}
The lepton-flavour-violating decays of a $Z$ boson into two charged leptons with different flavour constitute a clear signal for beyond the SM physics. In the presence of sterile neutrinos, such lepton-flavour-violating decays arise at the one loop level (as discussed in section \ref{sec:LFV}, LFV is absent at tree level). For the FCC-ee, this signal has been investigated in the context of sterile neutrino models in refs.~\cite{Abada:2014cca,Abada:2015zea}. Furthermore, also lepton-flavour violating Higgs boson decays into two charged leptons of different flavour may be measurable, see e.g.\ refs.\ \cite{Arganda:2004bz,Arganda:2014dta,Banerjee:2016foh}. 

\subsection{Electron-positron colliders: summary}
\label{sec:summary-ee}
In the above section we presented and discussed a complete list of the signatures for sterile neutrino searches at $e^- e^+$ colliders at leading order.
Here we summarize our findings, including results from previous works on the prospects of sterile neutrino searches at future $e^-e^+$ colliders (within the SPSS benchmark model). We have extended and updated the summary plot shown in fig.~\ref{fig:sensitivity_ee} for the CEPC and FCC-ee in the following ways:

We present first parton-level estimates for the lepton-dijet signature at $\sqrt s =$250~GeV and 350~GeV for the CEPC and FCC-ee, respectively, by the orange line.
We updated the sensitivities for the mono-Higgs signature from ref.~\cite{Antusch:2015gjw} for the dijet final state (at the reconstructed level)  according to the luminosity goals as given in fig.~\ref{fig:modioperandi}. This signature is shown for the CEPC and FCC-ee with the solid and dashed yellow lines for $\sqrt s =$250 and 350~GeV, respectively.
The sensitivity estimate for the conventional $Z$ pole search is shown by the dashed purple line, and it was obtained from a trivial rescaling (cf.\ ref.~\cite{Antusch:2015mia}) of the DELPHI results.
The sensitivity from the displaced vertex searches for sterile neutrinos at the $Z$ pole is taken from ref.~\cite{Antusch:2016vyf} and shown by the solid purple lines.
The estimates for the sensitivity of the indirect searches from the EWPOs, the Higgs boson branching ratios, and the sensitivity estimate for the dilepton final states at $\sqrt s =$250~GeV, are from ref.~\cite{Antusch:2015mia} .

We note that the $Z$ pole run of an $e^-e^+$ collider allows to test the active-sterile mixing parameter $|\theta|^2$, whereas the physics runs at higher energies, starting with the $WW$ threshold scan, are mainly sensitive to $|\theta_e|^2$.
The relative strength of the $|\theta_\alpha|$ can be inferred, e.g.\ from the lepton-dijet final states.

We find that the best sensitivity is given by the displaced vertex searches at the $Z$ pole, which can test $|\theta|^2$ as small as $\sim 10^{-8}$ and $\sim 10^{-11}$ at the CEPC and the FCC-ee, respectively, for heavy neutrino masses below $m_W$. It is worth noting that the sensitivity of the FCC-ee in this channel is even closing in on the active-sterile mixing that is expected from the na\"ive type-I seesaw relation.

Among the direct searches for the physics runs above the $Z$ pole, our estimate for the lepton-dijet signature shows a comparable sensitivity to the dilepton and the mono-Higgs signatures, and allow tests of $|\theta_e|^2$ down to $\sim 10^{-5}$ at the CEPC and FCC-ee, respectively. We remark that that the mono-$Z$ signature has not yet been investigated with respect to sterile neutrino searches, but we expect its sensitivity to be similar to the aforementioned direct searches.

The indirect searches for effects from sterile neutrinos via the electroweak precision observables allow to test the combination $|\theta_e|^2+|\theta_\mu|^2$ down to values slightly below $\sim 10^{-4}$ and $\sim 10^{-5}$ at the CEPC and FCC-ee, respectively, and they allow to test masses $M$ above the center-of-mass energy and well into the ${\cal O}(10)$ TeV range.

We remark that lepton-number-violating processes exist, that contribute to the lepton-dijet and to the dilepton final states. However, they always contain a light neutrino in the final state such that there is no unambiguous LNV signature. Nevertheless, using the different kinematics of signature and background final states with different light neutrino lepton number, it might be possible to find a signal of LNV at $e^-e^+$ colliders.

We have focused here on circular colliders, however we like to note that also linear colliders such as the ILC or CLIC can efficiently search for sterile neutrinos via the same signatures. The latter will also run at higher energies (cf.\ fig.\ \ref{fig:modioperandi}) and therefore their direct searches have an extended mass reach for testing heavy neutrinos, compared to the circular colliders.

\section{Searches at future $pp$ colliders}
Hadron colliders are designed to collide protons at highest center-of-mass energies.
The currently operating LHC can reach up to 14 TeV center-of-mass energy and with the high luminosity upgrade (HL-LHC), with its construction planned to begin in 2018, it is foreseen to collect 3~ab$^{-1}$ of data after 2030\cite{Apollinari:2015bam}. The currently discussed next generation hadron colliders, the Future Circular hadron Collider (FCC-hh)~\cite{Golling:2016gvc,Mangano:2016jyj,Contino:2016spe} and the Super proton-proton Collider (SppC)~\cite{Tang:2015qga}, envisage center-of-mass energies of up to 100~TeV and 70~TeV, respectively, with target integrated luminosities of around 20~ab$^{-1}$\cite{Hinchliffe:2015qma}.

Many authors have worked on the search for sterile neutrinos at the LHC, where they are often referred to as ``heavy neutral leptons'', see e.g.\ ref.\ \cite{Deppisch:2015qwa} and references therein for an overview.
In this section we discuss the signal channels of heavy neutrinos from the third column in fig.~\ref{fig:production_and_decay} and cast a first look at possible sensitivity prospects for sterile neutrino searches at the HL-LHC and FCC-hh/SppC, where we assume for the latter 100 TeV center-of-mass energy and 20 ab$^{-1}$ total integrated luminosity.

\subsection{Production mechanism}
\begin{figure}[t]
\begin{center}
{\bf\small Drell-Yan}

\includegraphics[width=0.3\textwidth]{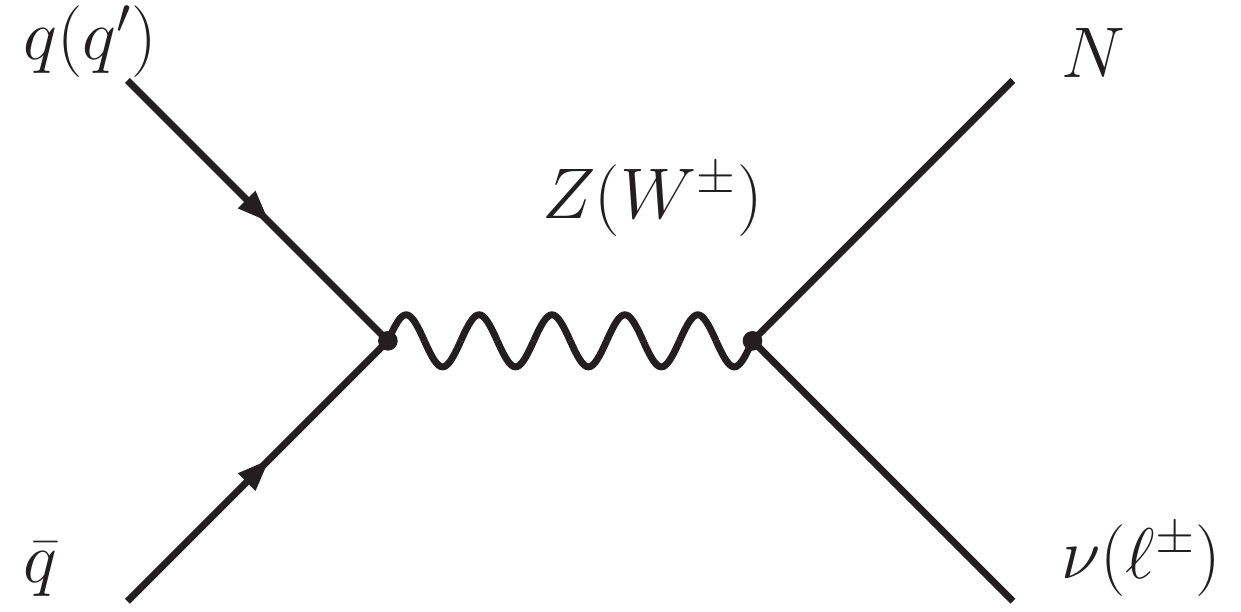}

production channel: $\mathbf{Z_s \;(W_s)}$
\end{center}
\caption{Dominant Feynman diagrams for the production of heavy neutrinos with masses below $\sim 1$ TeV in proton-proton collisions.
}
\label{fig:nproduction-hh}
\end{figure}

As discussed in sec.~\ref{sec:production processes}, heavy neutrinos can be produced in hadronic collisions e.g.\ from Drell-Yan processes (cf.\ fig.\ \ref{fig:nproduction-hh}), from Higgs boson decays, and in gauge boson fusion. The dominating production mechanism for smaller center-of-mass energies is Drell-Yan, while $W\gamma$ fusion becomes more important at higher center-of-mass energies and for larger $M$ \cite{Alva:2014gxa}.

The Drell-Yan production of a heavy neutrino yields an associated light neutrino or a (negatively or positively) charged lepton, for the production channel $\mathbf{Z_s}$ or $\mathbf{W_s}$, respectively, cf.~fig.~\ref{fig:production_and_decay}. The corresponding production cross sections for all the three possible associated leptons are quantitatively similar for the here considered center-of-mass energies and heavy neutrino masses. The exception to this statement is for 50 GeV $\leq M \leq 200$ GeV, where the production cross section $\sigma(pp\to \nu N)$ (via $\mathbf{Z_s}$) can be up to an order of magnitude larger than $\sigma(pp\to \ell^\pm N)$ (via $\mathbf{W_s}$).

\begin{figure}[t]
\begin{center}
\includegraphics[width=0.45\textwidth]{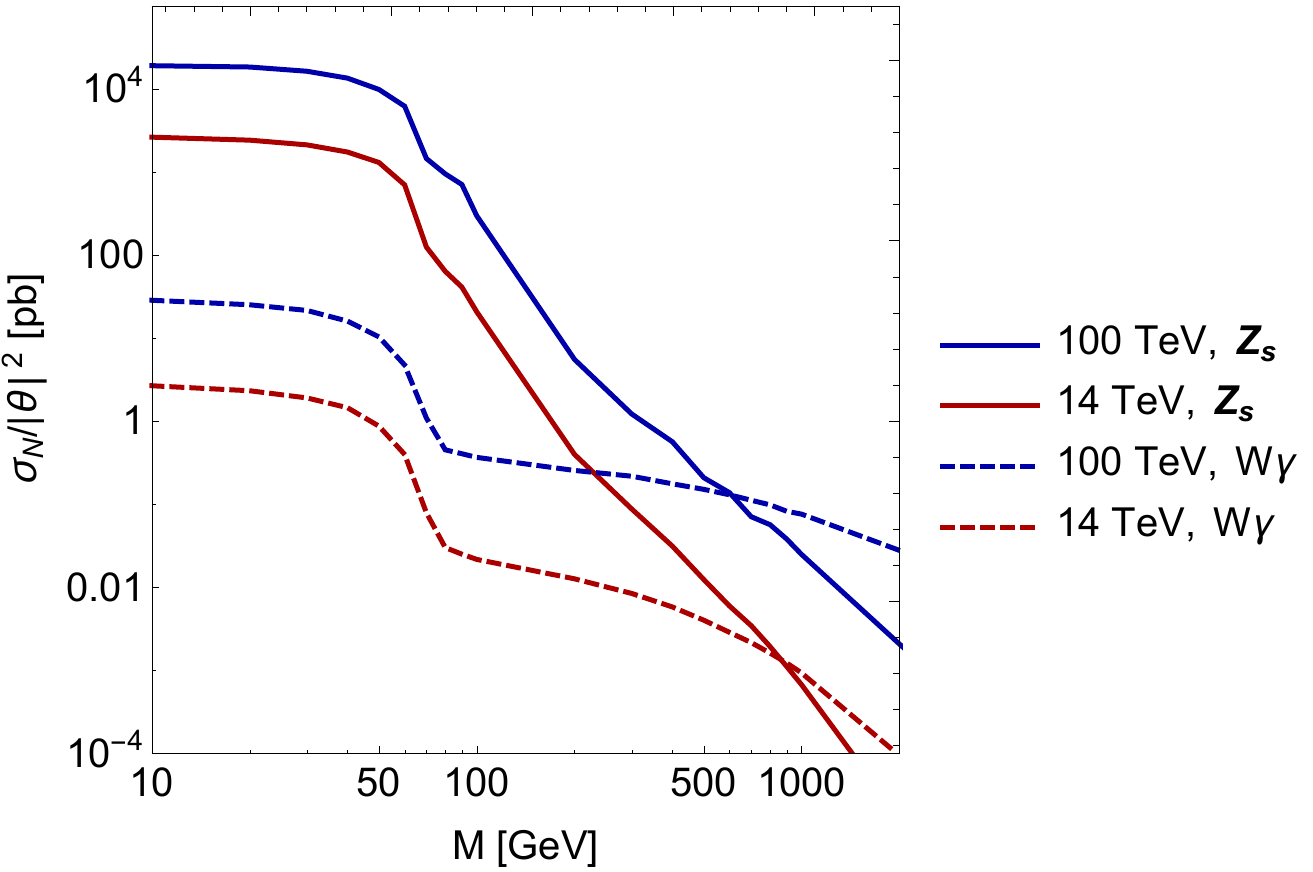}
\end{center}
\caption{Production cross sections $\sigma_{N}$, divided by $|\theta|^2$, for the process $p\,p \to N\,\nu$ via $\mathbf{Z_s}$ and $W\gamma$, with $p$ being a parton and $N$ a heavy neutrino. Production via $W\gamma$ fusion becomes relevant for larger heavy neutrino masses despite being higher order than $\mathbf{Z_s}$. The production cross section for the process $p\,p \to N\,\ell_\alpha$ via $\mathbf{W_s}$, with $\ell_\alpha$ being a charged lepton, is very similar to the one via $\mathbf{Z_s}$. }
\label{fig:cross-section-hh}
\end{figure}

We show the sterile neutrino production cross section in proton-proton collisions for $\mathbf{Z_s}$, and $\mathbf{W_s}$ respectively, and $W\gamma$ fusion in fig.\ \ref{fig:cross-section-hh}. We remark that $W\gamma$ fusion is included in the figure despite being higher order, because it becomes relevant for larger masses $M$ whereas the production cross section via $\mathbf{Z_s}$ and $\mathbf{W_s}$ are more strongly suppressed for large $M$ due to the high virtuality of the intermediate $Z$ and $W$ bosons \cite{Alva:2014gxa}. 
Furthermore, also the heavy neutrino production via the Higgs boson becomes increasingly relevant for large $M$, and has been shown to dominate over Drell-Yan at $\sqrt{s}=100$ TeV for $M > 1.5$ TeV, see refs.\ \cite{Hessler:2014ssa,Degrande:2016aje}.
Nevertheless, we will focus on the leading order signatures via $\mathbf{Z_s}$ and $\mathbf{W_s}$ in the following.

\subsection{Signatures and searches}
\begin{table*}[t]
\begin{center}
\begin{tabular}{|l|c|c|c|c|}
\hline
Name & Final State & Channel [production,decay] & $|\theta_\alpha|$ dependency & LNV/LFV \\ \hline\hline
dilepton-dijet & $\ell_\alpha\ell_\beta jj$ & [$\mathbf{W_s},W$] 
 & $\dfrac{|\theta_\alpha\theta_\beta|^2}{\theta^2}$
& \checkmark/\checkmark \\ \hline
trilepton
& $\ell_\alpha\ell_\beta\ell_\gamma\nu$ & [$\mathbf{W_s},\{W,Z (h)\}$]
 & $\left\{\dfrac{|\theta_\alpha\theta_\beta|^2}{\theta^2}^{(*)}
,{|\theta_\alpha|^2}^{(*)}\right\}$
&$\times$/\checkmark \\ \hline
lepton-dijet
& $\ell_\alpha\nu jj$ &  [$\mathbf{W_s}, Z (h)$], [$\mathbf{Z_s},W$]
 & $|\theta_\alpha|^2$ $\vphantom{\dfrac{{}^1}{{}^1}}$ 
&$\times$ \\ \hline
dilepton & $\ell_\alpha\ell_\beta \nu\nu$ & [$\mathbf{Z_s},\{W,Z (h)\}$]
 & $\left\{{|\theta_\alpha|^2}^{(*)},{|\theta|^2}^{(*)}\right\}$ $\vphantom{\dfrac{{}^1}{{}^1}}$ 
& $\times$ \\ \hline
mono-lepton & $\ell_\alpha \nu\nu\nu$ & [$\mathbf{W_s}, Z$]
 & $|\theta_\alpha|^2$ $\vphantom{\dfrac{{}^1}{{}^1}}$
& $\times$ \\ \hline
dijet & $\nu\nu jj$ & [$\mathbf{Z_s},Z (h)$] 
 & $|\theta|^2$ $\vphantom{\dfrac{{}^1}{{}^1}}$
& $\times$ \\ \hline
\end{tabular}
\end{center}
\caption{Signatures of sterile neutrinos at leading order for $pp$ colliders with their corresponding final states, production and decay channels (cf.\ section \ref{sec:production-and-decay}), and their dependency on the active-sterile mixing parameters.
A checkmark in the ``LNV/LFV'' column indicates that an unambiguous signal for LNV and/or LFV is possible (cf.\ discussion in sections \ref{sec:LNV} and \ref{sec:LFV}). \\
$^{(*)}:$ The dependency on the active-sterile mixing can be inferred when the origin of the charged leptons is known.}
\label{tab:signatures_pp}
\end{table*}

In this section, we discuss observable effects from sterile neutrinos at $pp$ colliders, which manifest themselves in specific final states with the related production and decay channels, cf.~fig.~\ref{fig:production_and_decay}, the dependency on the active-sterile mixing angles, lepton-number and lepton-flavour violation.
We refer to these effects as signatures and list them in tab.\ \ref{tab:signatures_pp}.

In the following we discuss these signatures one by one, including existing search strategies employed for sterile neutrino searches at the LHC. Moreover, we present ``first looks'' for the $1\sigma$ sensitivities of each channel to the sterile neutrino parameters at the parton level for the HL-LHC, and also for the FCC-hh/SppC.
We refer the reader to section \ref{app:distributions-pp} in the appendix on the calculation of the sensitivities.

\subsubsection{Dilepton-dijet}
\label{sec:Dilepton-dijet}
Heavy neutrino production via $\mathbf{W_s}$ together with their decays via the charged current gives rise to the semileptonic final state $\ell_\alpha \ell_\beta jj$. The hadronic jets $j$ stem from the intermediate $W$ boson, which is on-shell for $M>m_W$, leading to the dijet invariant mass $M_{jj} \sim m_W$.
The event rate is sensitive to the combination of $|\theta_\alpha|^2$ and $|\theta_\beta|^2/|\theta|^2$ through the production and decay channel, respectively, and the flavour indices $\alpha,\,\beta$ can be inferred from the charged leptons.

This signature includes ``same-sign'' dileptons, which violate lepton number by two units and has no SM background, apart from mis-identification (cf.\ however footnote \ref{fntLNV}).
It has been studied for the LHC 
 \cite{Han:2006ip,delAguila:2007qnc,delAguila:2008cj,Atre:2009rg,Chao:2009ef,Das:2012ze,Das:2015toa,Das:2016hof} and also for luminosities of 1 ab$^{-1}$ at the LHC and the FCC-hh in \cite{Alva:2014gxa}.
Furthermore, the question of how to infer the CP-phases, mixings, and mass ratios in the neutrino sector using di-lepton events has been discussed in \cite{Bray:2007ru,Gluza:2016qqv}.

The dilepton-dijet signature gives also rise to lepton-number conserving ``opposite-sign'' dileptons. 
We display our first estimate for a possible $1\sigma$ sensitivity of the HL-LHC and the FCC-hh/SppC to this signature with the green line in fig.~\ref{fig:sensitivity-hh}. Therein, we assumed one fixed flavour $\alpha$ with $\beta = \alpha$ and $|\theta_\delta|=0$ for all $\delta\neq \alpha$.

The final state $\ell_\alpha^\pm \ell_\beta^\mp j j$ for $\alpha\neq\beta$ yields an unambiguous signal for lepton flavour violation (at the parton level), as discussed in section \ref{sec:LFV} (cf.\ also \cite{Arganda:2015ija}, where this signature has been discussed for the ``Inverse Seesaw'' model at the LHC). We show our estimate for a possible $1\sigma$ sensitivity as the solid purple line in fig.~\ref{fig:sensitivity-hh}. Therein, we assumed $|\theta_\alpha|=|\theta_\beta|\neq 0$ and $|\theta_\delta|=0$ for $\delta\neq \alpha,\beta$.
We remark that although this signature has no background at the parton level, due to the finite resolution for the missing momentum at the detector level some SM processes with additional light neutrinos become relevant backgrounds. We consider here as background the dominant ditop production, as detailed in the appendix \ref{app:distributions-pp}. 
Without the backgrounds, sensitivities as small as $\sim {\cal O}(10^{-6})$ and $\sim {\cal O}(10^{-8})$ would be possible for the HL-LHC and the FCC-hh, respectively.

Moreover, we include a $1\sigma$ sensitivity estimate for the $\ell_\alpha^\pm \ell_\beta^\mp j j$ signature with $\alpha\neq\beta$
for the run 2 of the LHC with 50~fb$^{-1}$ by the purple dotted line in fig.~\ref{fig:sensitivity-hh}.

\subsubsection{Trilepton}\label{sec:trilepton}
The production channel $\mathbf{W_s}$ can give rise to the final state $\ell_\alpha \ell_\beta \ell_\gamma \nu$ for all the possible decay modes of the heavy neutrino when the $W,\,Z,$ and Higgs boson decay leptonically. For the LHC, this signature was, for instance, studied in refs.~\cite{Das:2012ze,delAguila:2008cj,delAguila:2008hw,Bray:2007ru,delAguila:2009bb,Das:2015toa}.

The leptonic decays of the $Z$ and Higgs boson yield $\ell_\beta \ell_\gamma$, with $\beta=\gamma$, and the invariant mass $M_{\ell_\beta \overline{\ell}_\gamma}$ is $\sim m_Z$ or $\sim m_h$, respectively, and for the Higgs boson only $\beta=\gamma=\tau$ has a sizeable branching ratio. This channel is sensitive to $|\theta_\alpha|^2$ given that the leptons from the decaying $Z$ or Higgs boson can be identified as such.

The decays of a heavy neutrino into a $W$ boson are accompanied by a charged lepton $\ell_\beta$, and the leptonic decays of the $W$ yield $\ell_\gamma\nu$. 
In this channel, the three lepton flavour indices are independent from each other, and the observable effects from heavy neutrinos are sensitive to $|\theta_\alpha \theta_\beta|^2/|\theta|^2$ given that the lepton stemming from the $W$ decay (i.e.\ $\ell_\gamma$) can be identified as such. 

We note that this signature may allow to test lepton-number violation, which may be separable from the lepton-number conserving channels using differences in the kinematic distributions of the leptons, as studied in ref.~\cite{Dib:2015oka,Dib:2016wge}.

Furthermore, the final state $\ell_e \ell_\mu \ell_\tau \nu$ yields an unambiguous signal of lepton flavour violation (at the parton level), with SM background only appearing at higher order with two additional light neutrinos in the final state.
This final state has been discussed in the context of a radiative neutrino mass model at the LHC in ref.\ \cite{Cherigui:2016tbm}.
We display a first estimate for the sensitivity within low scale seesaw models (using the SPSS as benchmark) for the HL-LHC and the FCC-hh/SppC by the blue line in fig.~\ref{fig:sensitivity-hh}. Therein we assumed $|\theta_\alpha|=|\theta_\beta|\neq 0$ and $|\theta_\delta|=0$ for $\delta\neq \alpha,\beta$ for simplicity (and treated the background as detailed in the appendix \ref{app:distributions-pp}).
Without the backgrounds, sensitivities as small as $\sim {\cal O}(10^{-5})$ and $\sim {\cal O}(10^{-7})$ would be possible for the HL-LHC and the FCC-hh, respectively.

\subsubsection{Lepton-dijet}
The effects from heavy neutrinos in the final state $\ell_\alpha \nu j j$ are sensitive to the active-sterile mixing $|\theta_\alpha|^2$, and the lepton flavour index can be inferred from the charged lepton. This signature can be created from heavy neutrinos via two different production and decay channels. 

The first possibility is the production of the heavy neutrino via $\mathbf{W_s}$ and its decay via a $Z$ or a Higgs boson into a neutrino and a dijet, with the invariant mass of the two jets $M_{jj} \sim m_Z$ or $m_h$, respectively.

The second possibility is given by $\mathbf{Z_s}$, with the subsequent decay of the heavy neutrino via the charged current into a charged lepton and two jets, where the invariant dijet mass $M_{jj}\sim m_W$, for $M>m_W$.
This channel features a resonance in the transverse mass distribution of the $\ell_\alpha^\pm jj$ system at the mass $M$.

We display first estimates for the sensitivities of the lepton-dijet final state at the HL-LHC and the FCC-hh/SppC by the red line in fig.~\ref{fig:sensitivity-hh}.

\subsubsection{Dilepton}
The final state $\ell_\alpha \ell_\beta \nu \nu$, with the neutrinos giving rise to missing transverse momentum, can be created from heavy neutrinos that are produced via $\mathbf{Z_s}$, and decay purely leptonically. 

In the case of the intermediate boson being a $Z$ or a Higgs the invariant mass $M_{\ell_\alpha \ell_\beta} = m_Z$ or $m_h$, respectively, and the lepton flavour indices are identical ($\alpha=\beta$). This channel may be referred to as a leptonic ``mono-Z'', or leptonic ``mono-Higgs'', which both are dependent on $|\theta|^2$.
In the case of the heavy neutrino decaying via the charged current, the leptonic $W$ decay can give rise to the dilepton final state with $\alpha\neq\beta$, which is dependent on $|\theta_\alpha|^2$.

We display first estimates for the sensitivity to $|\theta|^2$ of the dilepton final state at the HL-LHC and the FCC-hh/SppC by the cyan line in fig.~\ref{fig:sensitivity-hh}. The combined contribution from $W, Z$ and Higgs boson decays to the final states is obtained by summing over the flavour indices of the charged leptons. Here, for simplicity, only one of the $|\theta_\alpha|$ is taken to be non-zero.

\subsubsection{Monolepton}
The production of a heavy neutrino via $\mathbf{W_s}$ and its subsequent decay into three neutrinos, via a $Z$ boson, gives rise to the final state of a single charged lepton and missing energy. This signature is dependent on $|\theta_\alpha|^2$, which can be inferred from the charged lepton.
It is typically used in searches for dark matter, see e.g.~refs.~\cite{Bai:2012xg,Bell:2015rdw}. To the best of our knowledge, it has not been studied before with respect to searches for sterile neutrinos.

We display our own first estimates for the sensitivities of the monolepton final state at the HL-LHC and the FCC-hh/SppC by the orange line in fig.~\ref{fig:sensitivity-hh}.

\subsubsection{Dijet}
The production of a heavy neutrino via $\mathbf{Z_s}$ and its subsequent decay into two hadronic jets via an intermediate $Z$ or Higgs boson leads to the final state $jj\nu\nu$, with the invariant mass of the dijet $M_{jj}\sim m_Z$ or $m_h$, respectively. This process is dependent on the active-sterile mixing parameter $|\theta|^2$. 
It is often referred to as ``mono-$Z$'' or ``mono-Higgs'' depending on the invariant mass of the dijet system, and it is used for ``hidden sector'' searches at the LHC, see e.g.\ ref.~\cite{Petriello:2008pu}. 

To the best of our knowledge we are discussing the ``mono-$Z$'' signature here in the context of sterile neutrinos for the first time. The ``mono-Higgs'' signature at the LHC has been studied in \cite{Basso:2015aee}, however it was found that it is not very promising due to large top-related backgrounds.
We display our first estimates for the sensitivities of the mono-$Z$ signature with dijet final state at the HL-LHC and the FCC-hh/SppC by the yellow line in fig.~\ref{fig:sensitivity-hh}.

\subsubsection{Searches via Higgs boson properties}
As discussed in section \ref{sec:HiggsBranchingRatios} for the case of the proposed $e^- e^+$ colliders, searches for sterile neutrinos via Higgs boson branching ratios are also possible (although with less sensitivity) at future $pp$ colliders, as has been studied for the LHC in ref.~\cite{BhupalDev:2012zg,Cely:2012bz}. Effects of sterile neutrinos on measurements of the Higgs self-coupling have been discussed recently in ref.\ \cite{Baglio:2016ijw}.

\subsubsection{LFV signatures}
At the loop level, sterile neutrinos can induce lepton flavour violating decays of the Higgs and Z boson into charged leptons. Searches for heavy neutrinos via these signatures have been studied for the LHC in \cite{Arganda:2014dta,Banerjee:2016foh,Herrero-Garcia:2016uab,DeRomeri:2016gum}.

We note that the LFV signatures $\ell_\alpha \ell_\beta jj$ and $\ell_\alpha\ell_\beta\ell_\gamma\nu$ that can occur via heavy neutrinos at tree level are discussed above.

\subsubsection{Displaced vertex searches}
\label{sec:displacedvertex-pp}
\begin{figure}
\centering
\includegraphics[width=0.4\textwidth]{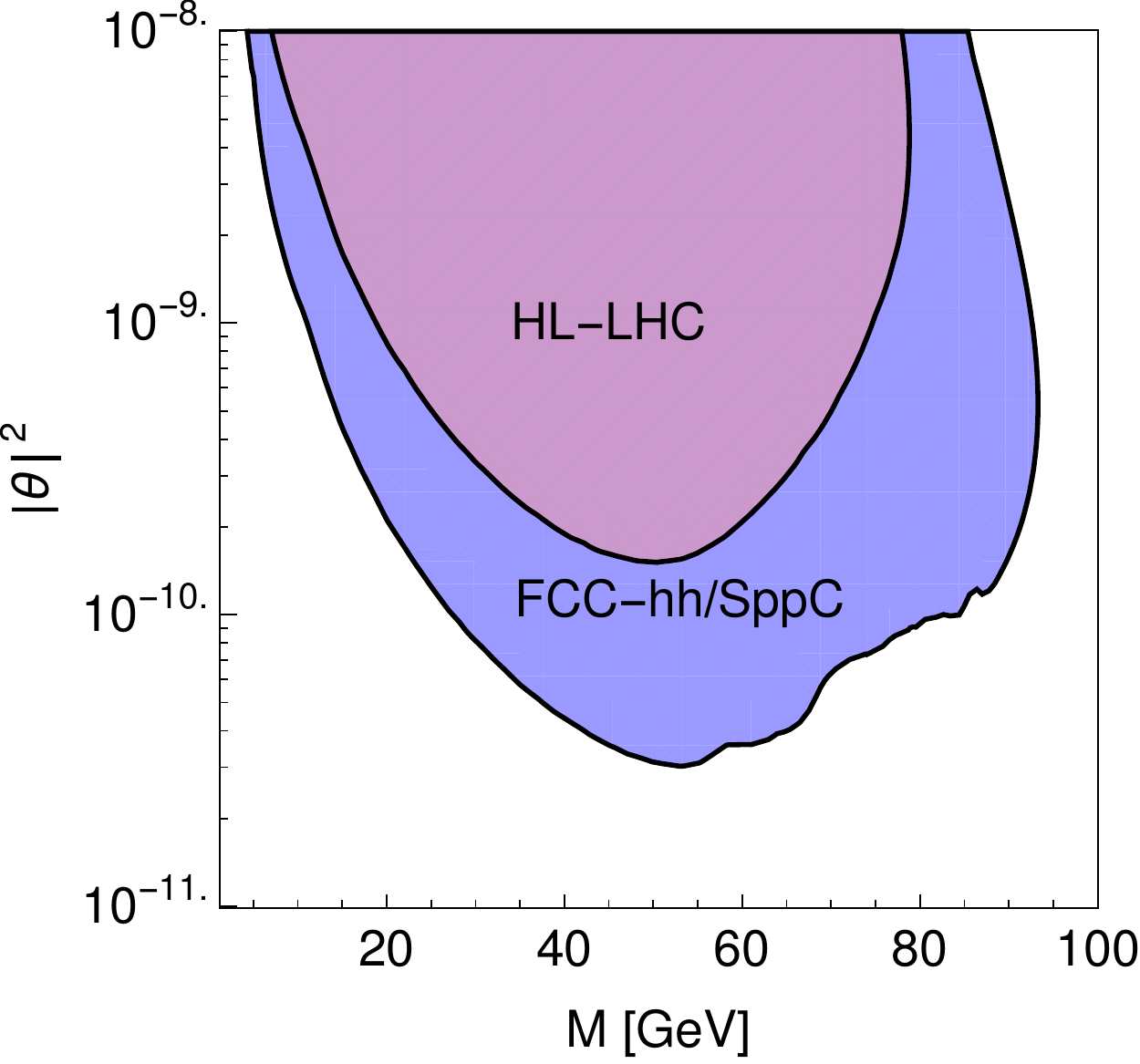}
\caption{First look at the sensitivity of the HL-LHC and the FCC-hh/SppC to sterile neutrinos via displaced vertex searches, where $|\theta|^2=\sum_\alpha |\theta_\alpha|^2$. For our estimate, we have considered vertex displacements between 1 mm and 1 m as background-free, 100\% signal efficiency, and an average Lorentz factor of 40 and 100 for the HL-LHC and the FCC-hh/SppC, respectively.}
\label{fig:displacedvertex-pp}
\end{figure}

Heavy neutrinos with $M\leq m_W$ and small active-sterile mixing can be ``long lived'' such that they decay with a measurable displacement from the interaction point. For integrated luminosities of 0.3 ab$^{-1}$ these searches at the HL-LHC have been studied in different theoretical frameworks in refs.~\cite{Graesser:2007pc,Perez:2009mu,Helo:2013esa,Gago:2015vma,Izaguirre:2015pga}. 
The FCC-hh produces a comparable number of heavy neutrinos to the FCC-ee, however, a dedicated analysis of the sensitivity has not been done yet.
For a first look at the possible sensitivity of the HL-LHC and the FCC-hh/SppC to sterile neutrinos via this signature, we proceed analogously to ref.\ \cite{Antusch:2016vyf}. We assume that vertices with a displacement from the interaction region of at least 1 mm and at most 1 m have no background  and can be measured with 100\% efficiency (as in \cite{Helo:2013esa}). Furthermore, from the kinematics of the heavy neutrinos we assume an average Lorentz factor of 40 and 100 from proton-proton collisions at 14 and 100 TeV, respectively. We show the estimated sensitivity of the $pp$ colliders to $|\theta|^2$ corresponding to at least four events in fig.\ \ref{fig:displacedvertex-pp}. We stress that due to the much more challenging experimental environment, we expect that the signal efficiency will be much lower than at $e^- e^+$ colliders. For a realistic estimate of the sensitivity a thorough study of the detector response and the backgrounds is required.

\begin{figure*}[t]
\begin{minipage}{0.34\textwidth}
\includegraphics[width=0.965\textwidth]{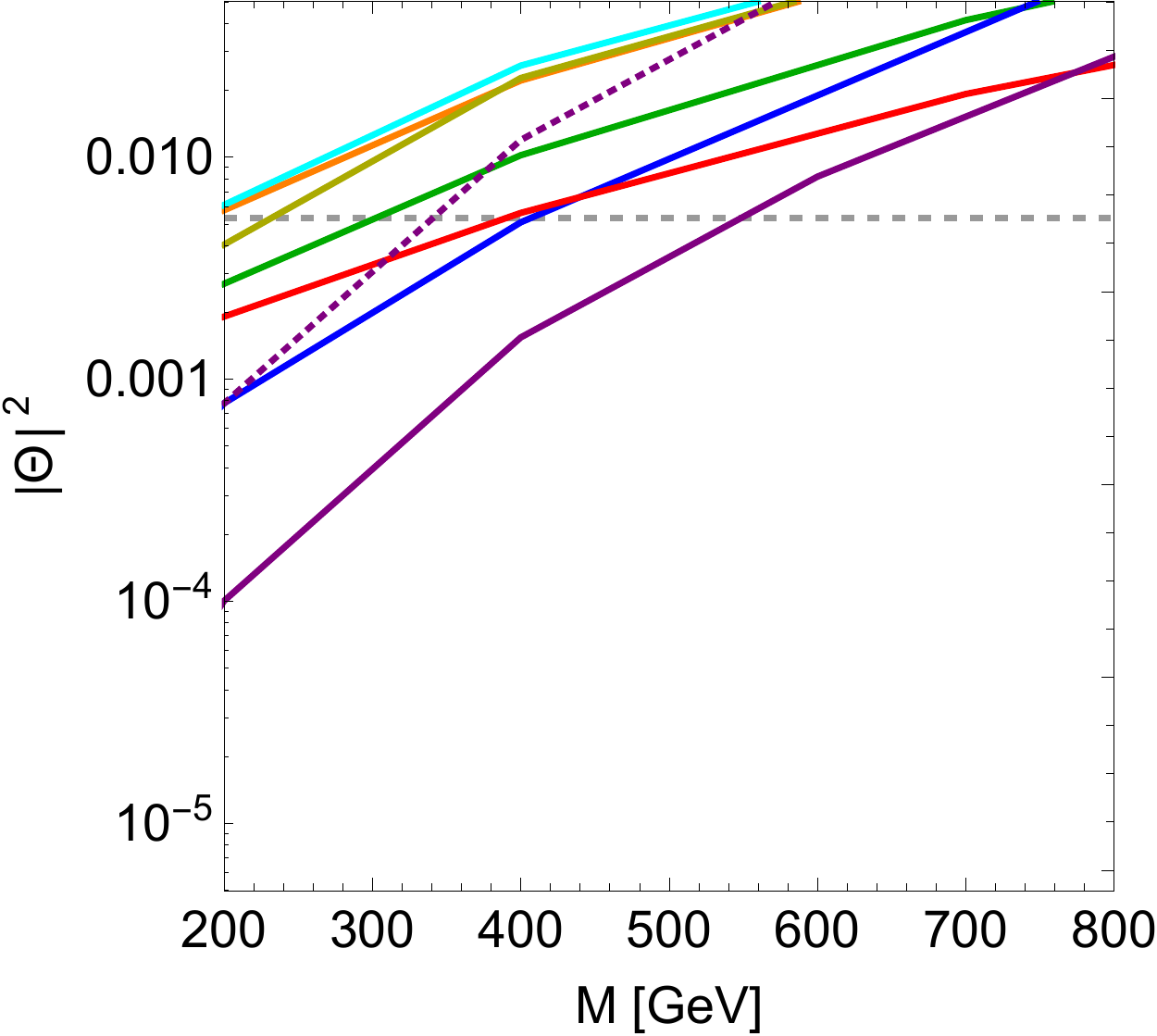}

\vspace{-50pt}\hspace{120pt}{\bf HL-LHC} 
\vspace{50pt}

\end{minipage}
\begin{minipage}{0.30\textwidth}
\begin{center}
\vspace{-30pt}
\includegraphics[width=\textwidth]{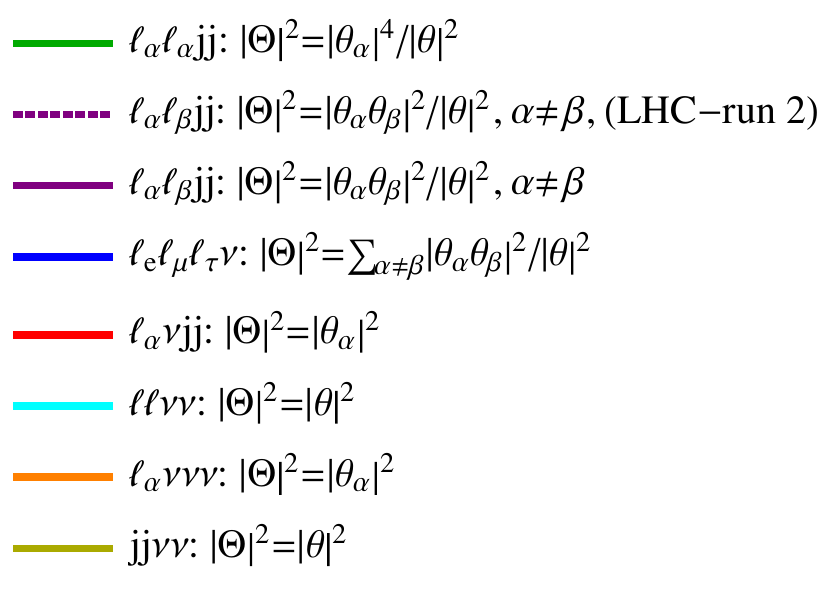}
\end{center}
\end{minipage}
\begin{minipage}{0.34\textwidth}
\includegraphics[width=1.015\textwidth]{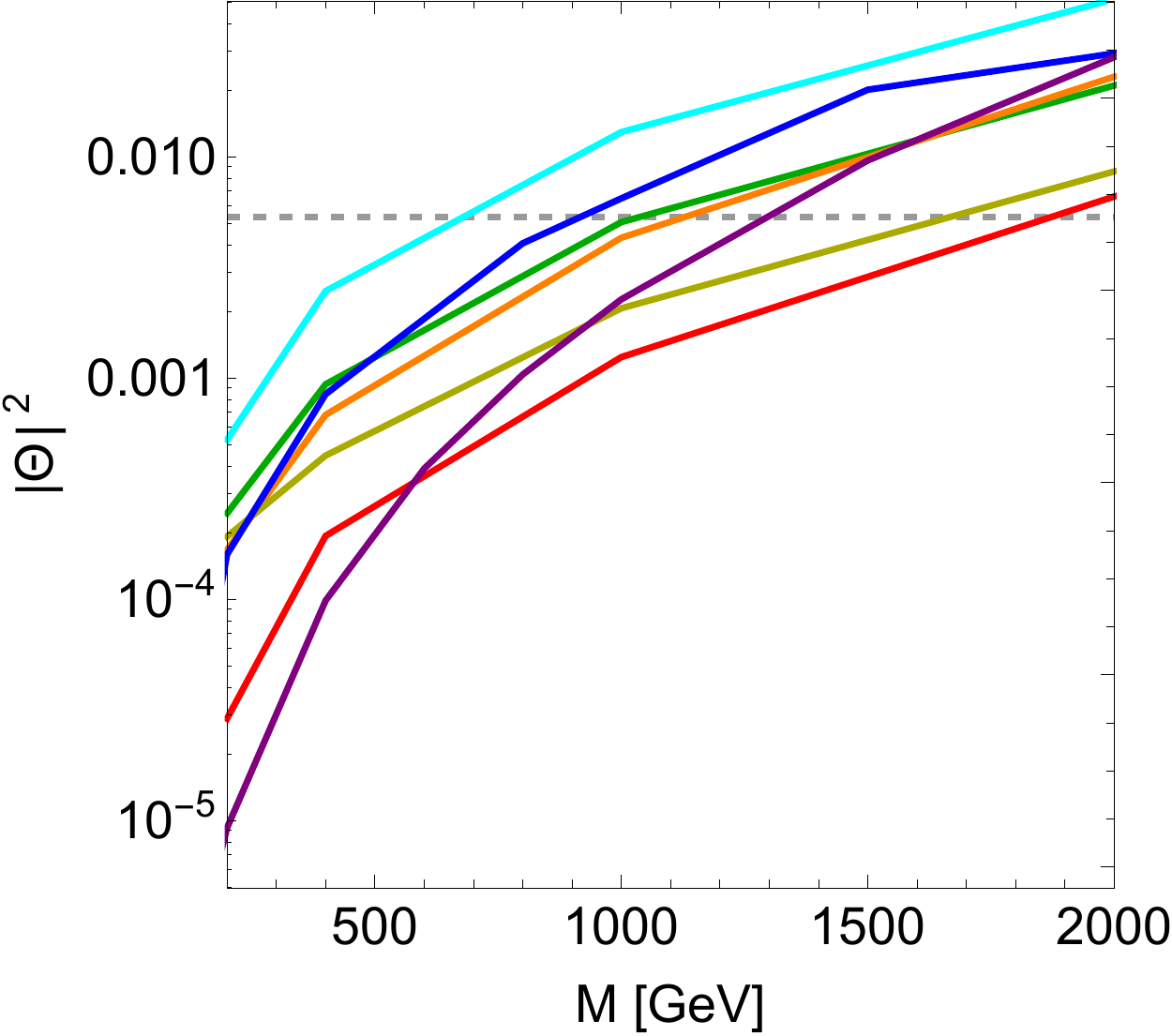}

\vspace{-50pt}\hspace{90pt}{\bf FCC-hh/SppC}
\vspace{50pt}
\end{minipage}
\caption{First look at the possible $1\sigma$ sensitivity of the lepton-number-conserving signatures (see tab.~\ref{tab:signatures_pp}) for sterile neutrino searches at $pp$ colliders. We consider an integrated total luminosity of 3 and 20 ab$^{-1}$ for the HL-LHC ($\sqrt s =$ 14 TeV) and the FCC-hh/SppC ($\sqrt s =$ 100 TeV), respectively.
The grey horizontal line denotes the present upper bound on the mixing angle $|\theta_\tau|^2$ at the 90$\%$ confidence level.
For details on the calculation of the sensitivities on the parton level, see section \ref{app:distributions-pp} in the appendix. }
\label{fig:sensitivity-hh}
\end{figure*}

\subsection{Proton-proton colliders: summary}
In this section we summarize our findings regarding the sensitivities for sterile neutrinos at future $pp$ colliders, for which we presented a complete list of signatures at leading order in the previous section. 

We present here a ``first look'' at the possible sensitivities of sterile neutrino searches via lepton-number-conserving final states and for sterile neutrino masses larger than 200~GeV in fig.~\ref{fig:sensitivity-hh}, assuming a total integrated luminosity of 3 ab$^{-1}$ and 20 ab$^{-1}$ for the HL-LHC and the FCC-hh/SppC, respectively.
We emphasize that our estimates are calculated at the parton level, and for all the new signatures a more thorough analysis on the reconstructed level should be done in the future.
In the figure, the grey dashed horizontal line denotes the present upper bound on the mixing angle $|\theta_\tau|^2$ at the 90$\%$ confidence level.

We note that the hadron colliders are sensitive to $|\theta_e|$ , $|\theta_\mu|$  and $|\theta_\tau|$ independently, and it is in principle possible to infer the relative strength of the $|\theta_\alpha|$ e.g.\ via the lepton-dijet final states.

We find that the HL-LHC can test sterile neutrinos with masses up to $\sim$ 450~GeV that are compatible with present constraints on active-sterile mixing. The FCC-hh/SppC enhances this mass reach to $\sim$ 2~TeV.
The best sensitivities for $M > 200$ GeV are given by the lepton-flavour violating dilepton-dijet final states $\ell_\alpha\ell_\beta jj$ for $\alpha\neq \beta$,  
which can test the active-sterile mixing combinations $|\theta_e\theta_\mu |^2/\theta^2$, $|\theta_e\theta_\tau |^2/\theta^2$ and $|\theta_\mu\theta_\tau |^2/\theta^2$ down to $\sim 10^{-4}$ and $\sim 10^{-5}$ at the HL-LHC and the FCC-hh, respectively, for $M \sim 200$ GeV. It is interesting to note that already run 2 at the LHC can provide sensitivities $\sim 10^{-3}$ via this channel. 
The increase in center-of-mass energy from 14~TeV to 100~TeV and in luminosity improves the sensitivities of all signatures.

As for the LHC, we expect that also at future $pp$ colliders the search via displaced vertices is possible for masses $M$ below $\sim 100$ GeV.  We presented a first look at the possible sensitivities of the HL-LHC and the FCC-hh/SppC in fig.\ \ref{fig:displacedvertex-pp}, which show that $| \theta |^2$ as small as $\sim 2\times10^{-10}$ and $\sim 3\times10^{-11}$ may yield a visible signal at the HL-LHC and the FCC-hh/SppC, respectively, given a signal efficiency of 100\%.

Furthermore, the lepton-number-violating final states give rise to exotic signals without SM backgrounds at the parton level, which may in principle provide good prospects for testing sterile neutrinos, but are suppressed by the (approximate) ``lepton number''-like protective symmetry. 

Furthermore, we expect that for $M$ above about $1$ TeV the lepton number conserving but lepton flavour violating dilepton-trijet signature via $W\gamma$ fusion could also have a competitive sensitivity.

\section{Searches at $e^-p$ colliders}\label{sec:ep_colliders}
Electron-proton colliders are hybrids between $e^-e^+$ and $pp$ colliders, which consist of a hadron ring with an intersecting electron beam. They allow for a cleaner collision environment compared to the $pp$ colliders and for higher center-of-mass energies than the $e^-e^+$ colliders. 

Currently, a future $e^-p$ collider is discussed as an upgrade of the LHC, the Large Hadron-electron Collider (LHeC), which comprises a 60 GeV electron beam and possible electron polarization of up to 80\% \cite{Klein:2009qt,AbelleiraFernandez:2012cc,Bruening:2013bga} that will collide with the 7 TeV proton beam inside the LHC tunnel.
The machine is planned to deliver up to 100 fb$^{-1}$ integrated luminosity per year at a center-of-mass energy of $\sim$ 1.0 TeV, collecting $\sim$1~ab$^{-1}$ over its lifetime.
A more ambitious design for an $e^-p$ collider is presently discussed among the Future Circular Collider design study, namely the Future Circular electron-hadron Collider (FCC-eh)\cite{Zimmermann:2014qxa}, which features a 60~GeV electron beam (higher energies are also possible) that is brought into collision with the 50~TeV proton beam from the FCC-hh.
This would result in center-of-mass energies up to $3.5$~TeV with comparable luminosities to the LHeC, cf.~ref.\ \cite{Klein:2016uwv}.

First studies of right-handed currents and heavy neutrinos in high-energy $e^-p$ collisions \cite{Buchmuller:1991tu,Buchmuller:1992wm} have been conducted for HERA at DESY, which was the first machine of this kind and operated from 1992 to 2007.
They were motivated by extended gauge sectors, such as left-right symmetric models, or quark-lepton unified gauge groups. 
The discussion of searches for heavy neutrinos at an LHeC-like collider started with ref.~\cite{Ingelman:1993ve} soon after the commissioning of HERA.
Recently, right-handed neutrino searches at $e^-p$ colliders were investigated in the context of seesaw models \cite{Liang:2010gm,Blaksley:2011ey,Mondal:2016kof}, effective field theories \cite{Duarte:2014zea}, and in left-right symmetric \cite{Mondal:2015zba,Lindner:2016lxq} theories.

\subsection{Production mechanism}
\begin{figure}
\begin{center}
\includegraphics[width=0.3\textwidth]{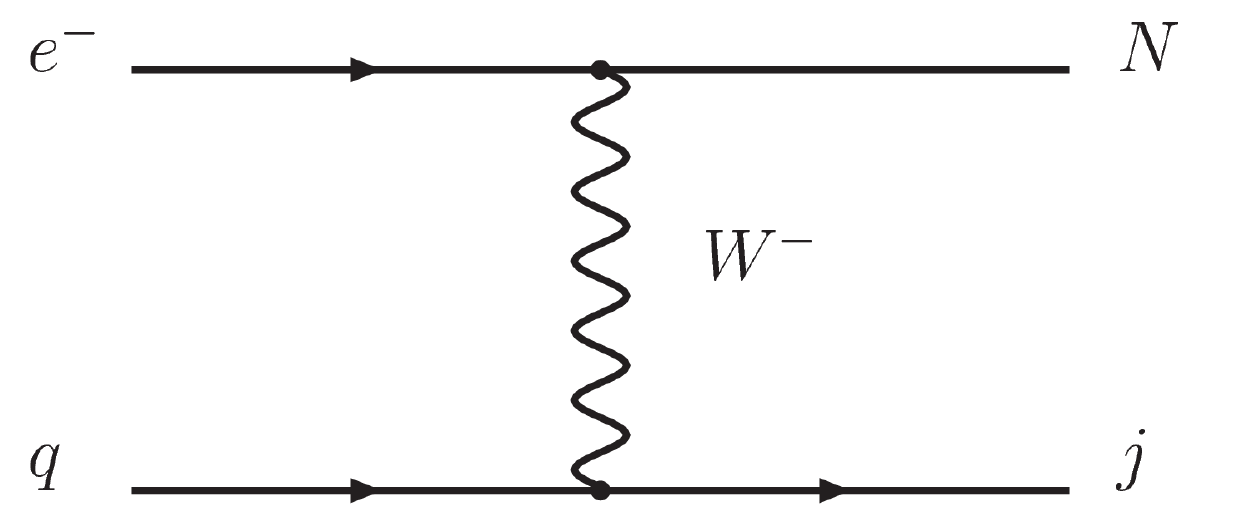}

production channel: $\mathbf{W_t}^{(q)}$
\vspace{10pt}

\includegraphics[width=0.3\textwidth]{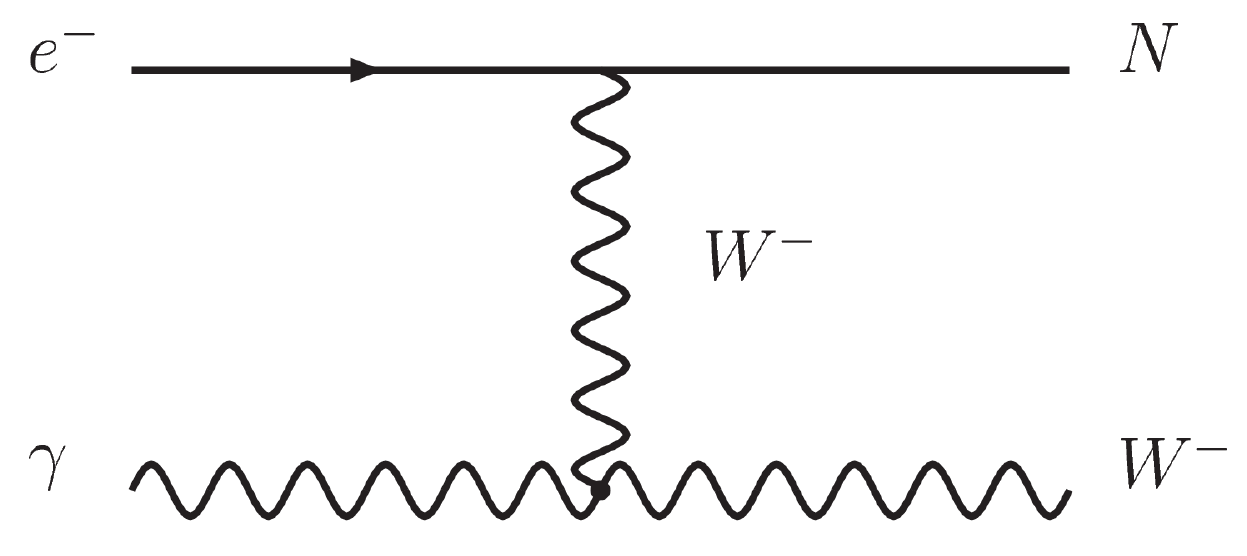}

production channel: $\mathbf{W_t}^{(\gamma)}$

\end{center}
\caption{Feynman diagrams denoting the production channels for heavy neutrinos in electron-proton scattering at the leading order. The dominant and suppressed production channel proceeds via $t$-channel $W$ boson exchange and gauge boson fusion, respectively.}
\label{fig:nproduction-eh}
\end{figure}

At $e^-p$ colliders the heavy neutrinos can be produced efficiently from the incident electron beam via the production channel $\mathbf{W_t}$, see also sec.~\ref{sec:production processes}. 
When the electron interacts with the quark current of the proton, the heavy neutrino is produced together with a quark jet and we label this channel $\mathbf{W_t}^{(q)}$ (see in fig.~\ref{fig:nproduction-eh} (top)). 
On the other hand, $W\gamma$-fusion gives rise to a heavy neutrino with a $W^-$ boson when the electron interacts with an initial state photon stemming from the proton. We label this channel $\mathbf{W_t}^{(\gamma)}$ (see in fig.~\ref{fig:nproduction-eh} (bottom)) and remark that it is suppressed by the parton distribution function of the photon.

\begin{figure}
\begin{center}
\includegraphics[width=0.45\textwidth]{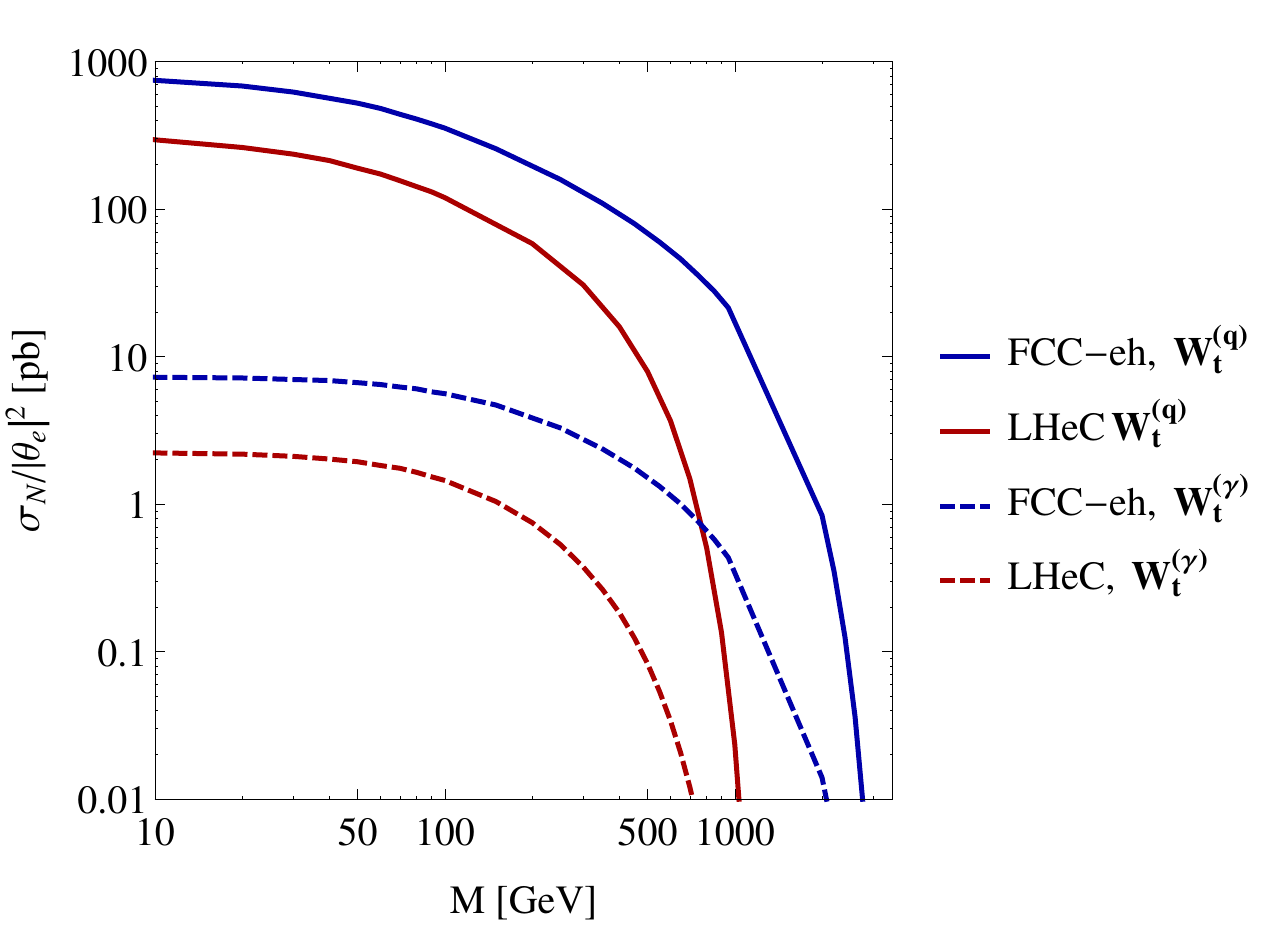}
\end{center}
\caption{Heavy neutrino production cross section $\sigma_{N}$ divided by the square of the active-sterile mixing parameter $|\theta_e|^2$ for the production channel $\mathbf{W_t}^{(q)}$ from the quark currents and $\mathbf{W_t}^{(\gamma)}$ from $W\gamma$ fusion for the LCeH and FCC-eh, respectively.
For the computation an angular acceptance of $1^\circ\leq \vartheta \leq 179^\circ$ has been assumed.}
\label{fig:cross-section-eh}
\end{figure}

Both production channels are dependent on the active-sterile mixing parameter $|\theta_e|$.  
We show the production cross section $\sigma_N$ divided by $|\theta_e|^2$ for heavy neutrinos via $\mathbf{W_t}^{(q)}$ and $\mathbf{W_t}^{(\gamma)}$, respectively, at the LHeC and the FCC-eh in fig.\ \ref{fig:cross-section-eh} as a function of the heavy neutrino mass $M$.

\subsection{Signatures and searches}
In this section, we discuss signatures from sterile neutrinos at $e^-p$ colliders, which manifest themselves in specific final states with the related production and decay channels, the dependency on the active-sterile mixing angles, and lepton-number and lepton-flavour violation.
For heavy neutrinos produced via the quark current of the proton, i.e.\ via the channel $\mathbf{W_t}^{(q)}$, we expect ``four-fermion final states'' while for heavy neutrinos produced via $W\gamma$ fusion, i.e.\ via the channel $\mathbf{W_t}^{(\gamma)}$, we expect ``five-fermion final states'' (at the parton level).
We list the signatures for the four-fermion final states in the upper part and the five-fermion final states in the lower part of tab.~\ref{tab:signatures_ep}.
In the following we discuss the signatures for all the four-fermion final states, and we discuss a few promising five-fermion signatures for comparison.
Furthermore, we present first estimates for the possible $1\sigma$ sensitivities to the sterile neutrino parameters of the LHeC and the FCC-eh at the parton level.
\begin{table*}[t]
\renewcommand*{\arraystretch}{2}
\begin{center}
\begin{tabular}{|l|c|c|c|c|}
\hline
Name & Final State & Channel [production,decay] & $|\theta_\alpha|$ dependency & LNV/LFV \\ \hline\hline
lepton-trijet  & $jjj \ell_\alpha$ & $[\mathbf{W_t}^{(q)}, W]$ & $\dfrac{|\theta_e\theta_\alpha|^2}{\theta^2}$ & $\checkmark$/$\checkmark$  \\ \hline
jet-dilepton  & $j \ell_\alpha^\pm \ell_\beta^\mp \nu$ & $[\mathbf{W_t}^{(q)}, \{W,Z(h)\}]$ & $\left\{\dfrac{|\theta_e\theta_\alpha|^2}{\theta^2}^{(*)}
,{|\theta_e|^2}^{(*)}\right\}$ & $\times$/$\checkmark$  \\ \hline
trijet  & $jjj \nu$ & $[\mathbf{W_t}^{(q)}, Z(h)]$ & $|\theta_e|^2$ & $\times$  \\ \hline
monojet & $j \nu\nu\nu$ & $[\mathbf{W_t}^{(q)}, Z]$ & $|\theta_e|^2$  & $\times$ \\ \hline
\multicolumn{1}{c}{} 
\\ 
\hline
lepton-quadrijet & $jjjj \ell_\alpha$ & $[\mathbf{W_t}^{(\gamma)}, W]$ & $\dfrac{|\theta_e\theta_\alpha|^2}{\theta^2}$  & $\checkmark$/$\checkmark$ \\ \hline
dilepton-dijet & $\ell_\alpha\ell_\beta \nu j j$ & $[\mathbf{W_t}^{(\gamma)}, \{W,Z(h)\}]$ & $\left\{\dfrac{|\theta_e\theta_\alpha|^2}{\theta^2}^{(*)}
,{|\theta_e|^2}^{(*)}\right\}$  & $\times$/$\checkmark$ \\ \hline
trilepton & $\ell_\alpha^- \ell_\beta^- \ell_\gamma^+ \nu\nu$ & $[\mathbf{W_t}^{(\gamma)}, \{W, Z (h)\}]$ & $\left\{\dfrac{|\theta_e\theta_\alpha|^2}{\theta^2}^{(*)}
,{|\theta_e|^2}^{(*)}\right\}$  &$\times$/$\checkmark$  \\ \hline
quadrijet & $jjjj \nu$ & $[\mathbf{W_t}^{(\gamma)}, Z (h)]$ & $|\theta_e|^2$  & $\times$ \\ \hline
lepton-dijet & $\ell_\alpha^- j j \nu \nu$ & $[\mathbf{W_t}^{(\gamma)},Z (h) ]$ & $|\theta_e|^2$  & $\times$\\ \hline
dijet & $j j \nu\nu\nu$ & $[\mathbf{W_t}^{(\gamma)}, Z]$ & $|\theta_e|^2$  & $\times$  \\ \hline
monolepton & $\ell_\alpha^- \nu\nu\nu\nu$ & $[\mathbf{W_t}^{(\gamma)}, Z]$ & $|\theta_e|^2$  & $\times$  \\ \hline
\end{tabular}
\end{center}
\caption{Signatures of sterile neutrinos at leading order for $e^-p$ colliders with their corresponding final states, production and decay channels (cf.\ section \ref{sec:production-and-decay}), and their dependency on the active-sterile mixing parameters.
A checkmark in the ``LNV/LFV'' column indicates that an unambiguous signal for LNV and/or LFV is possible (cf.\ discussion in sections \ref{sec:LNV} and \ref{sec:LFV}).
The upper and lower part of the table contains signatures where the heavy neutrino is produced via electron-quark scattering ($\mathbf{W_t^{(q)}}$) and $W\gamma$-fusion ($\mathbf{W_t^{(\gamma)}}$), respectively.\\
$^{(*)}:$ The dependency on the active-sterile mixing can be inferred when the origin of the charged leptons is known.}
\label{tab:signatures_ep}
\end{table*}

\subsubsection{Lepton-trijet}  \label{sec:Lepton-trijet}
After the heavy neutrino is produced via $\mathbf{W_t}^{(q)}$, its decays via the charged currents yield, with the hadronic decays of the $W$ boson, the final state $\ell_\alpha jjj$. Two of these jets have an invariant mass that is compatible with $m_W$, which might be useful for background rejection. 
Its production makes this signature dependent on $|\theta_e|^2$, its decays into $\ell_\alpha$ to ${|\theta_\alpha|^2}/{\theta^2}$ in the narrow width approximation.

An unambiguous signal for lepton-number-violating processes is given by the final states $\ell^+_\alpha jjj$ (where lepton number is violated by two units) for which there is no SM background. The lepton-number-violating final state with  $\alpha=e$ is studied by \cite{Liang:2010gm,Blaksley:2011ey,Mondal:2016kof,Duarte:2014zea,Mondal:2015zba,Lindner:2016lxq} for the LHeC, which are however suppressed by the (approximate) protective symmetry. 

The signature also includes the lepton-number-conserving case, given by the final states $\ell^-_\alpha jjj$, for which we give our estimates for the sensitivity to the neutrino parameters. The lepton-flavour-conserving case with $\alpha = e$ is studied in ref.~\cite{Mondal:2016kof} for the LHeC. We show our estimate for the  sensitivity of the LHeC and the FCC-eh via this channel by the green line in fig.\ \ref{fig:sensitivity-ep}, assuming $|\theta_\mu|=|\theta_\tau|=0$. 

For the case with $\alpha = \tau,\mu$ there exists a lepton-flavour-violating signature with no SM background at the parton level, which has been studied in ref.\ \cite{Liang:2010gm}. We remark that due to the finite resolution for the missing momentum at the detector level some SM processes with additional light neutrinos become relevant backgrounds, as detailed in the appendix \ref{app:distributions-ep}. In fig.~\ref{fig:sensitivity-ep} we show the sensitivity for $\alpha = \tau$ (to $|\theta_e\theta_\tau|^2/{\theta^2}$), assuming $|\theta_e|=|\theta_\tau|\neq 0$ and $|\theta_\mu|=0$, by the purple lines for the LHeC (left) and FCC-eh (right). The sensitivity is the same for $\alpha = \mu$, which depends on $|\theta_e\theta_\mu|^2/{\theta^2}$. 
Without the backgrounds, sensitivities as small as $\sim {\cal O}(10^{-7})$ and $\sim {\cal O}(10^{-8})$ would be possible for the LHeC and the FCC-eh, respectively.

\subsubsection{Jet-dilepton}\label{sec:Jet-dilepton}
All the decay channels from the heavy neutrino that is produced via $\mathbf{W_t^{(q)}}$ can contribute to the jet-dilepton final state $j\ell_\alpha\ell_\beta\nu$. 

Decays via the $Z$ and Higgs boson result in lepton-number-conserving final states with $\alpha=\beta$, which have an invariant mass of the dilepton system that is compatible with $m_Z$ and $m_h$, respectively. These channels are dependent on $|\theta_e|^2$ on the cross-section level.

Decays of the heavy neutrinos via the $W$ boson are accompanied by a charged lepton $\ell_\alpha$, and the leptonic decays of the $W$ yield $\ell_\beta\nu$.
This channel is dependent on $|\theta_e \theta_\alpha|^2/|\theta|^2$, however, one needs to distinguish the origin of the charged leptons, i.e.~identify $\ell_\alpha$ as the accompaniying lepton to the $W$ boson.
Moreover, it can proceed via lepton-number-conserving or violating processes. 
The lepton-number and lepton-flavour-conserving final state (with $\alpha = e$) is studied in ref.~\cite{Mondal:2016kof}.
It may be possible to infer the lepton-number-conserving and violating contributions to this signature via the kinematic distributions of the charged leptons as was discussed for the LHC in ref.~\cite{Dib:2016wge}.
Furthermore, the two lepton flavour indices are independent for this channel and can lead to lepton-flavour-violating final states.

The final state $j \ell_\alpha^- \ell_\beta^+ \nu$, with $\alpha\neq e$ and $\alpha\neq\beta$, i.e. when the negatively charged lepton is not an electron, provides a novel unambiguous signal for LFV at the parton level. However, as discussed in section \ref{sec:LFV}, SM backgrounds with a larger number of light neutrinos in the final state exist, which have to be taken into account. We include a first estimate of the $1\sigma$ sensitivity of the $j\ell_\mu\ell_\tau\nu$ final state by the red line in fig.\ \ref{fig:sensitivity-ep}, wherein we assumed $|\theta_e|=|\theta_\mu|$ and $|\theta_\tau|=0$ (and treated the SM background as discussed in the appendix \ref{app:distributions-pp}).
Without the backgrounds, sensitivities as small as $\sim {\cal O}(10^{-5})$ and $\sim {\cal O}(10^{-7})$ would be possible for the LHeC and the FCC-eh, respectively.

\subsubsection{Trijet}
The trijet final state $jjj \nu$, with the light neutrino giving rise to missing transverse momentum, results from $\mathbf{W_t^{(q)}}$ produced heavy neutrinos that decay hadronically via the $Z$ or Higgs boson. Consequently, the invariant mass of two jets is compatible with $m_Z$ or $m_h$, which may be helpful to reject some of the backgrounds.
Furthermore in the case of the Higgs decays, the jets are most likely to stem from b quarks and b-tagging may help to further separate the backgrounds.
This process is sensitive to $|\theta_e|^2$ and its sensitivity to the neutrino parameters via the Higgs decays is denoted by the the orange line in fig.~\ref{fig:sensitivity-ep}.

\subsubsection{Monojet}
The production of a heavy neutrino via $\mathbf{W_t^{(q)}}$ and its subsequent invisible decay into light neutrinos via an intermediate $Z$ boson leads to the final state $j\nu\nu\nu$, a jet and missing energy, referred to as monojet.
The Higgs boson can also decay into light neutrinos, which allows to test the Higgs invisible decay width in this channel, but it is suppressed by one further order in $|\theta_\alpha|$.
The monojet signature from heavy neutrinos is sensitive to $|\theta_e|^2$ and our estimate for the sensitivity is shown by the light blue line in fig.~\ref{fig:sensitivity-ep}.

\subsubsection{Five fermion final states from $W\gamma$-fusion}  \label{sec:Wgamma_fusion}
Heavy neutrinos produced via $W\gamma$-fusion ($\mathbf{W_t^{(\gamma)}}$) give rise to five-fermion final states at the leading order. These signatures are suppressed by the parton distribution function of the photon, however they can include final states with suppressed SM backgrounds.
For a complete list of these signatures, including the respective five-fermion final states, dependency on the active-sterile mixing parameters, and possible lepton-number and lepton-flavour violation, we refer the reader to the lower part of tab.~\ref{tab:signatures_ep}.

As an example of the lepton-number and flavour conserving $W\gamma$-fusion signatures we show our estimates for the sensitivity on $|\theta_e|^2$ of the lepton-dijet final state $\ell_e^-bb\nu\nu$, where the $b$-quarks stem from Higgs decays, by the blue line in fig.\ \ref{fig:sensitivity-ep}. We notice that the sensitivity improves strongly with increasing center-of-mass energy.

Furthermore, the following signatures include unambiguous signals of lepton-number and/or lepton-flavour violating final states (at the parton level): 
\begin{itemize}
\item Lepton-quadrijet in the lepton-number-violating final states $jjjj\ell_\alpha^+$ or in the lepton-flavour-violating final states $jjjj\ell_ \alpha$ for $\alpha\neq e$ provide signals for LNV and LFV.

\item The dilepton-dijet final states includes $\ell_\alpha^\pm\ell_\beta^\mp\nu jj$ and $\ell_\alpha^-\ell_\beta^-\nu jj$. Particularly interesting are the lepton-flavour-violating final states $\ell_\alpha^+\ell_\beta^-\nu jj$ with $\beta\neq e$ and $\alpha\neq\beta$, and $\ell_\mu^-\ell_\tau^-\nu jj$. 

\item Trilepton final states arise from the charged current decays of the heavy neutrino can lead to the lepton-number-violating final state $\ell_\alpha^-\ell_\beta^-\ell^+_e \nu \nu$ for $\alpha,\beta\neq e$. 
\end{itemize}
The above final states have been studied for the LHeC in ref.~\cite{Mondal:2016kof} for the case $\alpha = e$.

\subsubsection{LFV signatures}
As mentioned earlier, at the loop level the sterile neutrinos can induce lepton-flavour-violating decays of the Higgs and Z boson into charged leptons. Such searches are in principle also possible at the $e^- p$ collider. 

We note that novel LFV signatures are discussed above in section \ref{sec:Lepton-trijet} on the lepton-trijet, \ref{sec:Jet-dilepton} on the jet-dilepton and \ref{sec:Wgamma_fusion} for the lepton-quadrijet signature.

\subsubsection{Displaced vertex searches}
\begin{figure}[t]
\centering
\includegraphics[width=0.35\textwidth]{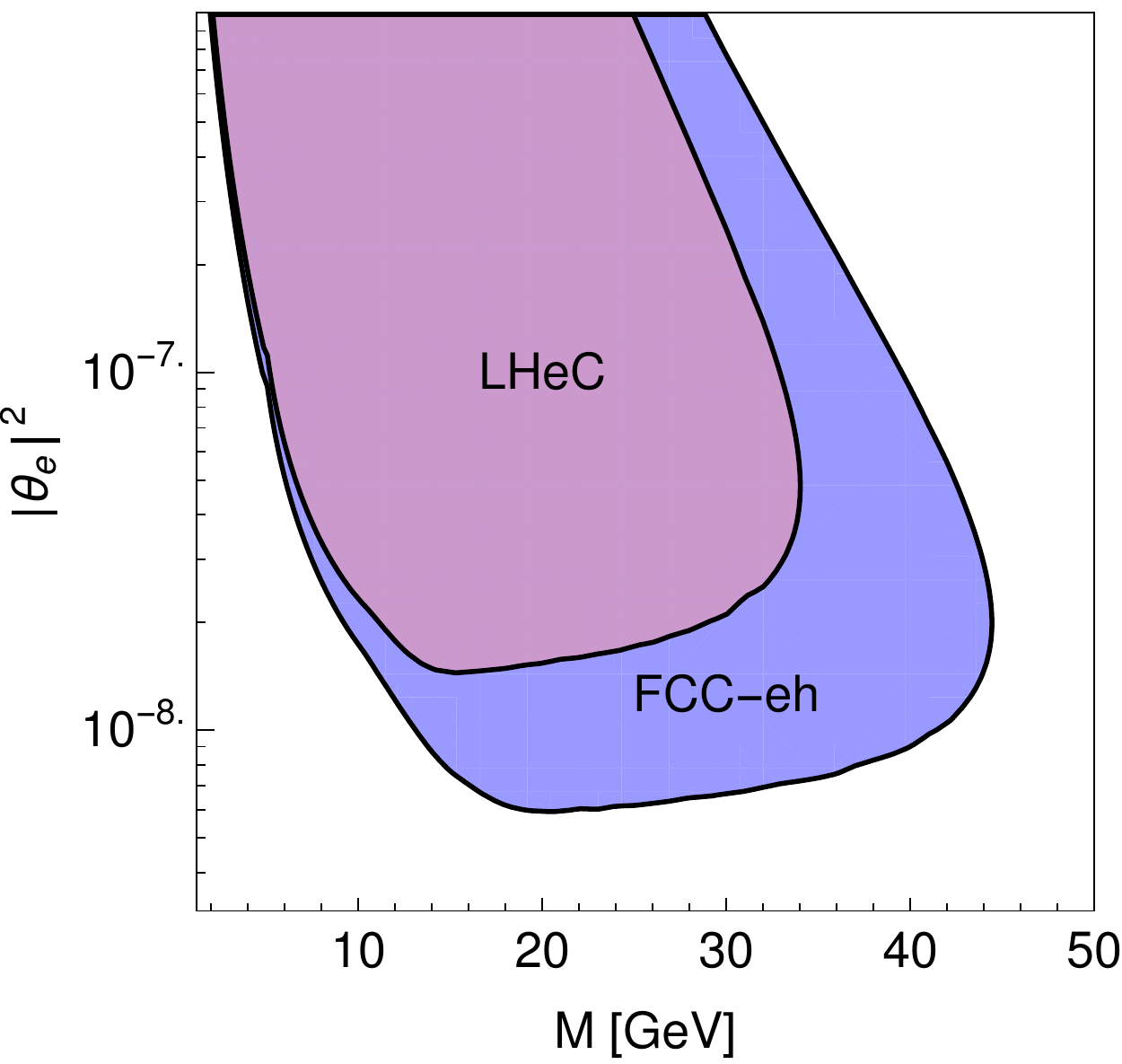}
\caption{First look at the sensitivity of the LHeC and the FCC-eh to sterile neutrinos via displaced vertices. For our estimate, we have considered vertex displacements between 1 mm and 1 m as background-free, 100\% signal efficiency, and an average Lorentz factor of 3 and 5 for the LHeC and the FCC-eh, respectively.}
\label{fig:displacedvertex-ep}
\end{figure}

For a first look at the possible sensitivity of the LHeC and the FCC-eh to sterile neutrinos via this signature, we make the same assumptions regarding the visibility of displacement and the signal efficiency (assumed to be 100\%) as in sec.\ \ref{sec:displacedvertex-pp}. Furthermore, from the kinematics of the heavy neutrinos we assume a Lorentz factor of 3 and 5 from electron-proton collisions at the LHeC and the FCC-eh, respectively. We show the sensitivity corresponding to at least four events in fig.\ \ref{fig:displacedvertex-ep}. We remark that for a realistic estimate of the sensitivity a thorough study of the detector response and the backgrounds is required.

\begin{figure*}
\begin{minipage}{0.33\textwidth}
\includegraphics[width=\textwidth]{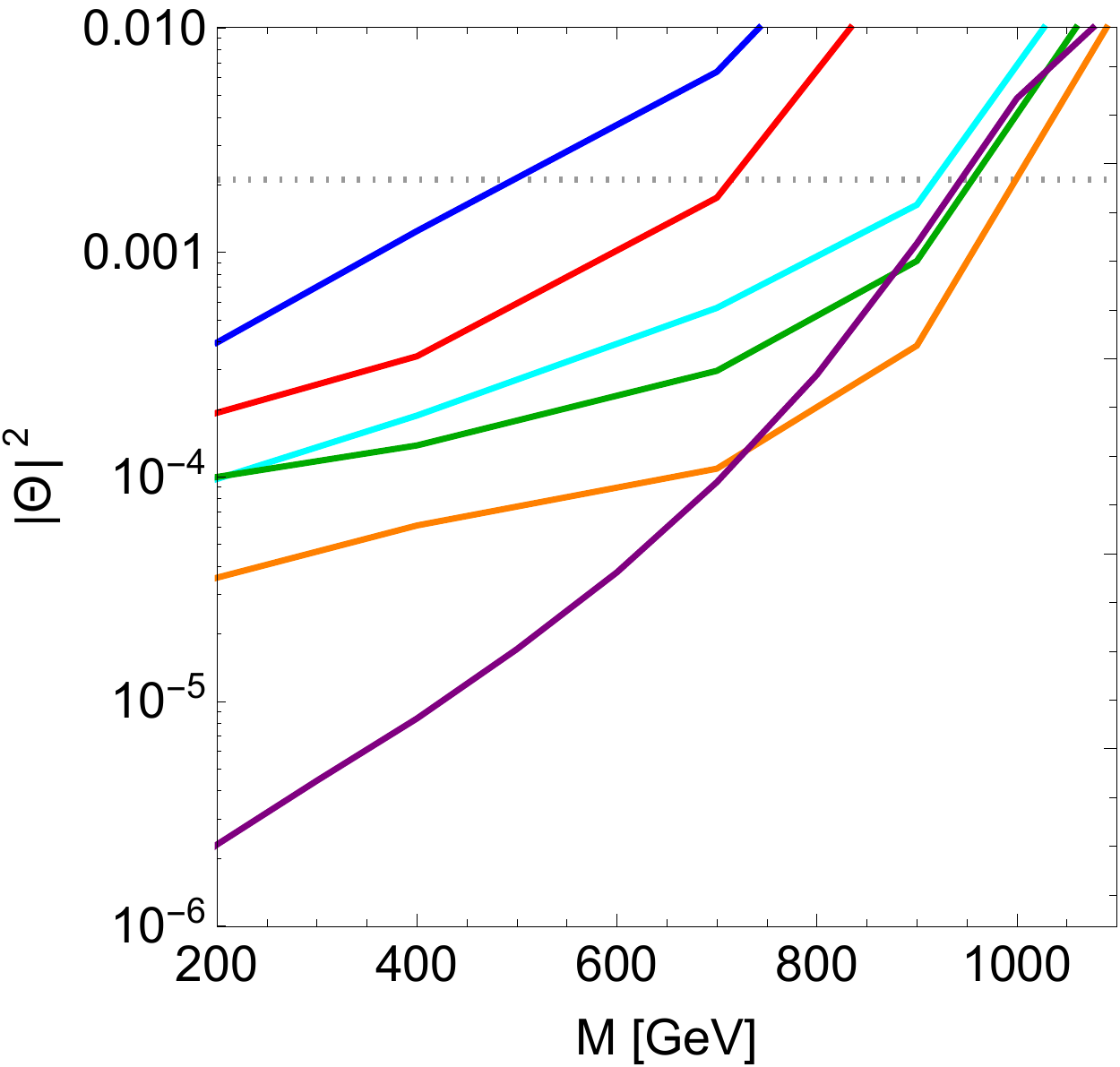}

\vspace{-50pt}\hspace{120pt}{\bf LHeC}
\vspace{50pt}
\end{minipage}
\begin{minipage}{0.30\textwidth}
\begin{center}
\vspace{-30pt}
\includegraphics[width=0.7\textwidth]{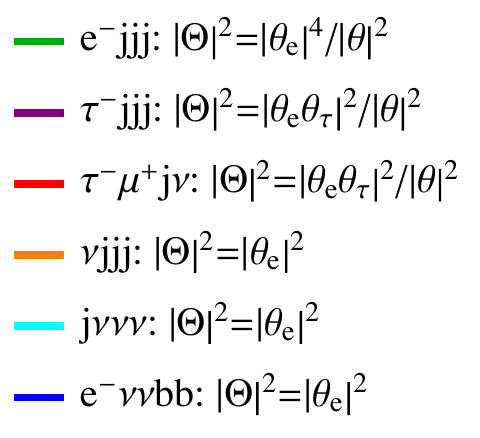}
\end{center}
\end{minipage}
\begin{minipage}{0.33\textwidth}
\includegraphics[width=\textwidth]{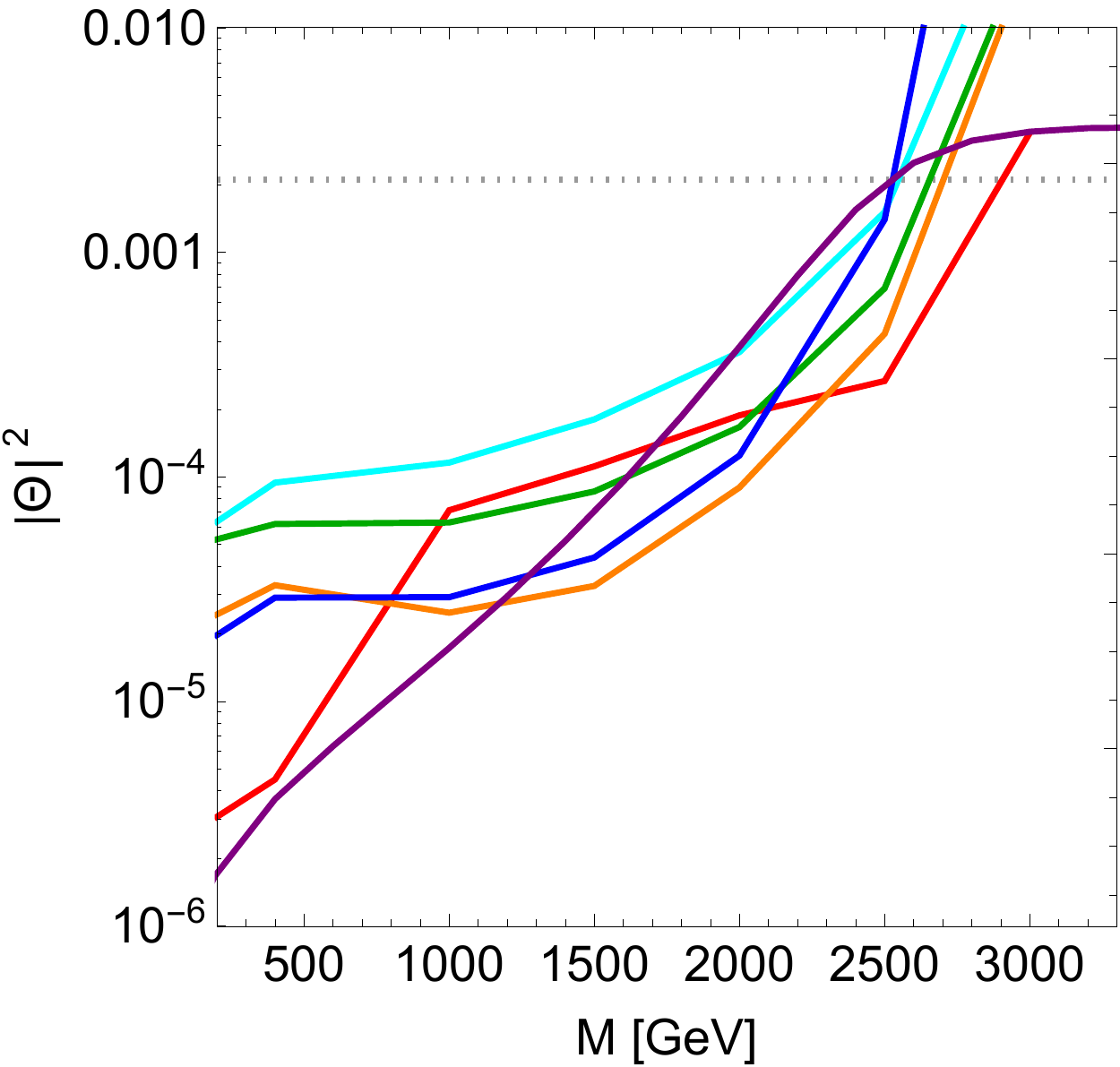}

\vspace{-50pt}\hspace{120pt}{\bf FCC-eh}
\vspace{50pt}
\end{minipage}
\caption{First look at the possible $1\sigma$ sensitivity of the lepton-number-conserving signatures (see tab.~\ref{tab:signatures_ep}) for sterile neutrino searches at $e^-p$ colliders. We consider an integrated total luminosity 1 ab$^{-1}$ for the LHeC (7 TeV proton beam energy) and the FCC-eh (50 TeV proton beam energy), respectively.
For details on the calculation of the sensitivities on the parton level, see section \ref{app:distributions-ep} in the appendix. 
Note that the the purple line becomes flat for large $M$ because the process probes the non-unitarity of the PMNS matrix in this limit.}
\label{fig:sensitivity-ep}
\end{figure*}

\subsection{Electron-proton colliders: summary}
In this subsection, we summarize our findings regarding the sensitivities for sterile neutrinos at future $e^-p$ colliders. In the previous section we have presented a complete list of signatures for sterile neutrinos, which were produced at leading order either from interactions of the electron with the quark current of the proton or from $W\gamma$-fusion.

We present a ``first look'' at the $1\sigma$ sensitivities for lepton-number-conserving final states, at the LHeC and the FCC-eh with an integrated luminosity of 1~ab$^{-1}$ each, in fig.~\ref{fig:sensitivity-ep} for heavy neutrino masses above 200~GeV.
We emphasize that our estimates are calculated on the parton level, and for all the new signatures a more thorough analysis at the reconstructed level should be done in the future.
In the figure the grey dotted horizontal line denotes the present upper bound on $|\theta_e|$ at the 90\% Bayesian confidence level.

We note that the heavy neutrino production cross section is dependent on $|\theta_e|^2$ only, and the relative strength of the $|\theta_\alpha|^2$ can be inferred via the relative strength e.g.\ via the lepton-trijet, and the lepton-quadrijet signature, which are both depending on $|\theta_e \theta_\alpha|^2/|\theta|^2$. 

We find that sterile neutrinos with mixings close to the present upper bound can be tested for masses up to $\sim 1$ TeV and $\sim 2.7$ TeV, at the LHeC and FCC-eh, respectively. The comparison of the sensitivities of the LHeC and the FCC-eh for the individual signatures (cf.\ fig.~\ref{fig:sensitivity-ep}) shows, that an increased proton beam energy improves the sensitivities of the lepton-flavour-conserving signatures, and, especially the maximum sensitivity of the LFV jet-dilepton improves by about two orders of magnitude.
We expect that (at least some of) the other $W\gamma$-fusion signatures should have sensitivities that might be comparable to the ones shown in the figure.

Although some of the lepton-number-violating signatures do not have SM backgrounds at the parton level, their cross sections are suppressed by the protective symmetry, and we therefore do discuss them here in more detail.

The best sensitivity for masses above $200$ GeV is obtained from the lepton-flavour-violating lepton-trijet final state, which could test squared active-sterile mixings as small as $\sim 10^{-6}$ for both, the LHeC and the FCC-eh, assuming $|\theta_e|=|\theta_\alpha|,\,\alpha=\mu,\tau$.
For masses below $m_W$, the best sensitivity can be achieved via displaced vertex searches.  We presented a first look at the possible sensitivities of the LHeC and the FCC-eh in fig.\ \ref{fig:displacedvertex-ep}, which show that $| \theta_e |^2$ as small as $\sim 2\times10^{-8}$ and $\sim 6\times10^{-9}$ may be tested at the LHeC and the FCC-eh, respectively, given a signal efficiency of 100\%.

\section{Conclusions}
In this paper we have systematically analyzed the possibilities of future electron-positron, proton-proton and electron-proton colliders for searching sterile neutrinos. 
We discussed the production and decay channels and provided a complete list of the leading order signatures. Among other things, we discussed several novel search channels and presented first looks at the possible sensitivities for the active-sterile mixings $|\theta_\alpha|$ and the heavy neutrino mass $M$. 
Our discussion of the production and decay channels comprises their dependencies on the active-sterile neutrino mixing parameters, lepton-number-violating (LNV) and lepton-flavour-violating (LFV) effects, vertex displacement and non-unitarity effects.

For our sensitivity estimates we have considered low scale seesaw scenarios with a protective symmetry (using the SPSS as a benchmark) and focused on lepton number conserving signatures, since lepton number violating channels are suppressed by the protective symmetry.

The signatures at the different types of colliders have different dependencies on the $\theta_\alpha$: In particular, the $e^-e^+$ colliders are sensitive to $|\theta |^2 = \sum_\alpha |\theta_\alpha |^2$ at the $Z$ pole and to $|\theta_e|^2$ at higher energy runs, the $pp$ colliders can individually test the $|\theta_\alpha|^2$, and the production of heavy neutrinos at $e^-p$ colliders is always proportional to $|\theta_e|^2$.  

\begin{figure}[t]
\begin{center}
\includegraphics[width=0.4\textwidth]{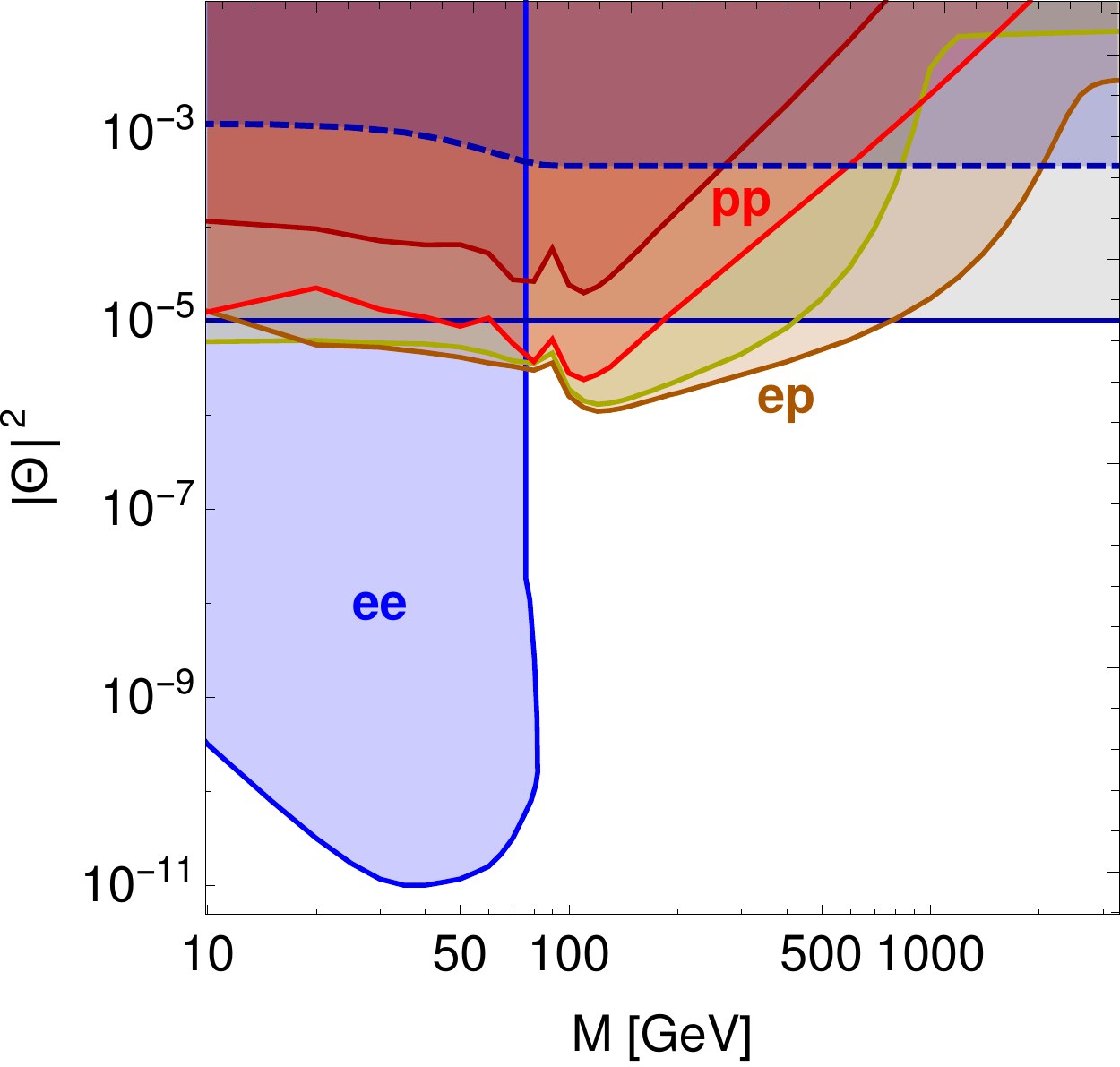}
\end{center}
\caption{Summary of selected estimated sensitivities of the FCC-ee, -hh, and -eh colliders, including the HL-LHC and the LHeC.
The best sensitivity for heavy neutrino masses $M<m_W$ is obtained from the displaced vertex searches at the Z pole run of the FCC-ee shown by the blue line, which are sensitive to $|\theta|^2=|\theta_e|^2 + |\theta_\mu|^2+|\theta_\tau|^2$.
For heavy neutrino masses above $m_Z$ the $pp$ colliders (red:~FCC-hh, dark-red:~HL-LHC) and $e^-p$ colliders (brown:~FCC-eh, yellow:~LHeC) have the best prospects for discovering sterile neutrinos via the LFV signatures.
The lepton-dijet signature at the $pp$ colliders with final states $\ell_\alpha^\pm \ell_\beta^\mp jj,\, \alpha\neq \beta$ yields sensitivities to the active-sterile mixing parameter combinations $|\theta_\alpha\theta_\beta|^2/|\theta|^2$, and it is shown by the  red lines.
The lepton-trijet signature at the $e^-p$ colliders with final states $\ell^-_\alpha jjj,\,\alpha\neq e$ is sensitive to $|\theta_e\theta_\alpha|^2/|\theta|^2$, and it is shown by the brown lines.
Finally, for very large heavy neutrino masses the best sensitivity is given by the EWPO measurements at the FCC-ee. The solid and dashed horizontal blue line denotes the sensitivity to $|\theta_e|^2+|\theta_\mu|^2$ and $|\theta_\tau|^2$, respectively.}
\label{fig:complementarity}
\end{figure}

A summary and comparison of selected estimated sensitivities is shown in fig.~\ref{fig:complementarity}.
We have found that the best sensitivity to $|\theta |^2$ can be reached by electron-positron colliders such as the FCC-ee via displaced vertex searches, where it is sensitive to $|\theta |^2 \sim 10^{-11}$ for $M$ below $m_W$. 
For the sensitivities of the $pp$ and $ep$ colliders via displaced vertex searches, see figs.\ \ref{fig:displacedvertex-pp} and \ref{fig:displacedvertex-ep}.
 Above the $W$ mass, the best reach can be achieved by $e^- p$ colliders such as the FCC-eh via the LFV lepton-trijet signatures $\mu^- jjj$ and $\tau^- jjj$, which are sensitive to $|\theta_e\theta_\mu|^2/\theta^2$ and $|\theta_e\theta_\tau |^2/\theta^2$ down to $\sim 10^{-6}$ for $M\sim 200$ GeV. 
$pp$ colliders like the FCC-hh can additionally test $|\theta_\mu\theta_\tau |^2/\theta^2$ and reach $\sim 10^{-5}$ for $M\sim 200$ GeV with the LFV dilepton-dijet signature $\ell_\alpha \ell_\beta jj$. It is also interesting to note that already run 2 at the LHC might provide sensitivities $\sim 10^{-3}$ for the parameter combinations $|\theta_e\theta_\mu |^2/\theta^2$, $|\theta_e\theta_\tau |^2/\theta^2$ and $|\theta_\mu\theta_\tau |^2/\theta^2$ via this channel. 

Regarding the sensitivity reach to $M$, we compared the sensitivity of the signatures with the (weakest) relevant present constraints on the mixing parameters, i.e. with the bounds on $|\theta_\tau |$ for the FCC-hh and on $|\theta_e |$ for the FCC-eh. The highest mass reach is possible for the FCC-ee with the indirect searches via the EWPOs, which could strongly improve the present sensitivities on $|\theta_e |$ and $|\theta_\tau |$ and use them to probe $M$ up to values up to about $60$ TeV (for $|y_{\nu_\alpha}| = {\cal O}(1)$). Via the direct searches the HL-LHC and FCC-hh can test heavy neutrinos with masses up to $\sim 450$ GeV and $2$ TeV, respectively. 
The electron-proton colliders LHeC and FCC-eh can extend the mass reach of the proton-proton colliders and test heavy neutrinos masses up to $\sim 1$ TeV and $2.7$ TeV, respectively.

If deviations from the SM are found in the respective final states in one or more channels, one can start to attack the next challenge, which would be to test the sterile neutrino properties, i.e.\ their active-sterile mixings and mass. For various of the discussed channels sterile neutrinos cause a ``bump'' in the relevant kinematic distributions, which can give direct access to the mass of the heavy neutrino. Different combinations of the mixing angles $|\theta_\alpha |$ can be probed via the various signatures. This also shows the great complementarity between the different collider types, which not only allow to probe different mass ranges, but also different mixing angle combinations. 

\subsection*{Acknowledgements}
This work has been supported by the Swiss National Science Foundation and by the ``Fund for promoting young academic talent'' from the University of Basel under the internal reference number DPA2354. O.F. and S.A. are thankful to the organizers and participants of the FCC week in Rome and the FCC-ee brainstorming workshop at CERN for the stimulating atmosphere and many useful discussions. 

\appendix

\vspace{0.7cm}

\noindent {\bf\Large Appendix}
\section{Sensitivity estimates}
For our sensitivity estimate of the different heavy neutrino signatures to the active-sterile mixing parameters at $e^- e^+$ (only lepton-dijet), $pp$, and $e^-p$ colliders we considered on-shell $Z,W,$ and Higgs bosons as ``intermediate states'', and we use the narrow width approximation. 
We generated $10^5$ events for each benchmark mass for the signal distribution and $10^5$ to $10^6$ events for the background distributions. For the simulation we used the Monte Carlo event generator WHIZARD 2.2.8 \cite{Kilian:2007gr,Moretti:2001zz}.
We emphasize that for a first look at the sensitivities we have restricted ourselves to parton-level estimates. 
On the one hand, we are likely to underestimate the backgrounds due to detector, interference, and hadronization effects. On the other hand, in some channels one might gain sensitivity by optimizing cuts on other kinematic variables.

\subsection{Calculation of the sensitivity}
\label{app:calculation_sensitivity}
When several processes $x_i\ (i>1)$ with individual sensitivities $s_{x_i}$ to the sterile-active parameters contribute to the same final state, a combined sensitivity of the final state, labelled $s_{x}$, is obtained in the following way:
\begin{equation}
s_{x}^{-1} = \sqrt{\sum_i s_{x_i}^{-2}}\,.
\label{eq:combined-sensitivity}
\end{equation}
This implies the underlying assumption that the contributions from each process $x_i$ allow for a separate test of the signal-strength.

In order to obtain the individual sensitivity $s_{x_i}$ for each process $x_i$, we consider the transverse momentum\footnote{For the $e^-e^+$ colliders we consider the invariant mass distribution instead.} ($P_t$) of a lepton if one is present, and otherwise that of a $Z$ and Higgs boson for both, the signal and the background. 
We then identify an interval in the $P_t$ distribution that maximizes the significance 
\begin{equation}
n = N_S/\sqrt{N_B+N_S},
\label{eq:signigicance}
\end{equation}
with $N_S$ being the number of signal events that is proportional to $|\Theta|^2$ (where the parameter $|\Theta|^2$ depends on the considered production-and-decay channel, cf.\ tab.\ \ref{tab:dependency}) and $N_B$ being the number of background events, for a given luminosity and a benchmark value for $|\Theta|^2$.
The resulting significance corresponds to a heavy neutrino signal at the $n\sigma$ level.
The $1\sigma$ sensitivity $s_{x_i}$ of the process $x_i$ to the sterile neutrino parameters is the value of $|\Theta|^2$ that corresponds to a significance of $n=1$ and it is obtained by solving eq.\ \eqref{eq:signigicance} for $|\Theta|^2$.

The considered heavy neutrino signatures given in tab.~\ref{tab:signatures_ee},~\ref{tab:signatures_pp}, and~\ref{tab:signatures_ep} include the production-and-decay channels $x_i$, which are given by the production of a heavy neutrino via $\mathbf{Z_s}$, $\mathbf{W_t}$, or $\mathbf{W_s}$ and its subsequent decay into $\nu Z,\,\ell W,\,$ and $\nu h$.

The cross sections of the considered SM backgrounds at $pp$ and $e^-p$ colliders are listed tab.\ \ref{tab:pp}, and \ref{tab:ep}, respectively, where for the lepton-flavour-conserving signatures we only consider the electron flavour for simplicity.
For a given final state that only includes charged leptons, neutrinos, and jets, the background process' cross section is multiplied by the appropriate branching ratios, for which we use:
\begin{equation}
\begin{array}{lcc}
{\rm Br}(W\to jj) & = & 0.67,\\
{\rm Br}(W\to \nu \ell) &  = & 0.33, \\
{\rm Br}(Z\to jj) & = & 0.7, \\
{\rm Br}(Z\to \ell\ell) & = & 0.1,\\
{\rm Br}(Z\to \nu\nu) & = & 0.2 \\
{\rm Br}(h\to jj) & = & 0.7, \\
{\rm Br}(h\to \ell\ell) & = & 0.06.
\end{array}
\label{eq:branchings}
\end{equation}

\subsection{Electron-positron collisions}
\label{app:distributions-ee}
For $e^- e^+$ colliders we calculate a new sensitivity estimate for the lepton-dijet signature, as given in tab.~\ref{tab:signatures_ee}. It receives contributions from the decays of the heavy neutrino via $\ell W$, together with the hadronic decays of the $W$ boson into two jets.

We simulate the inclusive final state $e^- \nu u \bar{d}$ for a heavy neutrino benchmark mass of 95, 145, 195, and 245 GeV for $\sqrt{s}=250$ GeV, and additionally for $M= 295$, and 345 GeV for $\sqrt{s}=350$ GeV, using a benchmark value for the active-sterile mixing.
Furthermore, we considered the integrated luminosities according to the modi operandi given in fig.~\ref{fig:modioperandi}.
\begin{figure}
\centering
\includegraphics[width=0.5\textwidth]{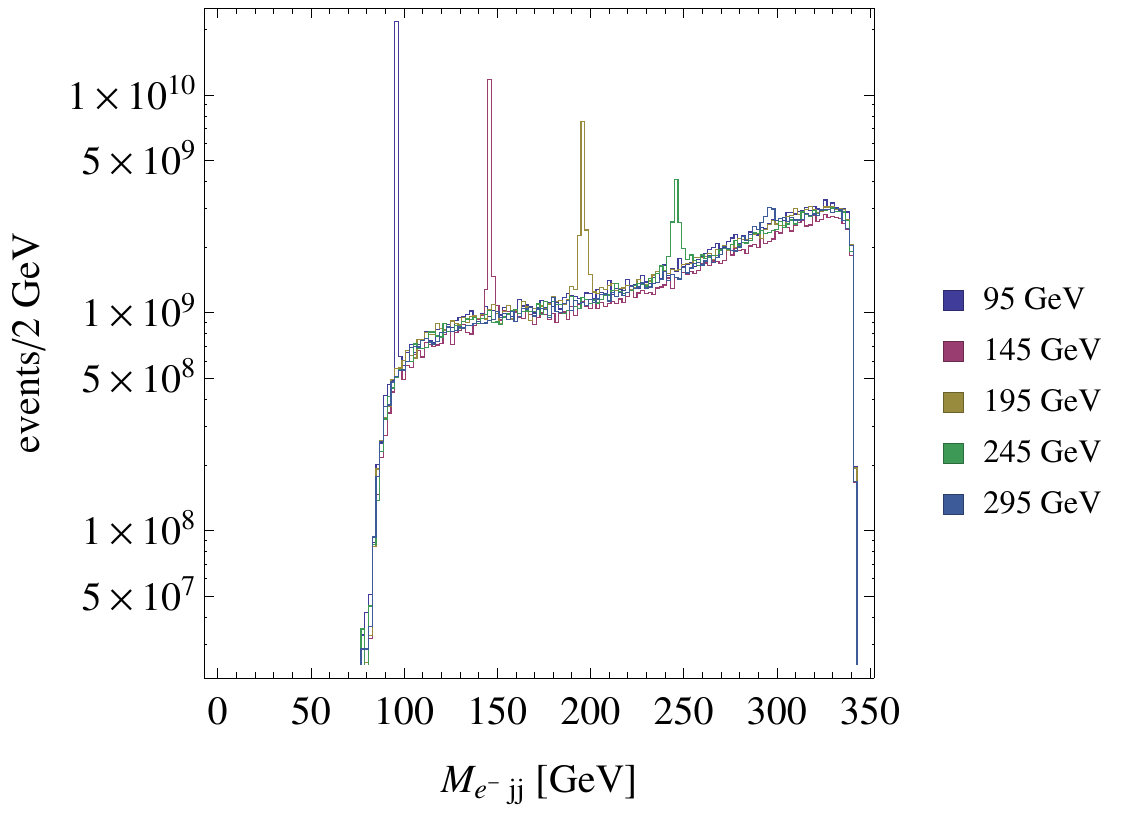}
\caption{Distribution of the invariant mass $M_{e^-jj}$ from the process $e^-e^+\to \nu e^-jj$ at the center-of-mass energy of 350 GeV, for a bin size of 2 GeV. The process was simulated with WHIZARD 2.2.8 and it includes heavy neutrinos with different masses according to the color code. The y axis has been scaled to the expected number of events for 1.5 ab$^{-1}$.}
\label{fig:distribution-ee}
\end{figure}

We considered the invariant mass distribution of the $e^- jj$ system, $M_{e^-jj}$ with a bin size of 2 GeV. We show some example distributions for $\sqrt{s}=350$ GeV in fig.~\ref{fig:distribution-ee}, which generally feature a peak in the number of events per bin for $M_{e^-jj}\simeq M$.
We identify the bins that correspond to the mass $M$, from which we get the number of signal and background events, $N_{S+B}$.
We obtain the number of background events $N_B$ by assuming a smooth background distribution and interpolating it around the signal bins. With the number of signal events $N_S=N_{S+B}-N_B$ we obtain the sensitivity as described in section \ref{app:calculation_sensitivity}. To account for the processes $e^- \nu s \bar{c},e^+ \nu d \bar{u},$ and $e^+ \nu s \bar{c}$ we assume that they all have identical distributions and cross sections, and multiply both, $N_S$ and $N_B$ with a factor of four.

For the heavy neutrino masses below $\sim m_W$, the background distribution of $M_{e^- jj}$ is effectively zero. We have therefore assumed that for $M=10$ GeV no background exists, and use the number of signal events $\left(\sigma_{M=10\text{ GeV}}-\sigma_{SM}\right) {\cal L}_{\rm int}$, with the cross sections $\sigma$ and the integrated luminosity ${\cal L}_{\rm int}$.
We show the cross sections of the processes for the benchmark value of the active-sterile mixing $|\theta_e|^2 = 0.042, |\theta_\mu|=|\theta_\tau|=0$ in tab.~\ref{tab:app-ee}.

\begin{table}
\centering
\begin{tabular}{|c|c|c|}
\hline
$M$	&	250 GeV & 350 GeV \\
\hline\hline
10	& 	680	&	631	\\
95	&	646	&	587	\\
145	&	614	&	559	\\
195	&	590	&	544	\\
245	&	576	&	528	\\
295	&	--	&	518	\\
345	&	--	&	514	\\
\hline
SM	&	576	&	506 \\
\hline
\end{tabular}
\caption{Cross section in fb for the four-fermion final state $e^-\nu u\bar{d}$ at $e^-e^+$ colliders for the center-of-mass energies of 250 and 350 GeV, the numerical error is $\sim 2$ fb. The calculation of the cross section includes initial state radiation and a cut on the tranverse momentum of all final state fermions, $P_t > 5$ GeV.}
\label{tab:app-ee}
\end{table}

\subsection{Proton-proton collisions}
\label{app:distributions-pp}
\begin{table}
\begin{center}
\begin{tabular}{|c|cc|}
\hline
Background	& pp14 [fb] & pp100 [fb] \\
\hline\hline
$e^+ \nu Z$		&	770		& 7500	\\
$e^- \nu Z$		&	480		& 5850	\\
$\nu\nu Z$		&	680		& 7400	\\
\hline
$e^+\nu h$		&	33		& 330	\\
$e^-\nu h$		&	21		& 250	\\
$\nu\nu h$	&	19		& 220	\\
\hline
$e^- e^+ W^-$ 		&	175 	& 2150 \\
$e^+ e^- W^+$ 	 	&	290 	& 2650 \\
$\nu \,e^- W^+$ 			&	3450		& 34600 \\
$\nu \,e^+ W^-$ 			&	3450		& 35000 \\
\hline
$t \bar t$		& 35000	& 410000 	\\
\hline
\end{tabular}
\end{center}
\caption{Parton-level cross sections for the considered Standard Model background processes for heavy neutrino searches in proton-proton collisions.
The cross sections include a cut $P_t \geq 20$~GeV for all the fermions in the final state, except for the $t\bar t$. For the evaluation we used WHIZARD with CTEQ6L and CT10 parton distribution functions.}
\label{tab:pp}
\end{table}

The list of signatures for which we calculate the sensitivities is given in tab.\ \ref{tab:signatures_pp}. Therein, each production-and-decay channel $x_i$ contributes separately, and we simulate signal and background events to obtain the individual sensitivities $s_{x_i}$, as discussed in sec.\ \ref{app:calculation_sensitivity}.

We are working in the narrow width approximation where it is sufficient to consider $Z,\,W,$ and Higgs bosons that are produced on the mass shell, together with the two accompanying leptons $\ell \ell, \ell \nu$, or $\nu\nu$ as ``intermediate states''.
For the simulation at the HL-LHC and the FCC-hh/SppC we used the built-in PDF CTEQ6L that are included in the WHIZARD package. 
For the heavy neutrinos we considered the following benchmark values: 
200, 400, 700, 1300 GeV for the HL-LHC, and 
200, 400, 1000, 2000 GeV for the FCC-hh/SppC.
For the SM background cross sections we used the values shown in tab.\ \ref{tab:pp}, where we limit ourselves to the electron flavour for the lepton-flavour-conserving signatures for simplicity.

\begin{figure*}

\includegraphics[width=0.3\textwidth]{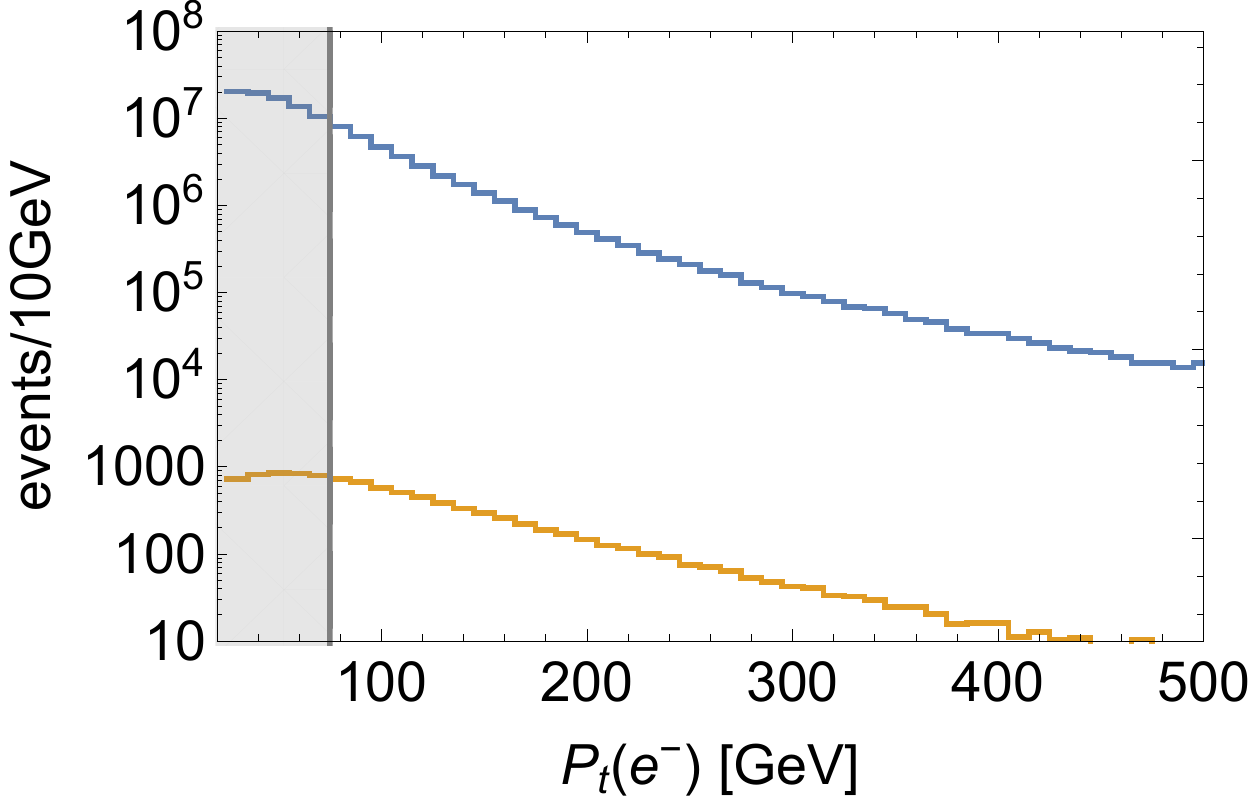}
\includegraphics[width=0.3\textwidth]{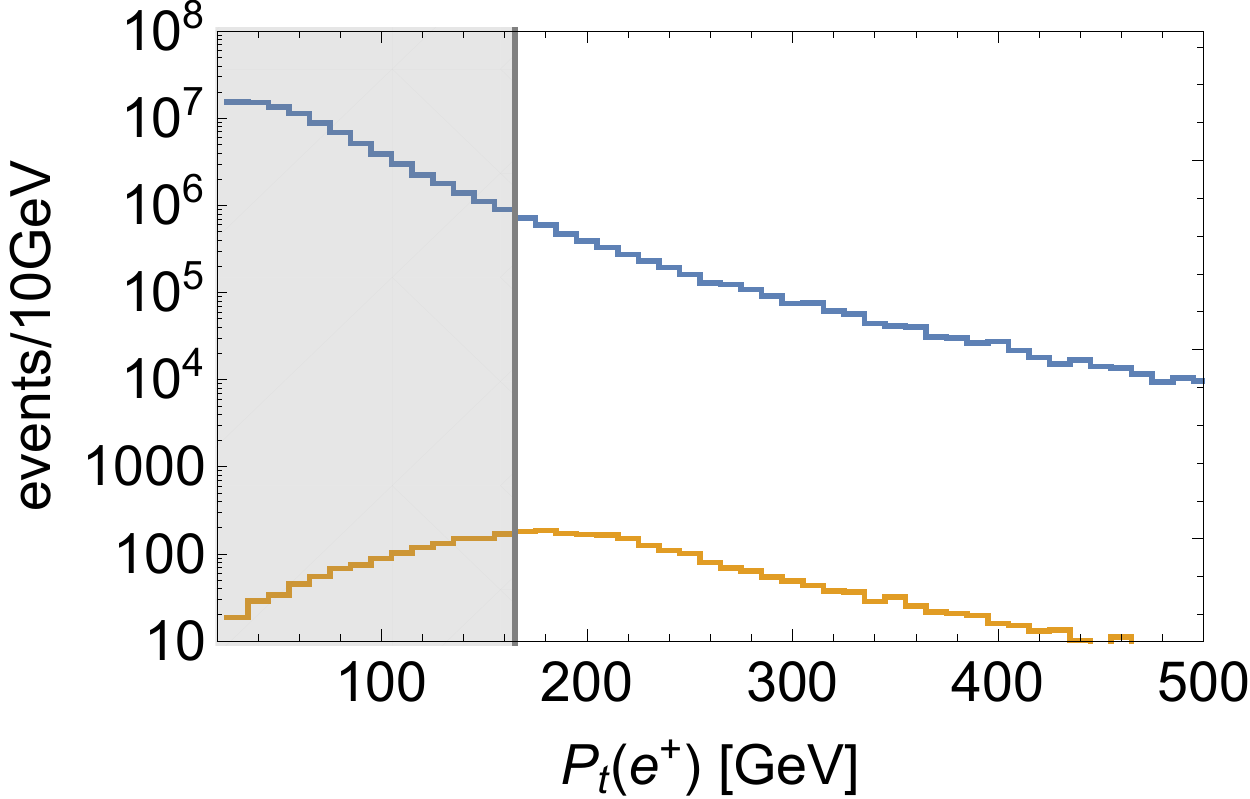}
\includegraphics[width=0.3\textwidth]{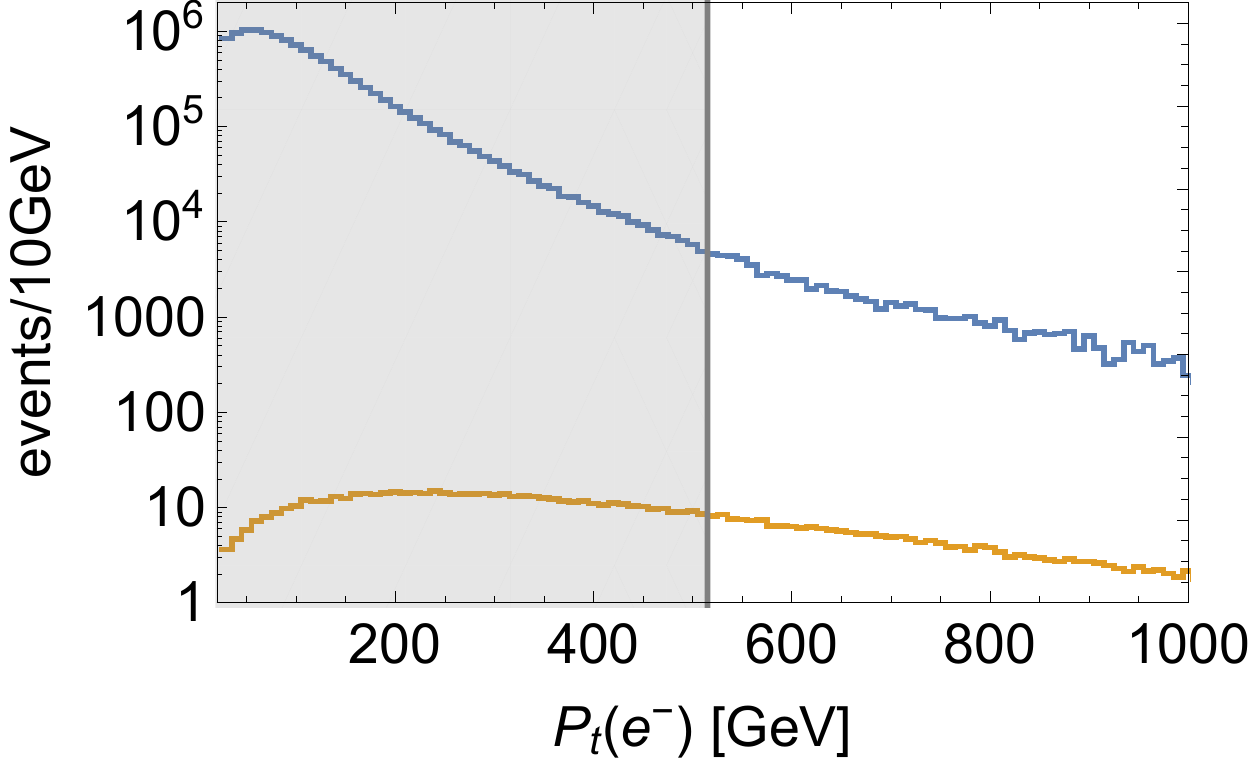}

\begin{center}
$M=200$ GeV \hspace{0.175\textwidth} $M=400$ GeV \hspace{0.175\textwidth} $M=1000$ GeV  \\

$p\,p \to e^-\nu Z$ \hspace{0.185\textwidth} $p\,p \to e^-e^+ W^- $ \hspace{0.185\textwidth} $p\,p \to e^-\nu h$  \\
\end{center}

\vspace{20pt}
\includegraphics[width=0.3\textwidth]{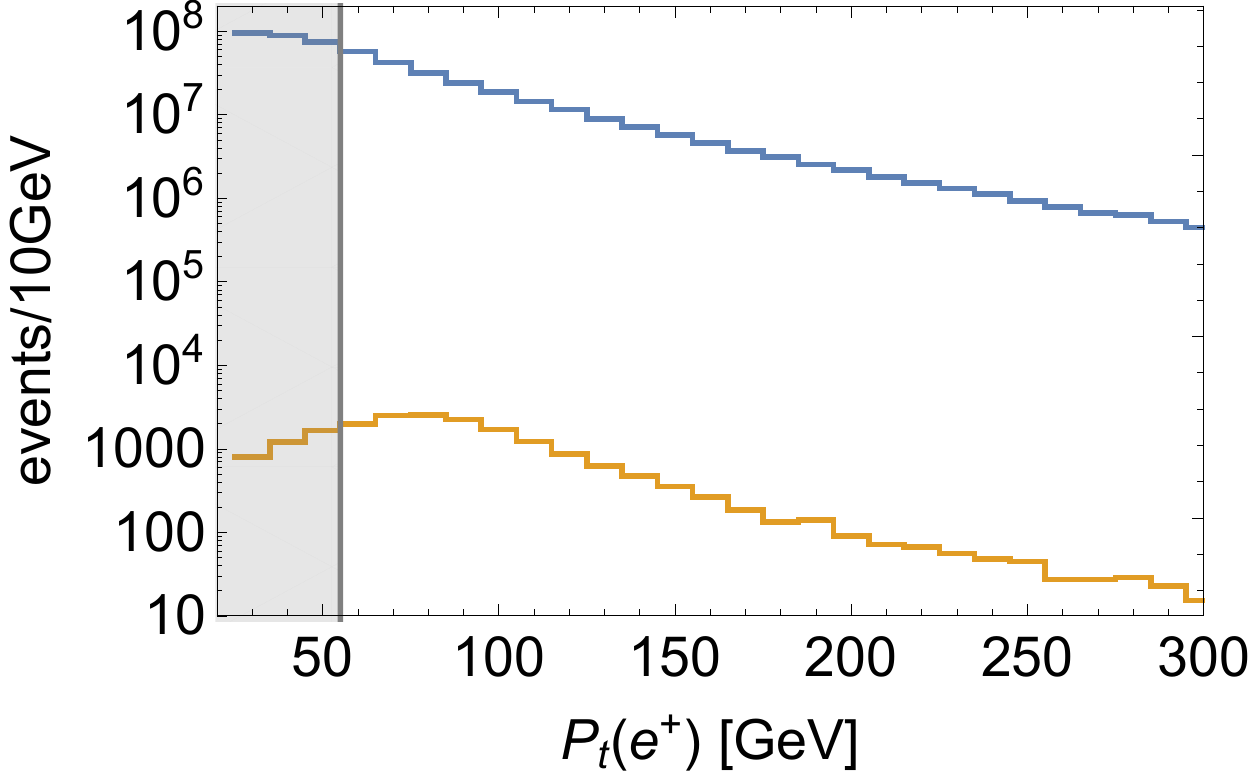}
\includegraphics[width=0.3\textwidth]{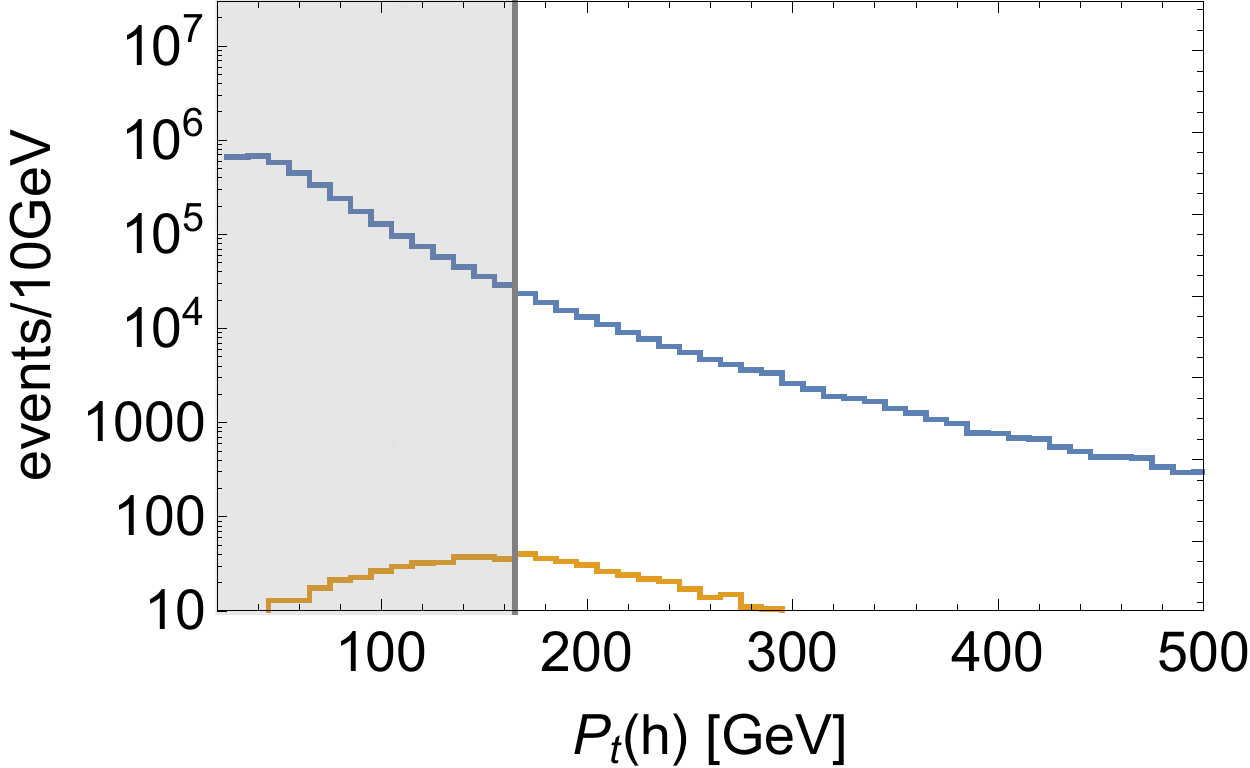}
\includegraphics[width=0.3\textwidth]{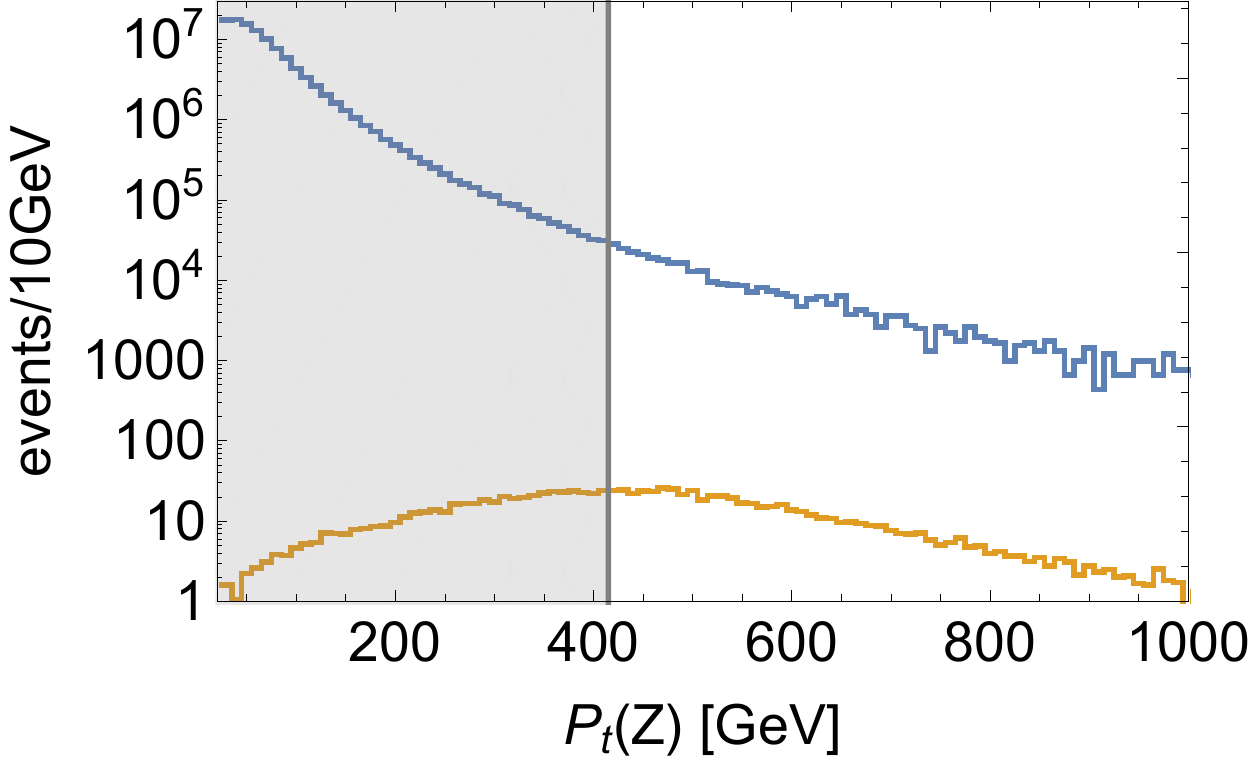}

\begin{center}
$M=200$ GeV \hspace{0.175\textwidth} $M=400$ GeV \hspace{0.175\textwidth} $M=1000$ GeV  \\

$p\,p \to \nu e^+ W^-$ \hspace{0.19\textwidth} $p\,p \to \nu \nu h$ \hspace{0.21\textwidth} $p\,p \to \nu \nu Z$  \\
\end{center}
\caption{Example transverse momentum distributions from processes contributing to the production of $Z,W$ and Higgs boson in $pp$ collisions. The blue and orange line denote the SM background and the signal, respectively, the gray shaded areas indicate the cuts. The upper and lower panel correspond to the final states from the heavy neutrino production channel $\mathbf{W_s}$ and $\mathbf{Z_s}$, respectively. The y axis is scaled to the expected number of events for a total integrated luminosity of 1 ab$^{-1}$.}
\label{fig:distribution-pp}
\end{figure*}

From the intermediate states we maximise the significance by cuts on the transverse momentum of a charged lepton if one is present. For the intermediate states $\nu\nu Z,$ and $\nu\nu h$ we optimise the transverse momentum of the $Z$, or Higgs boson.
We show some example distributions for the SM background and the signal in fig.\ \ref{fig:distribution-pp} for the benchmark masses $M=200,\,400$ and 1000 GeV at the FCC-hh/SppC with an integrated luminosity of 20 ab$^{-1}$.
Therein, the active-sterile mixing angle is tuned to the value that results in a significance of 1 after the kinematic cuts, which are shown by the shaded areas.
We display a list of the optimised cuts on the considered transverse momentum of the charged lepton (or the $Z$ or Higgs boson if none is present) in tab.\ \ref{tab:cuts-pp}.

To obtain the event numbers of the final state from the intermediate state, we multiply with the appropriate branching ratios in eq.\ \eqref{eq:branchings}.

We discuss the lepton-number-conserving signature lepton-dijet (plus missing energy) at a $pp$ collider as an example. It receives contributions from the production-and-decay channels $\nu \ell W$, $\nu \ell Z$, and $\nu \ell h$, with individual sensitivities $s_{\nu\ell W}$, $s_{\nu \ell Z}$, and $s_{\nu \ell h}$, respectively.
We combine their sensitivities using the hadronic branching ratios of the $W$ and $Z$ boson:
\begin{equation}
s_{\ell\nu jj}^{-1} = \sqrt{\frac{\rm{Br}(W\to jj)}{s_{\nu \ell W}^2}+\frac{\rm{ Br}(Z\to jj)}{s_{\nu \ell Z}^2}+\frac{\rm{ Br}(h\to jj)}{s_{\nu \ell h}^2}}\,.
\end{equation}

\begin{figure}
\includegraphics[height=0.2\textwidth]{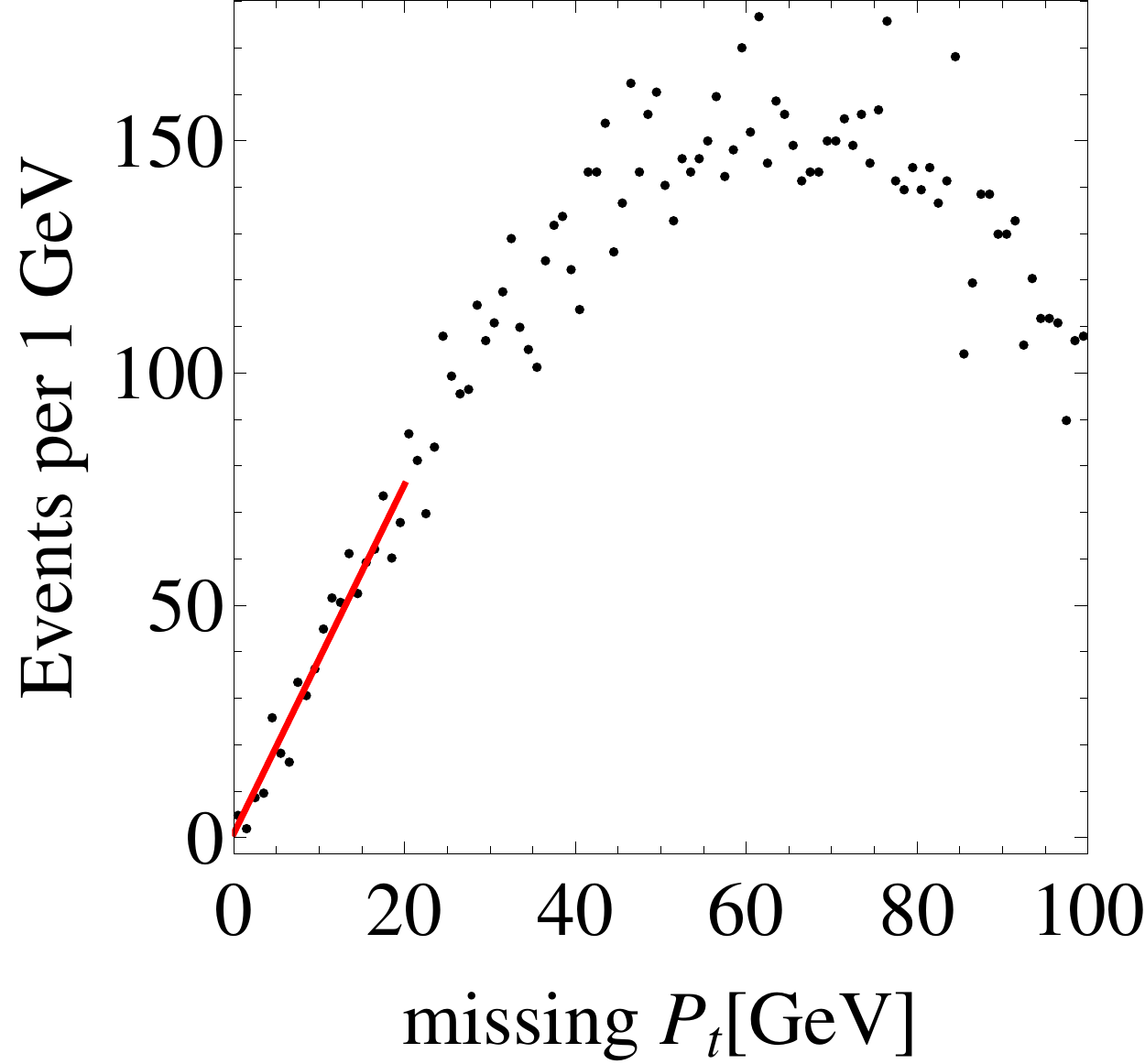}
\includegraphics[height=0.205\textwidth]{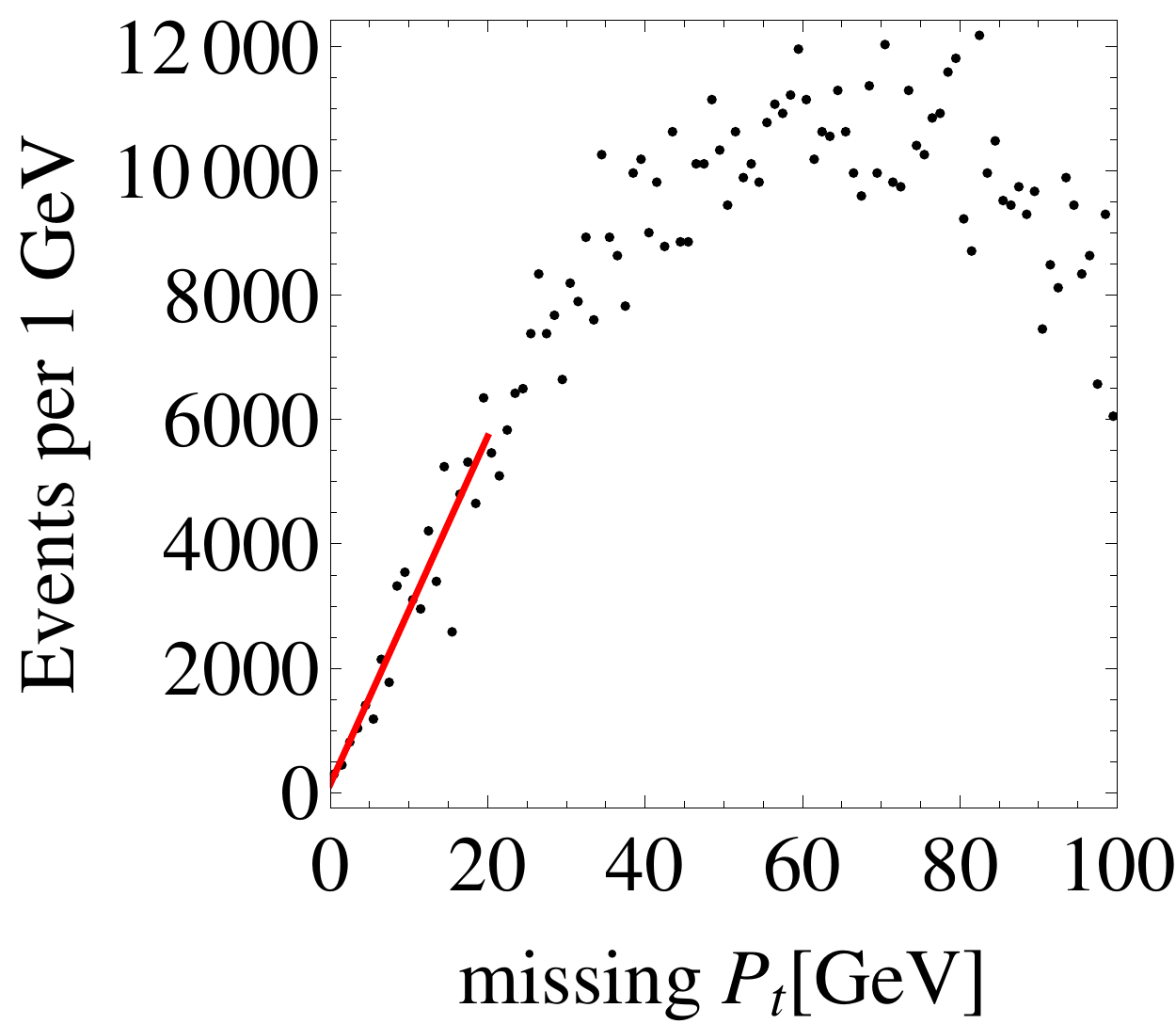}

\vspace{-40pt}\hspace{80pt} LHC \hspace{60pt} FCC-hh/SppC \vspace{40pt}
\caption{Distribution of the missing transverse momentum of the visible fermions from the process $pp \to t \bar t \to \ell \nu \ell \nu b \bar b$, at 14 and 100 TeV, from samples with $10^6$ events generated with WHIZARD. The y axis is scaled to the expected number of events per 1 GeV for 3 and 20 ab$^{-1}$, respectively.
The red line denotes the linear fit of the distribution up to 20 GeV.}
\label{fig:lljj-ditop}
\end{figure}
The sensitivity of the lepton-number-conserving and lepton-flavour-violating final state $\ell_\alpha^\pm \ell_\beta^\mp jj$ is obtained without optimising kinematic cuts. As irreducible background we consider the process $pp\to t \bar t$, with the top quarks yielding b-jets and $W$ bosons. We multiply the production cross section with the leptonic branching ratio of the $W$ bosons, and consider a b-tag veto with an efficiency of 0.3 per jet.
We simulate the final state $\ell\ell \nu \nu b b$, of which we display the missing transverse momentum in fig.\ \ref{fig:lljj-ditop} for the LHC and the FCC-hh/SppC. We veto all the events with missing $P_t> 20$ GeV, which results in $\sim 800$ and $\sim 6 \times 10^4$ events for the LHC and the FCC-hh/SppC, respectively.

\begin{table*}
\centering
$\sqrt{s}=14$~TeV

\begin{tabular}{|c|c|c|c|c|c|c|}
\hline
$M$ & $e^-\nu Z$ & $e^-\nu h$ & $e^- e^+ W^-$ & $\nu\nu Z$ & $\nu\nu h$ & $\nu e^+ W^-$ \\
\hline\hline
200	& $>$	90	& $>$	60	& $>$	80 & $>$	90	& $>$	70	& $>$	70\\
400	& $>$	160	& $>$	160	& $>$	170& $>$	170	& $>$	160	& $>$	160\\
700	& $>$	300	& $>$	300	& $>$	310& $>$	310	& $>$	300	& $>$	300\\
1300	& $>$	370	& $>$	570	& $>$	570	& $>$	520	& $>$	600	& $>$	600\\
\hline
\end{tabular}

\medskip

\centering
$\sqrt{s}=100$~TeV

\begin{tabular}{|c|c|c|c|c|c|c|}
\hline
$M$ & $e^-\nu Z$ & $e^-\nu h$ & $e^- e^+ W^-$ & $\nu\nu Z$ & $\nu\nu h$ & $\nu e^+ W^-$ \\
\hline\hline
200	& $>$	80	& $>$	30 & $>$		60	& $>$	70	& $>$	30 	& $>$	60	\\
400	& $>$	220	& $>$	200 & $>$	170	& $>$	140	& $>$	120 & $>$	120	\\
1000	& $>$	470	& $>$	520 & $>$	440	& $>$	420	& $>$	420 & $>$	420	\\
2000	& $>$	720	& $>$	970 & $>$	920	& $>$	820	& $>$	820 & $>$	820	\\
\hline
\end{tabular}

\caption{Momentum cuts for the signatures at $pp$ colliders. Selected intervals for the transverse momentum for a lepton, where one is present, and on the $Z$ and Higgs boson (candidates), where none is present. All numbers are in units of GeV.}
\label{tab:cuts-pp}
\end{table*}

\subsection{Electron-proton collisions}
\label{app:distributions-ep}

The list of signatures for which we calculate the sensitivities is given in tab.\ \ref{tab:signatures_ep}. Therein, each production-and-decay channel $x_i$ contributes separately, and we simulate signal and background events to obtain the individual sensitivities $s_{x_i}$, as discussed in sec.\ \ref{app:calculation_sensitivity}.

In the narrow width approximation it is sufficient to consider $Z,\,W,$ and Higgs bosons that are produced on the mass shell as ``intermediate states''.

We consider an unpolarised electron beam at 60 GeV and a proton beam  with an energy of 7 and 50 TeV for the LHeC and FCC-eh, respectively. We use the built-in PDFs CTEQ6L and MRST2004QEDP for $\mathbf{W_t^{(q)}}$ and $\mathbf{W_t^{(\gamma)}}$, respectively. 

For the LHeC we consider benchmark values for the masses of 200, 400, 700, 900 and 1100 GeV. For the FCC-eh, we consider the benchmark masses of 200, 400, 1000, 1500, 2000, 2500, and 3000 GeV. The cross sections of the SM background processes are listed in tab.\ \ref{tab:ep}.

The significance is maximized by cuts on the transverse momentum of the intermediate bosons. Example distributions for the signal and background are shown in fig.~\ref{app:distributions-ep} for benchmark masses $M=200,\,400$ and 700 GeV for the LHeC, and 400, 1000, and 1500 GeV for the FCC-eh.
The active-sterile mixing angle is tuned to the value that results in a significance of 1 after the kinematic cuts, which are shown by the shaded areas.
The optimzed cuts are displayed in tab.~\ref{tab:cuts-ep}.

\begin{figure}
\includegraphics[height=0.2\textwidth]{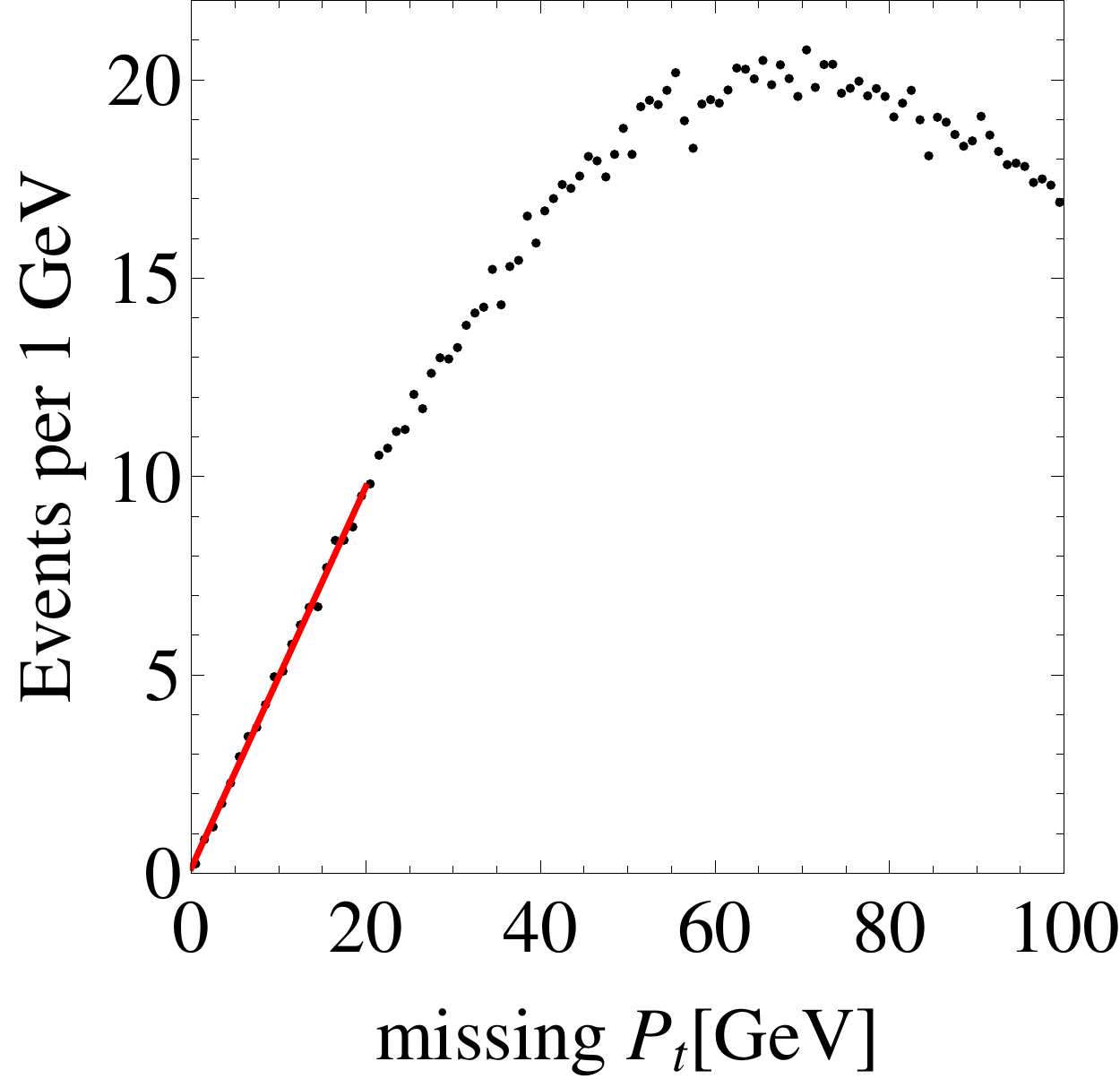}
\includegraphics[height=0.21\textwidth]{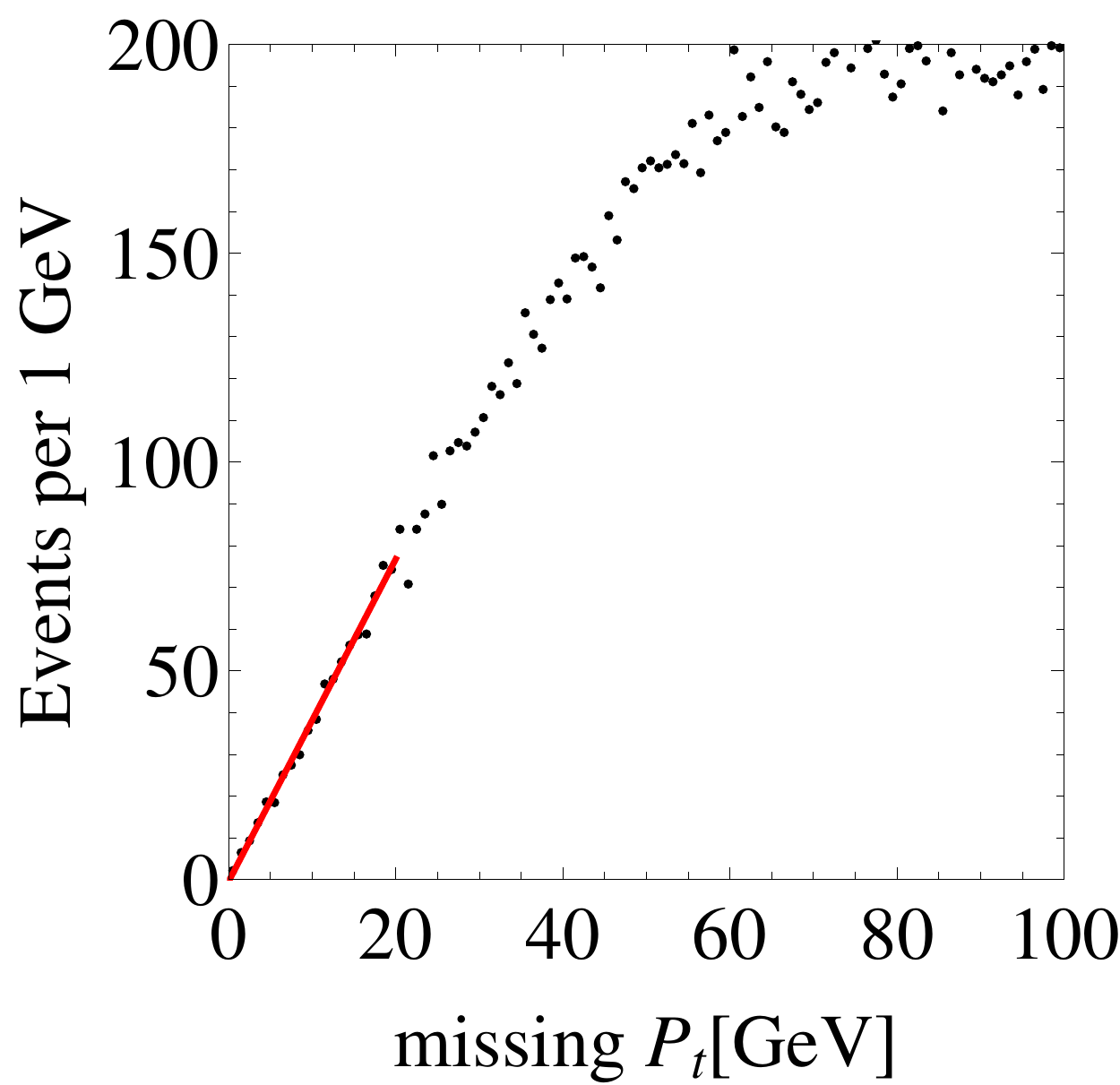}

\vspace{-40pt}\hspace{70pt} LHeC \hspace{80pt} FCC-eh\vspace{40pt}
\caption{Distribution of the missing transverse momentum of the visible fermions from the process $e^-p \to j \ell_\alpha^- jjj \nu \nu,\, \alpha\neq e$, for proton beam energies of 7 and 50 TeV, from samples with $10^6$ events generated with WHIZARD. The y axis is scaled to the expected number of events per 1 GeV for 1 ab$^{-1}$ in both panels.
The red line denotes the linear fit of the distribution up to 20 GeV.}
\label{fig:ljjjSM}
\end{figure}
For the lepton-number-conserving and lepton-flavour-violating final state $j \mu^- \tau^+ \nu$ we considered the SM background to be given by the higher order process $e^- p\to j \mu^- \tau^+ \nu\nu\nu$. We optimized its sensitivity via the transverse momentum of the $\mu^-$.
The sensitivity for the lepton-number-conserving lepton-flavour-violating signature $\ell_\alpha^- jjj, \, \alpha \neq e$ is obtained without kinematic cuts. As irreducible backgrounds we consider the processes $e^- p \to j W^- V \nu$, with the leptonic decays of the $W^-$ and the hadronic decays of the vector boson $V=W^\pm,\,Z$.
We simulated $10^6$ events for the three backgrounds and display the resulting distribution of the missing transverse momentum in fig.\ \ref{fig:ljjjSM}.
We veto all the events with missing $P_t> 20$ GeV, which results in $\sim 100$ and $\sim 800$ events for the LHeC and the FCC-eh, respectively.

\begin{table}
\begin{center}
\begin{tabular}{|c|c|c|}
\hline
Background		& LHeC [fb]	&	FCC-eh [fb]	\\
\hline\hline
$\nu\, Z\,j$ 	& 490.		&	2550.	\\
\hline
$\nu\, h\,j$ 	& 86.0		&	490.		\\
\hline
$e^- W^+j$ 		& 1285.		&	6000.	\\
\hline
$j \mu^-\tau^+ \nu\nu\nu$ & 0.81	& 3.5	\\
\hline
$j W^- V \nu$	& 7.1		&	103	\\
\hline
\end{tabular}
\end{center}
\caption{Background cross sections for the intermediate states to the sterile neutrino signatures at $e^-p$ colliders.
For the simulations final state electrons are required to have transverse momenta above 10 GeV, and the momentum transfer between initial state parton and final state jet is required to be larger than 20 GeV (to avoid numerical singularities). The SM processes $j \mu^-\tau^+ \nu\nu\nu$ and $j W^- V \nu,$ with $V=W^\pm,Z$ constitute the background for the lepton-number-conserving lepton-flavour-violating final states $j \mu^-\tau^+ \nu$ and $\mu^-jjj$, respectively.}
\label{tab:ep}
\end{table}

\begin{figure*}

\includegraphics[width=0.3\textwidth]{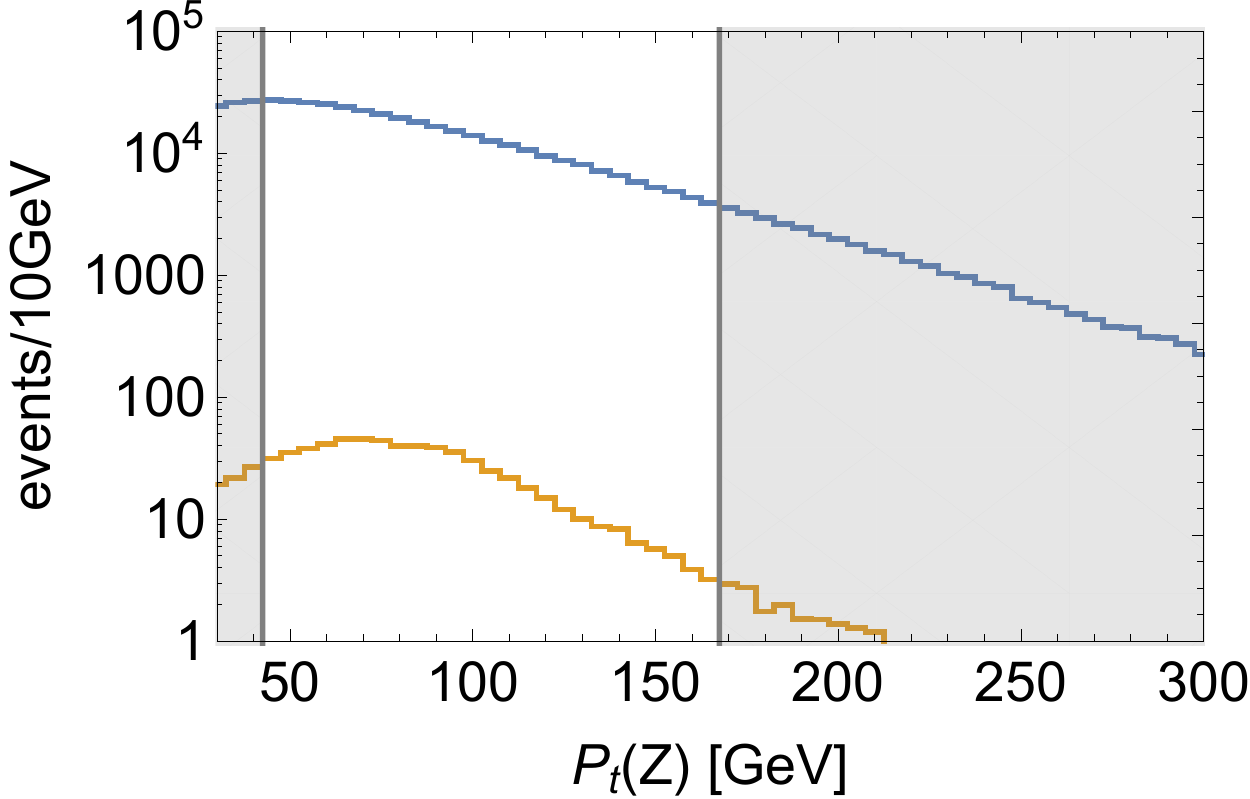}
\includegraphics[width=0.3\textwidth]{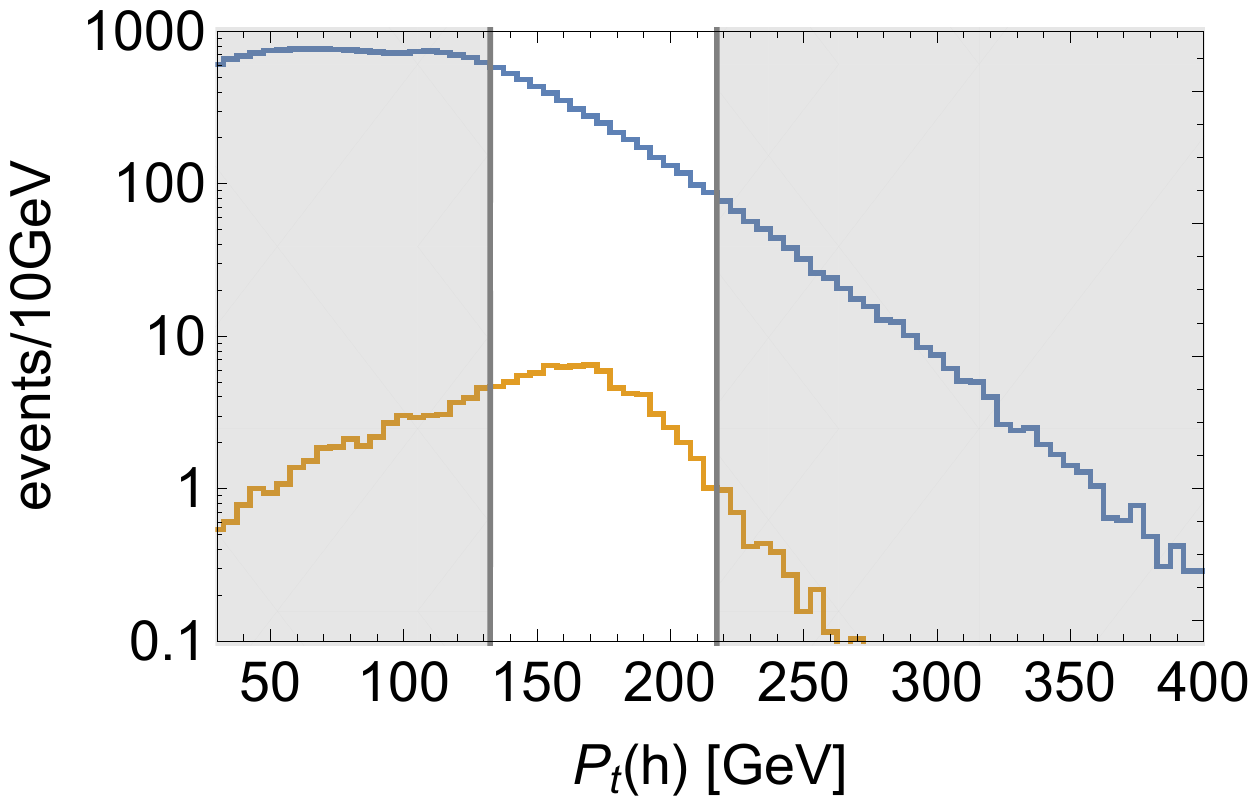}
\includegraphics[width=0.3\textwidth]{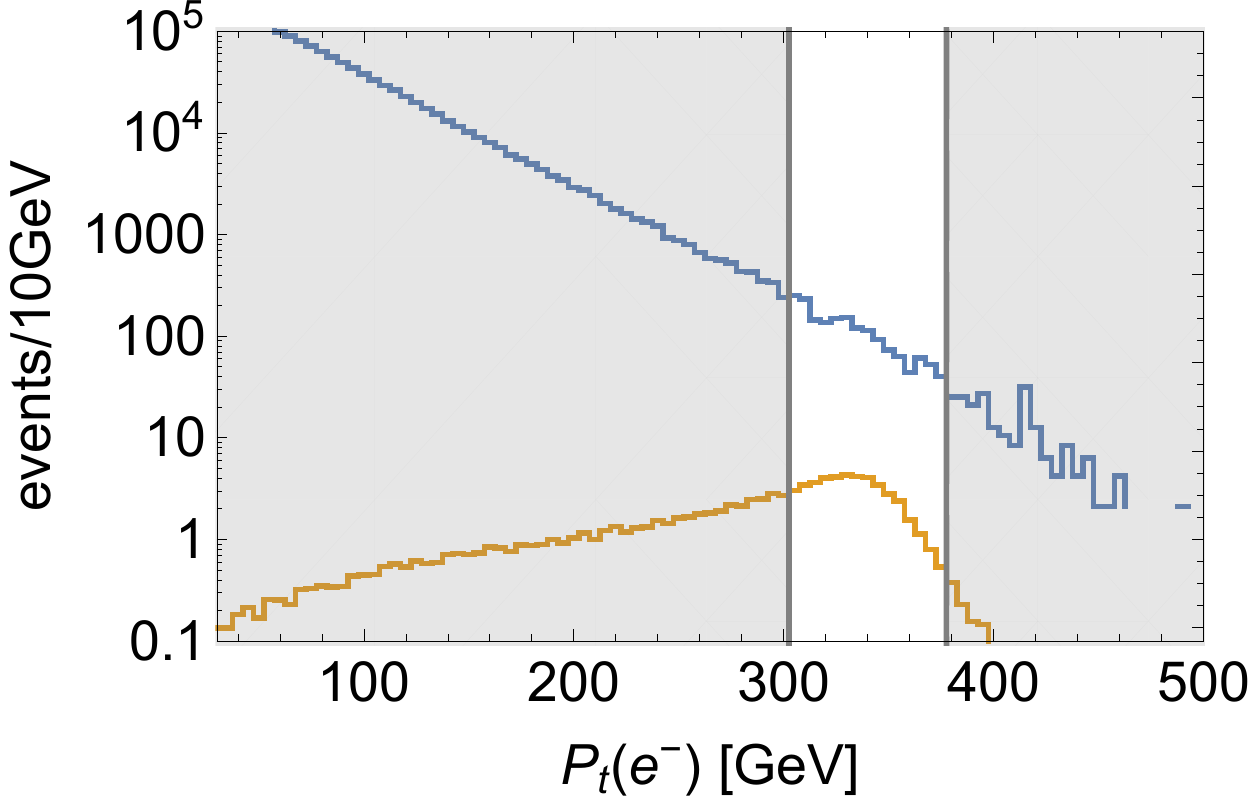}

\begin{center}
$M=200$ GeV \hspace{0.175\textwidth} $M=400$ GeV \hspace{0.175\textwidth} $M=700$ GeV  \\

$e^-\,p \to \nu Z\,j$ \hspace{0.195\textwidth} $e^-\,p \to \nu h\,j$ \hspace{0.18\textwidth} $e^-\,p \to e^- W^+\,j$  \\
\end{center}

\vspace{20pt}
\includegraphics[width=0.3\textwidth]{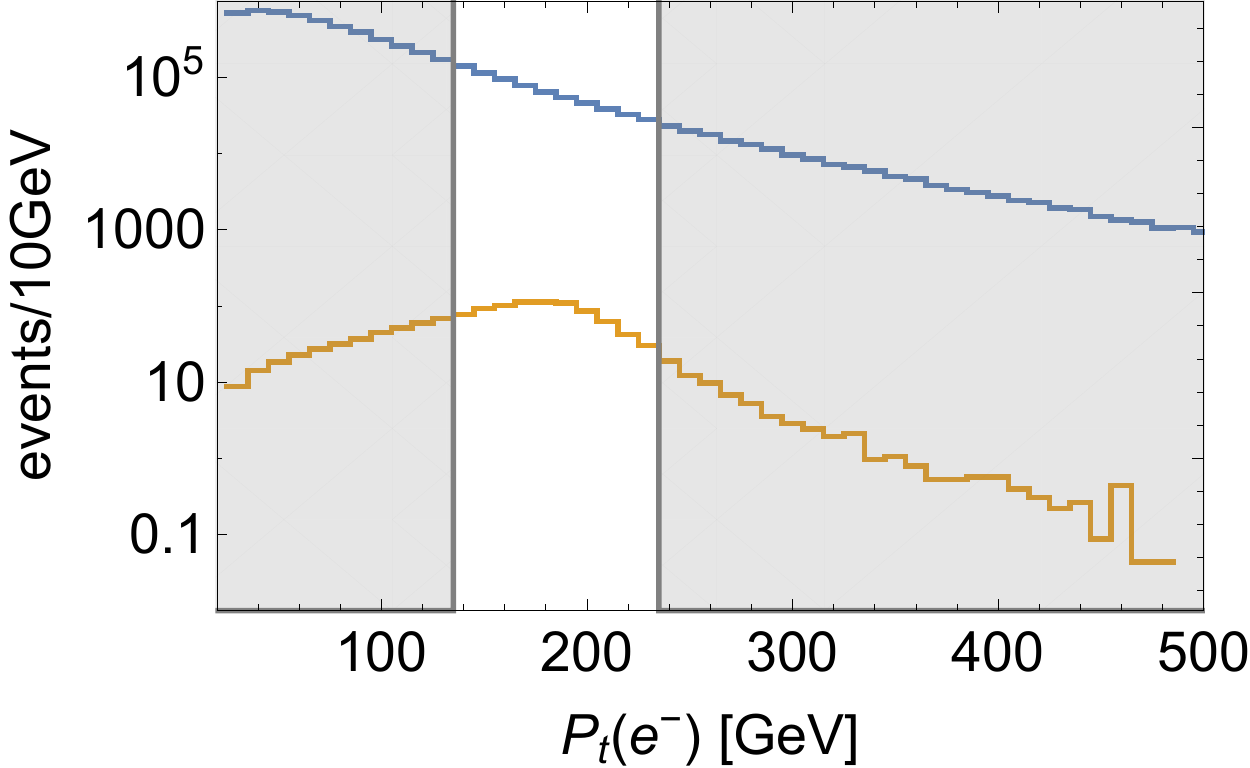}
\includegraphics[width=0.3\textwidth]{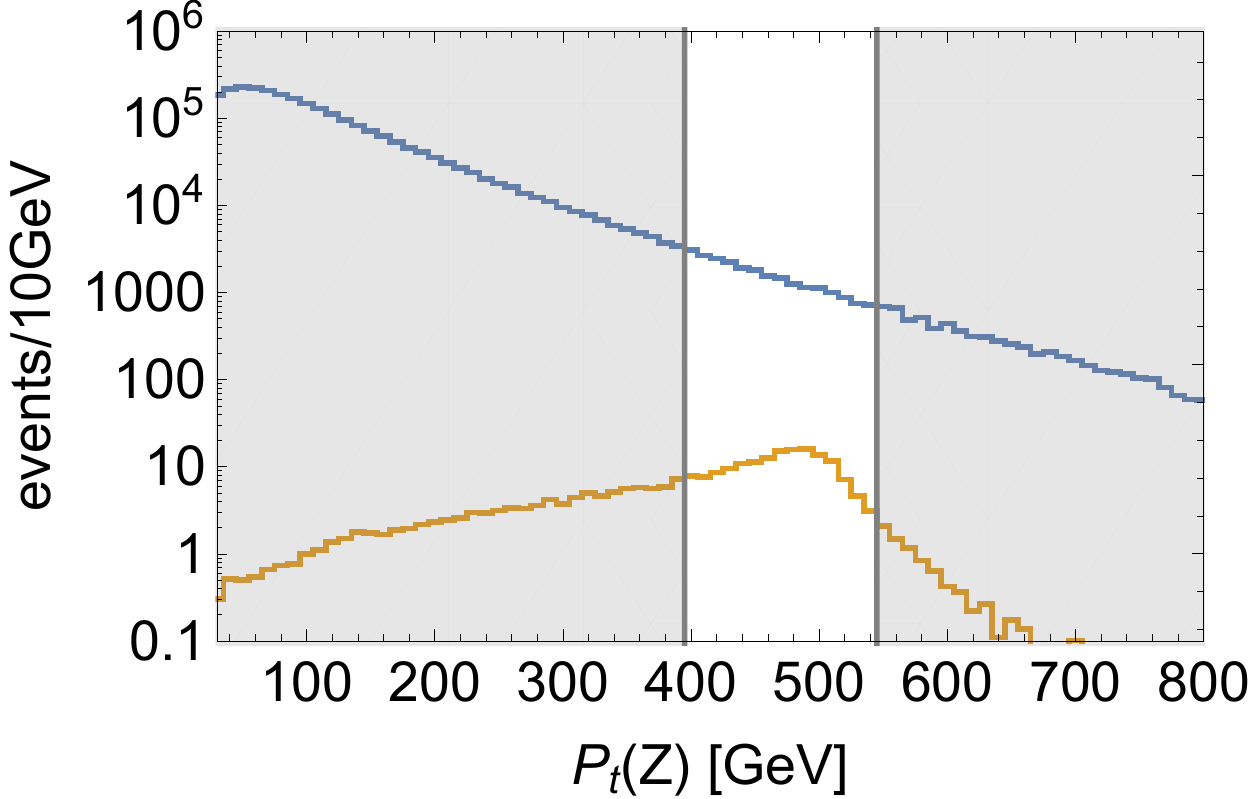}
\includegraphics[width=0.3\textwidth]{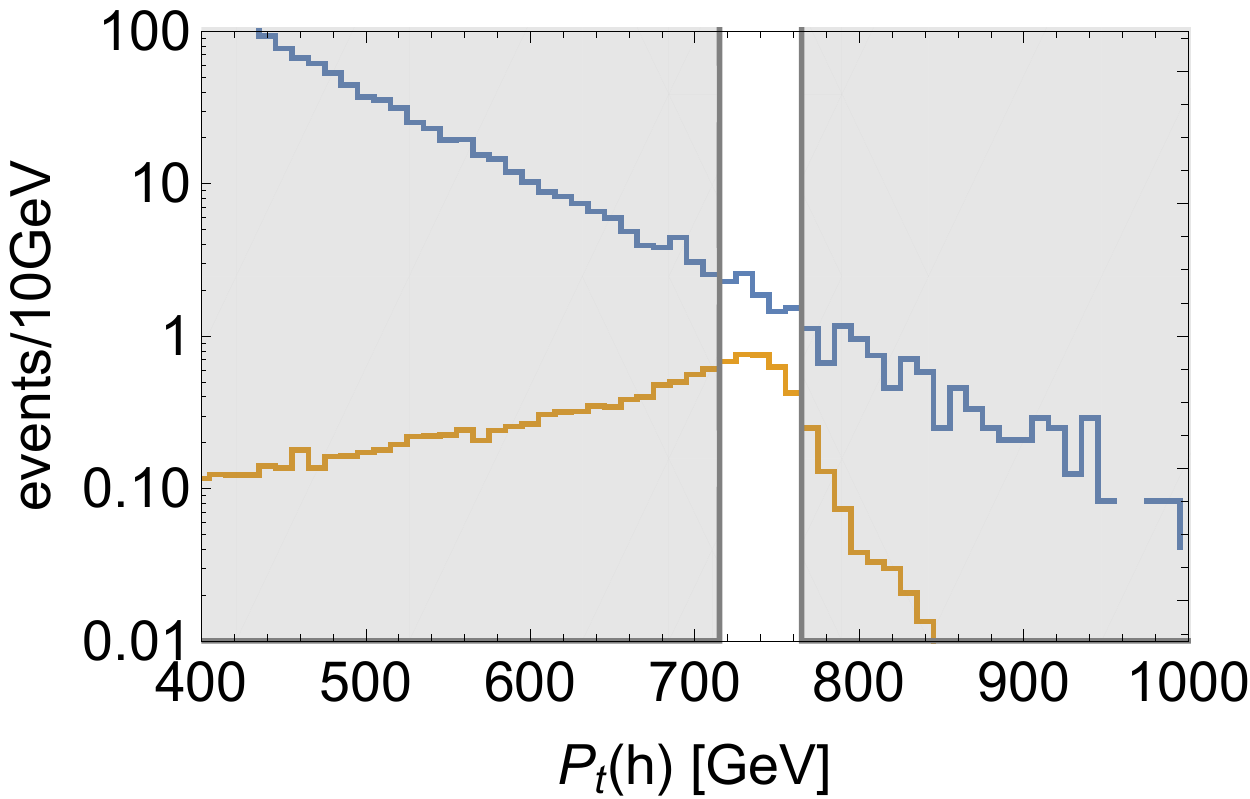}

\begin{center}
$M=400$ GeV \hspace{0.175\textwidth} $M=1000$ GeV \hspace{0.175\textwidth} $M=1500$ GeV  \\

$e^-\,p \to e^- W^+\,j$ \hspace{0.19\textwidth} $e^-\,p \to \nu Z\,j$ \hspace{0.20\textwidth} $e^-\,p \to \nu h\,j$  \\
\end{center}
\caption{Example transverse momentum distributions from processes that contribute to the production of $Z,W$ and Higgs boson in electron-proton collisions. The blue and orange line denote the SM background and the signal, respectively, the gray shaded areas indicate the cuts. The upper and lower panels correspond to the LHeC and the FCC-eh, respectively. The y axis is scaled to the expected number of events for a total integrated luminosity of 1 ab$^{-1}$.}
\label{fig:distribution-ep}
\end{figure*}

\begin{table*}
\centering
LHeC

\begin{tabular}{|c|c|c|c|c|c|}
\hline
$M$ &	$j\nu Z$	& $j\nu h$	& $je^- W^+$ & $\mu^-\tau^+j \nu$ & $e^-\nu\nu h$	\\
\hline\hline
200	& [	70	,	320	] & [	30	,	130	] & [	60	,	210	]  & [50 ,140 ] & [0,120]\\
400	& [	270	,	470	] & [	250	,	420	] & [	270	,	470	]  & [140 , 240] & [130,320]\\
700	& [	520	,	770	] & [	600	,	740	] & [	590	,	740	]  & [ 320, 440] & [270,420]\\
900	& [	820	,	980	] & [	780	,	920	] & [	820	,	920	]  & [ 420, 620] & [370,520]\\
\hline
\end{tabular}

\medskip
\centering
FCC-eh

\begin{tabular}{|c|c|c|c|c|c|}
\hline
$M$ &	$j\nu Z$	& $j\nu h$	& $je^- W^+$ & $\mu^-\tau^+j \nu$ & $e^-\nu\nu h$	\\
\hline\hline
200	 & [	30	,	180	] & [	30	,	80	] & [	30	,	230	] & [70 , 150] & [0,140]\\
400	 & [	140	,	260	] & [	110	,	220	] & [	140	,	240	]  & [150 , 240] & [140,270]\\
1000	 & [	400	,	550	] & [	440	,	520	] & [	400	,	580	]  & [400 , 580] & [440,570]\\
1500	 & [	620	,	870	] & [	720	,	770	] & [	620	,	820	]  & [620 , 820] & [600,$\infty$]\\
2000	 & [	870	,	1070	] & [	970	,	1020	] & [	870	,	1070	]  & [870 , 1220] & [600,$\infty$]\\
2500	 & [	1120	,	1320	] & [	970	,	1370	] & [	1120	,	1320	]  & [1120 , 1320] & [600,$\infty$]\\
3000 & [	1320	,	1520	] & [	1220	,	1620	] & [	1320	,	1620	]  & [1320 , 1620] & [600,$\infty$]\\
\hline
\end{tabular}
\caption{Momentum cuts for the signatures at $e^-p$ colliders. Selected intervals for the transverse momentum for a lepton, where one is present, and on the $Z$ and Higgs boson (candidates), where none is present. All numbers are in units of GeV.}
\label{tab:cuts-ep}
\end{table*}

\newpage\mbox{}
\newpage

\bibliographystyle{unsrt}

\end{document}